\documentclass[aps,prc,twocolumn,superscriptaddress,showpacs,amsmath,amssymb,nofootinbib]{revtex4}	%,nobibnotes,,nofootinbib,preprintnumbers,amsmath,amssymb
\usepackage{graphicx,epsfig,longtable}
\usepackage{mathptmx}
\usepackage[pdftex]{hyperref}

\begin{document}

\title{Analysis of possible systematic errors in the Oslo method}

\author{A.~C.~Larsen}
\email{a.c.larsen@fys.uio.no}
\affiliation{Department of Physics, University of Oslo, N-0316 Oslo, Norway}
\author{M.~Guttormsen}
\affiliation{Department of Physics, University of Oslo, N-0316 Oslo, Norway}
\author{M.~Krti\u{c}ka}
\affiliation{Institute of Particle and Nuclear Physics, Charles University, Prague, Czech Republic}
\author{E.~B\v{e}t\'{a}k}
\affiliation{Institute of Physics SAS, 84511 Bratislava, Slovakia}
\affiliation{Faculty of Philosophy and Science, Silesian University, 74601 Opava, Czech Republic}
\author{A.~B\"{u}rger}
\affiliation{Department of Physics, University of Oslo, N-0316 Oslo, Norway}
\author{A.~G\"{o}rgen}
\affiliation{Department of Physics, University of Oslo, N-0316 Oslo, Norway}
%\affiliation{Dapnia/SPhN, CEA-Saclay, France}
\author{H.~T.~Nyhus}
\affiliation{Department of Physics, University of Oslo, N-0316 Oslo, Norway}
\author{J.~Rekstad}
\affiliation{Department of Physics, University of Oslo, N-0316 Oslo, Norway}
\author{A.~Schiller}
\affiliation{Department of Physics and Astronomy, Ohio University, Athens, Ohio 45701, USA}
\author{S.~Siem}
\affiliation{Department of Physics, University of Oslo, N-0316 Oslo, Norway}
\author{H.~K.~Toft}
\affiliation{Department of Physics, University of Oslo, N-0316 Oslo, Norway}
\author{G.~M.~Tveten}
\affiliation{Department of Physics, University of Oslo, N-0316 Oslo, Norway}
\author{A.~V.~Voinov}
\affiliation{Department of Physics and Astronomy, Ohio University, Athens, Ohio 45701, USA}
\author{K.~Wikan}
\affiliation{Department of Physics, University of Oslo, N-0316 Oslo, Norway}

\date{\today}

\begin{abstract}
In this work, we have reviewed the Oslo method,  
which enables the simultaneous extraction of level density and $\gamma$-ray transmission 
coefficient from a set of particle-$\gamma$ coincidence data. 
Possible errors and uncertainties have been investigated. Typical data sets from 
various mass regions as well as simulated data have been tested against the assumptions behind 
the data analysis.

\end{abstract}

% PACS Numbers: 21.10.Ma (level density), 
%               21.10.-k (Properties of nuclei; nuclear energy levels), 
%               21.60.Jz (HF & QRPA)
% 		25.20.Lj (Photoproduction reactions)
%		24.30.Gd Other resonances
%		25.55.Hp	3He transfer reactions
%		25.40.Ep	Inelastic proton scattering
%		27.40.+z	39 ² A ² 58
%		27.50.+e	59 ² A ² 89
% From Voinov's PRL: 25.40.Lw, 25.20.Lj, 25.55.Hp, 27.40.+z

\maketitle

\section{Introduction}
\label{sec:intro}

Nuclear level densities and $\gamma$-ray strength functions are two 
indispensable quantities for many nuclear structure studies and 
applications. 
The approach developed by the nuclear physics group at the Oslo Cyclotron Laboratory 
(OCL), was first described in 1983~\cite{rekstad1}, 
and has during a period of nearly 30 years been refined and extended to the 
sophisticated level presently referred to as the Oslo method. 
The method is designed to extract the nuclear level 
density and the $\gamma$-ray transmission coefficient (or $\gamma$-ray 
strength function) up to the neutron (proton) threshold from particle-$\gamma$ 
coincidence data. Typical reactions that have been utilized are transfer 
reactions such as ($^3$He, $\alpha \gamma$) and ($p,t\gamma$), and 
inelastic scattering reactions, e.g., ($^3$He, $^3$He$^{\prime}\gamma$) 
and (p,p$^{\prime}\gamma$). 

The measurements have been very successful and have shed new light on 
important issues in nuclear structure, such as the $M1$ scissors mode~\cite{SchillerM1,undraa,bagheri,hildes_DyRSF}, 
and the sequential breaking of Cooper pairs~\cite{melb0,Sn_pairbreak}. For the Sn isotopes,
a resonance-like structure that may be due to the so-called $E1$ pygmy resonance has been 
observed~\cite{Sn_RSF,Heidi_Sn}. 
Also, a new, unpredicted low-energy increase in the $\gamma$-ray strength function for $E_\gamma \leq 3 $ MeV 
of medium-mass nuclei has been discovered by use of this method~\cite{Fe_Alex,Fe_Emel,Mo_RSF,V,Sc,magne_46Ti}. This 
enhancement might have a non-negligible impact on stellar reaction rates relevant for the 
nucleosynthesis~\cite{larsen_goriely}. At present, this structure is poorly understood.

In this work, we have investigated the possible systematic errors that can occur in the Oslo method 
due to experimental limitations and the assumptions made in the various steps of the method. 
We have studied experimental data with high statistics (and thus low statistical errors) so that
systematic errors can be revealed. Also, we have used simulated data to enable better control on 
the input parameters (level density and $\gamma$-ray strength function). 
In Sec.~\ref{sec:expmethod} we give a short overview of the experiments and the various steps in the method. 
In Sec.~\ref{sec:err} we present the possible systematic errors for each main step of the method. 
Finally, a summary and concluding remarks are given in Sec.~\ref{sec:sum}.

\section{Experimental procedure and the Oslo method}
\label{sec:expmethod}
The Oslo method is in fact a set of methods and analysis techniques, which together make it
possible to measure level density and $\gamma$-ray strength from particle-$\gamma$ coincidence data. 
In this section, these techniques and methods will be described. 

\subsection{Experimental details}
\label{subsec:exp}

The experiments were performed at the OCL using a light-ion beam delivered by the MC-35 Scanditronix cyclotron. 
Typically, $^3$He beams with energy $30-45$ MeV has been used. Recently, also proton beams with energy $15-32$ 
MeV have been applied. Self-supporting targets enriched to $\approx 95$\% in the isotope of interest, and with 
a thickness of $\approx 2$ mg/cm$^2$ were placed in the center of the multi-detector array CACTUS~\cite{Cactus}. 
CACTUS consists of 28 collimated NaI(Tl) $\gamma$-ray detectors with a total efficiency of 15.2(1)\% 
for $E_{\gamma} = 1332$ keV. Usually also a 60\% Ge detector has been applied to monitor the populated spin range 
($\sim 0 - 8 \hbar$) and possible target contaminations. The experiments were typically run for a period of 1$-$2 weeks 
with beam currents of $\sim 1$ nA.
%-----------------------------------------------------------------%
\begin{figure}[htb]
\begin{center}
\includegraphics[clip,width=\columnwidth]{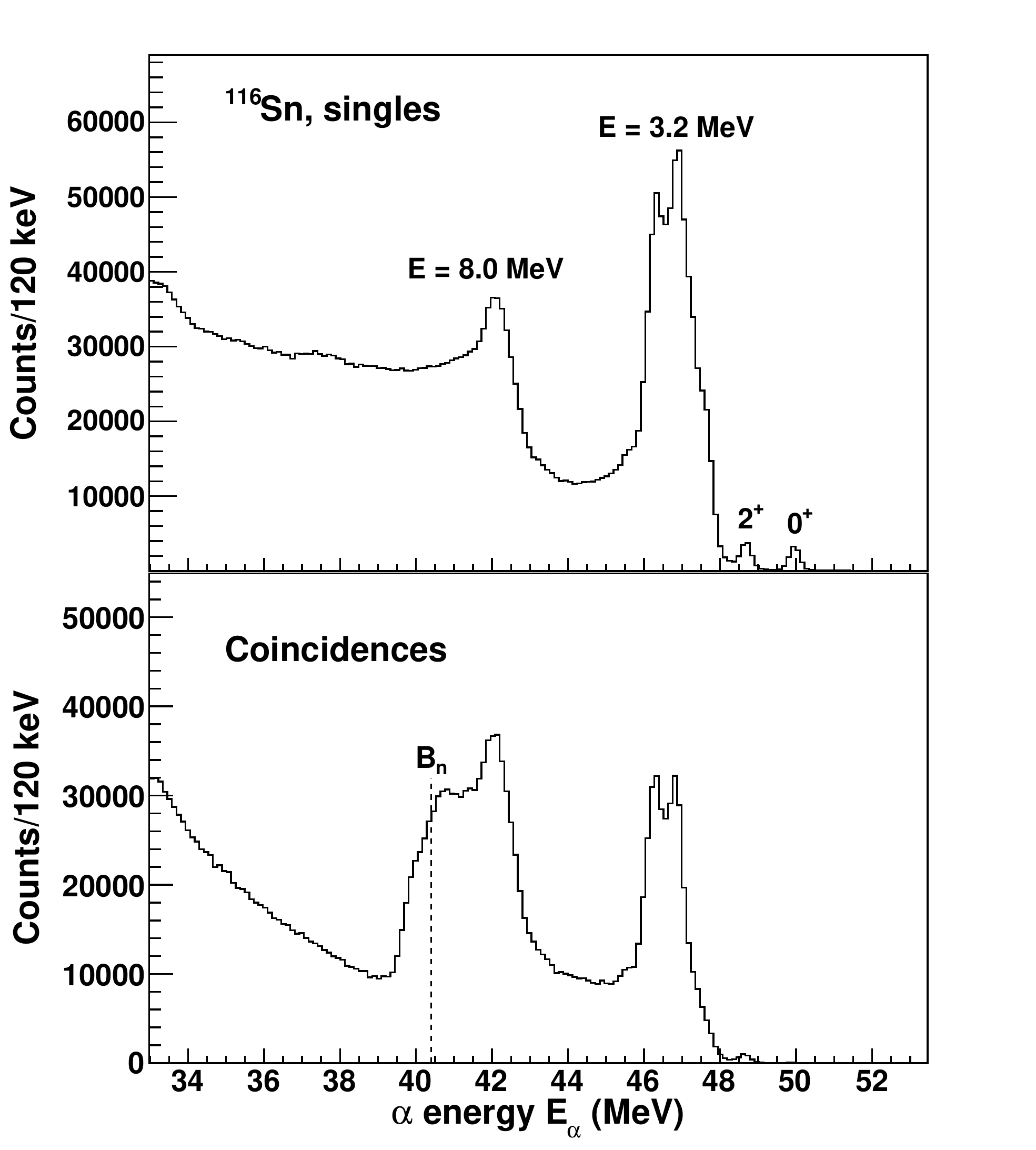}
\caption{Singles $\alpha$-particle spectrum (upper panel), and 
 $\alpha$-$\gamma$ coincidence spectrum (lower panel) from the
$^{117}$Sn($^3$He, $\alpha$)$^{116}$Sn reaction. The data are taken from the experiment
 presented in Ref.~\cite{Sn_pairbreak}.}
\label{fig:alphaSn}
\end{center}
\end{figure}
%-----------------------------------------------------------------%

\begin{figure}[htb]
\includegraphics[clip,width=\columnwidth]{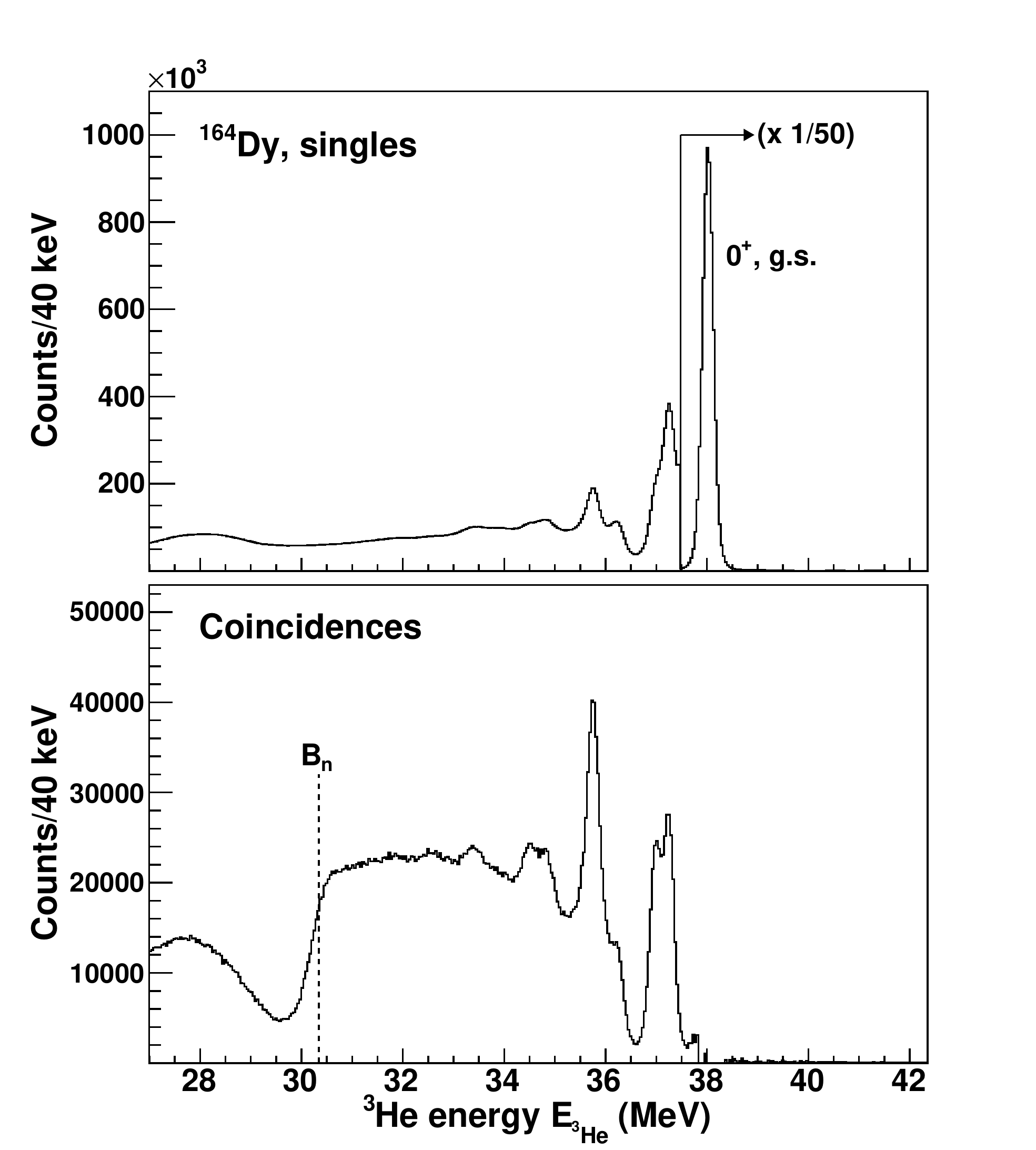}
\caption{Singles $^3$He spectrum (upper panel), and 
 $^3$He-$\gamma$ coincidence spectrum (lower panel) from the
$^{164}$Dy($^3$He,$^3$He$^{\prime}$)$^{164}$Dy reaction. The data are taken from the experiment
 presented in Ref.~\cite{hildes_DyRSF}.}
\label{fig:he3Dy}
\end{figure}

\begin{figure}[htb]
\includegraphics[clip,width=\columnwidth]{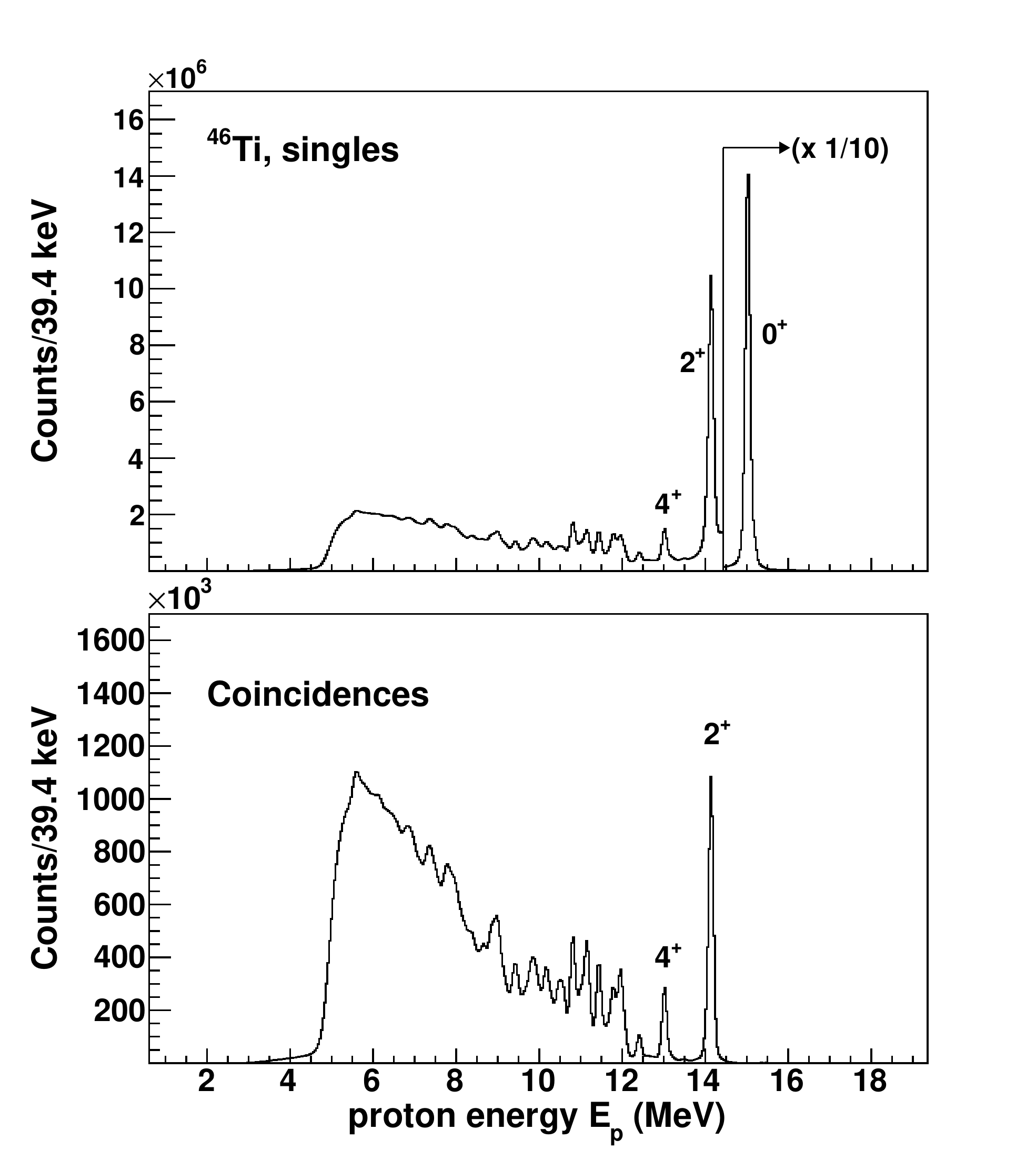}
\caption{Singles proton spectrum (upper panel) and 
 $p$-$\gamma$ coincidence spectrum (lower panel) from the
$^{46}$Ti($p$, $p^{\prime}$)$^{46}$Ti reaction. The data are taken from the experiment
 presented in Ref.~\cite{magne_46Ti}.}
\label{fig:pTi}
\end{figure}

Inside the CACTUS array, eight collimated Si particle detectors were used for detecting the charged 
ejectiles from the nuclear reactions. The particle detectors were placed at 45$^{\circ}$ relative to the beam 
line in forward direction. The detectors were of $\Delta E - E$ type with a thin ($\sim 140$ $\mu$m) front detector 
and a thick ($\sim 1500$ $\mu$m) end detector where the charged ejectiles stop. The particle telescopes enable a 
good separation between the various charged-ejectile species. The energy resolution of the particle spectra 
ranges from $150-300$ keV depending on the beam species, the mass of the target nucleus and the size of the collimators. 
Both singles and coincidence events were measured for the ejectiles. In Figs.~\ref{fig:alphaSn},~\ref{fig:he3Dy} 
and~\ref{fig:pTi} the singles and particle-$\gamma$ coincidence spectra are shown for the reactions 
$^{117}$Sn($^3$He,$\alpha$)$^{116}$Sn, $^{164}$Dy($^3$He,$^3$He$^{\prime}$)$^{164}$Dy 
and $^{46}$Ti($p$, $p^{\prime}$)$^{46}$Ti, respectively. 
%-----------------------------------------------------------------%
%-----------------------------------------------------------------%
%-----------------------------------------------------------------%
%-----------------------------------------------------------------%

\begin{figure*}[htb]
\includegraphics[clip,totalheight=7cm]{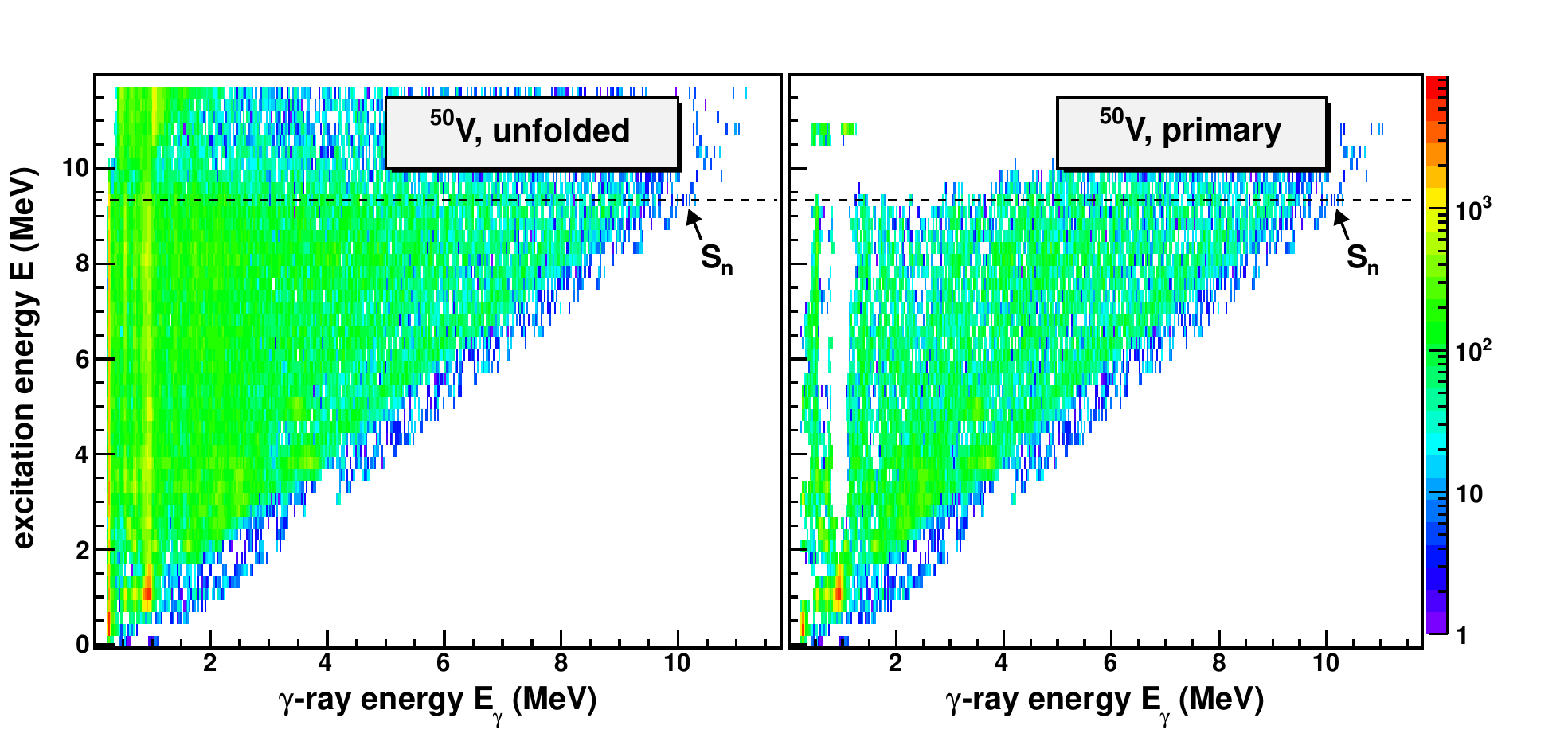}
\caption{(Color online). Excitation-energy vs. $\gamma$-ray energy matrix (left), 
and primary $\gamma$-ray matrix
(right) for $^{50}$V. The data are taken from the experiment
 presented in Ref.~\cite{V}.}
\label{fig:matrix50V}
\end{figure*}

The ejectile energy can easily be transformed to the excitation energy of the residual nucleus using the 
reaction $Q$-values and kinematics. Thus, an excitation energy vs. $\gamma$-ray energy matrix 
can be built from the coincidence events. An example of such a matrix is shown in Fig.~\ref{fig:matrix50V} 
(left panel), 
where the $\gamma$-ray spectra have been corrected for the known response functions of the 
CACTUS array~\cite{Gut96}. The correction (or unfolding) method is described in detail in Ref.~\cite{Gut96}. 
One of the main advantages with this method is that the fluctuations of the original spectra are preserved 
without introducing additional, spurious fluctuations. 
%-----------------------------------------------------------------%
%-----------------------------------------------------------------%

\subsection{Extracting primary $\gamma$ rays}
\label{subsec:firstgen}

As $\gamma$ decay from highly excited states often involves a cascade of transitions, it is necessary 
to isolate the $\gamma$ rays that are emitted in the first decay step of all the possible
decay routes, since information on the level density and the $\gamma$-ray strength function
can be extracted from the distribution of these primary $\gamma$ rays (also called first-generation $\gamma$ rays). 
Therefore, a method has been developed
in order to separate the primary $\gamma$-ray spectra from the $\gamma$ rays that origin from 
the later steps in the decay cascades at each
excitation energy. This method will hereafter be referred to as the \textit{first-generation method} 
and is described in detail in Ref.~\cite{Gut87}. This method is very important, since the correctness
of the further analysis is completely dependent on correctly determined primary $\gamma$-ray spectra. 
The first-generation method shares several features with another subtraction technique developed by Bartholomew
\textit{et al.}~\cite{Bartholomew}.  
The main features of the method will be outlined in the following.

From the $E$ vs. $E_\gamma$ matrix, where $\gamma$-ray spectra for each excitation energy are 
contained, the primary $\gamma$-ray spectra are extracted through an iterative subtraction technique. 
The unfolded spectra $f_i$ are made of all generations of $\gamma$ rays from all possible cascades 
decaying from the excited levels within the excitation-energy bin $i$. Now, we utilize 
the fact
that the spectra 
$f_{j<i}$ for all the underlying energy bins $j$ contain the same $\gamma$-transitions as $f_i$ 
\textit{except} the first $\gamma$ rays emitted\footnote{This is only true if the 
$\gamma$-decay pattern is the same regardless of whether the 
states in the bin were populated from the direct reaction, or from decay from above-lying bins. 
See Sec.~\ref{subsec:fgerr}.}, since they will bring the nucleus from the states 
in energy bin $i$ to underlying states in the bins $j$. Thus, the primary $\gamma$-ray spectrum $h_i$
for each bin $i$ can be found by
\begin{equation}
h_i = f_i - g_i,
\label{eq:fg}
\end{equation}
where $g_i$ is a weighted sum of all spectra
\begin{equation}
g_i = n_{i1}w_{i1}f_1 + n_{i2}w_{i2}f_2 + \ldots + n_{ij}w_{ij}f_j = \sum_j n_{ij}w_{ij}f_j.
\end{equation}
Here, the unknown coefficients $w_{ij}$ (with the normalization $\sum_j w_{ij} = 1$) represent the probability of the decay 
from states in bin $i$ to states in bin $j$. In other words, the $w_{ij}$ values make up the weighting function 
for bin $i$, and contain the 
distribution of branching ratios as a function of $\gamma$-ray energy. 
Therefore, the weighting function $w_{i}$ corresponds directly to 
the primary $\gamma$-ray spectrum $h_i$ for bin $i$. 

The coefficients $n_{ij}$ are correcting factors for the different cross sections of populating levels in bin $i$ 
and the underlying levels in bin $j$. They are determined so that the total area of each spectrum $f_i$ multiplied 
by $n_{ij}$ corresponds to the same number of cascades. This can be done in two ways~\cite{Gut87}:
\begin{itemize}
\item{Singles normalization. The singles-particle cross section is proportional to 
the number of events populating levels in a specific bin, and thus to the number of 
decay cascades from this bin. We denote the number of counts measured 
for bin $i$ and $j$ in the singles spectrum $S_i$ and $S_j$, respectively. The normalization factor $n_{ij}$ 
that should be applied to the spectrum $f_j$ is then given by
\begin{equation}
n_{ij} = \frac{S_i}{S_j}.
\end{equation}
}
\item{Multiplicity normalization. The average $\gamma$-ray multiplicity $\langle M \rangle$ can be obtained 
in the following way~\cite{Rekstad}: Assume an $N$-fold population of an excited level $E$. The decay from 
this level will result in $N$ $\gamma$-ray cascades, where the $i$th cascade contains $M_i$ $\gamma$ rays. 
The average $\gamma$-ray energy $\langle E_{\gamma} \rangle$ is equal to the total energy carried by the 
$\gamma$ rays divided by the total number of $\gamma$ rays:
\begin{equation}
\langle E_{\gamma} \rangle = N \cdot \frac{E}{\sum_{i=1}^N M_i} = \frac{E}{\frac{1}{N}\sum_{i=1}^N M_i} = \frac{E}{\langle M \rangle}.
\end{equation}
Then, the average $\gamma$-ray multiplicity is simply given by
\begin{equation}
\langle M \rangle = \frac{E}{\langle E_{\gamma} \rangle}.
\label{eq:avmul}
\end{equation}
The average $\gamma$-ray multiplicity $\langle M_i \rangle$ can thus easily be calculated for each 
excitation-energy bin $i$. Let the area (or total number of counts) of the $\gamma$-ray spectrum $f_i$ be denoted by $A(f_i)$. 
Then the singles particle cross section $S_i$ is proportional to the ratio $A(f_i)/\langle M_i \rangle$, and 
the normalization coefficient $n_{ij}$ that should be applied to bin $i$ when subtracting bin $j$ is 
\begin{equation}
n_{ij} = \frac{A(f_i)/\langle M_i \rangle}{A(f_j)/\langle M_j \rangle} = \frac{\langle M_j \rangle A(f_i)}{\langle M_i \rangle A(f_j)}.
\end{equation}
}
\end{itemize}
The two normalization methods give normally the same results within the experimental error bars\footnote{In case of the presence 
of isomeric states, the multiplicity method must be used to get the correct normalization.}, see also Sec.~\ref{subsec:fgerr}. 
The resulting primary $\gamma$-ray 
matrix of $^{50}$V is shown in Fig.~\ref{fig:matrix50V} (right panel), using the singles normalization method. 

In cases where the multiplicity is well determined, an area consistency check can be applied to Eq.~(\ref{eq:fg}). 
Assume that a small correction is introduced by substituting $g_i$ by $\delta g_i$, where $\delta$ is 
close to unity. The area of the first-generation $\gamma$ spectrum is then 
\begin{equation}
A(h_i) = A(f_i) - \delta A(g_i),
\label{eq:area1}
\end{equation}
and this corresponds to a $\gamma$-ray multiplicity of one unit. Since the number of primary $\gamma$ rays in 
the spectrum $f_i$ equals $A(f_i)/\langle M_i \rangle$, $A(h_i)$ is also given by 
\begin{equation}
A(h_i) = A(f_i)/\langle M_i \rangle.
\label{eq:area2}
\end{equation}
Combining Eqs.~(\ref{eq:area1}) and (\ref{eq:area2}) yields
\begin{equation}
\delta = (1 - 1/\langle M_i \rangle)\frac{A(f_i)}{A(g_i)}.
\end{equation}
The $\delta$ parameter can be varied to get the best agreement of the areas $A(h_i)$, $A(f_i)$ and $A(g_i)$ 
within the following restriction: $\delta = 1.00\pm 0.15$; that is, the correction should not exceed 15\%. 
If a larger correction is necessary, then improved weighting functions should be determined instead.

As mentioned before, the weighting coefficients $w_{ij}$ correspond directly to the first-generation 
spectrum $h_i$, and this close relationship makes it possible to determine $w_{ij}$ (and thus $h_i$)  
through a fast converging iteration procedure~\cite{Gut87}:
\begin{enumerate}
\item{Apply a trial function for $w_{ij}$.}
\item{Deduce $h_i$.}
\item{Transform $h_i$ to $w_{ij}$ by giving $h_i$ the same energy calibration as 
	$w_{ij}$, and normalize the area of $h_i$ to unity.}
\item{If $w_{ij}$(new) $\approx$ $w_{ij}$(old), convergence is 
	reached, and the procedure is finished. Otherwise repeat from step 2.}
\end{enumerate}
The first trial function could be the unfolded spectrum $f_i$, or a theoretical estimate based on a
model for the level density and $\gamma$-ray transmission coefficient, or a constant function;
it turns out that the resulting first-generation spectra are not sensitive to the starting trial function. 
Also, previous tests of the convergence properties of the procedure have shown that excellent agreement 
is achieved between the exact solution (from simulated spectra) and the trial function 
$w_{ij}$ already after three iterations~\cite{Gut87}. Usually, about $10-20$ iterations 
are performed on experimental spectra.

\subsection{Determining level density and $\gamma$-ray strength}
\label{subsec:nld_rsf}

Once the primary $\gamma$-ray spectra are obtained for each excitation energy, the first-generation matrix
$P(E,E_{\gamma})$ is used for the determination of level density and $\gamma$-ray strength. For statistical
$\gamma$-decay, the decay probability (given by $P(E,E_{\gamma})$) of a $\gamma$-ray with energy $E_{\gamma}$ 
decaying from a specific excitation energy $E$ is proportional to the level density $\rho (E_{\mathrm{f}})$ 
at the final excitation energy $E_{\mathrm{f}}=E-E_{\gamma}$, and the $\gamma$-ray transmission coefficient 
$\mathcal{T}(E_{\gamma})$:
\begin{equation}
P(E, E_{\gamma}) \propto  \rho (E_{\mathrm{f}}) {\mathcal{T}}  (E_{\gamma}).
\label{eq:brink}
\end{equation}
The above relation holds for decay from \textit{compound states}, which means that the relative probability 
for decay into any specific set of final states is independent on how the compound nucleus was formed. Thus,
the nuclear reaction can be described as a two-stage process, where a compound state is first formed before 
it decays in a manner that is independent of the mode of formation~\cite{BM,hend1}. This is believed to be 
fulfilled at high excitation energy, even though direct reactions are used, as already discussed previously. 
Equation~(\ref{eq:brink}) can also be compared with Fermi's golden rule:
\begin{equation}
\lambda = \frac{2\pi}{\hbar} \left| \left< \mathrm{f}\left|\hat{H}_{\mathrm{int}}\right| \mathrm{i} \right> \right|^2 \rho(E_{\mathrm{f}}),
\end{equation}
where $\lambda$ is the decay rate of the initial state $\left|\mathrm{i} \right>$ to the final state $\left|\mathrm{f} \right>$, 
and $\hat{H}_{\mathrm{int}}$ is the transition operator. In Eq.~(\ref{eq:brink}), an ensemble of initial and final states within 
each excitation-energy bin is considered, and thus we obtain here the average decay properties of a set of initial states to a set of final states. 
Note, however, that in contrast to Fermi's golden rule where the matrix element is strictly dependent on the initial and final state,
the transmission coefficient ${\mathcal{T}}$ is only dependent on the $\gamma$-ray energy and neither the initial nor the final excitation
energy. This is in accordance with the Brink hypothesis~\cite{brink}, which states that the collective giant dipole mode 
built on excited
states has the same properties as if built on the ground state. In its generalized form, this hypothesis includes 
all types of collective decay modes. Assuming that this hypothesis holds, the first-generation matrix $P(E,E_{\gamma})$ is separable into
two functions $\rho$ and ${\mathcal{T}}$ as given in Eq.~(\ref{eq:brink}). 

To extract the level density and the $\gamma$-ray transmission coefficient, an iterative procedure~\cite{schi0} is applied 
to the first-generation matrix $P(E, E_{\gamma})$. The basic idea of this method is to minimize
\begin{equation}
\chi^{2} = \frac{1}{N_{\mathrm{free}}}\sum_{E=E^{\mathrm{min}}}^{E^{\mathrm{max}}}
\sum_{E_{\gamma}=E_{\gamma}^{\mathrm{min}}}^{E} 
\left( \frac{P_{\mathrm{th}}(E, E_{\gamma}) - P(E, E_{\gamma})}{\Delta P(E, E_{\gamma})} \right)^{2},
\label{eq:chisquared}
\end{equation}
where $N_{\mathrm{free}}$ is the number of degrees of freedom, and $\Delta P(E, E_{\gamma})$ is the uncertainty 
in the experimental first-generation $\gamma$-ray matrix. 
The fitted first-generation $\gamma$-ray matrix $P_{\mathrm{th}}(E, E_{\gamma})$ can theoretically be approximated by 
\begin{equation}
P_{\mathrm{th}}(E, E_{\gamma}) = \frac {\rho (E -E_{\gamma}) 
{\mathcal{T}}  (E_{\gamma})}{\sum_{E_{\gamma}=E_{\gamma}^{\mathrm{min}}}^{E} \rho (E -E_{\gamma}) {\mathcal{T}}  (E_{\gamma})}.
\label{eq:theory}
\end{equation}
The experimental matrix of first-generation $\gamma$ rays 
is normalized~\cite{schi0} such that for every excitation-energy bin $E$, the sum over all $\gamma$ energies 
$E_{\gamma}$ from some minimum value  $E_{\gamma}^{\mathrm{min}}$ to the maximum value $E_{\gamma}^{\mathrm{max}}=E$ 
at this excitation-energy bin is unity:
\begin{equation}
\sum_{E_{\gamma}=E_{\gamma}^{\mathrm{min}}}^{E} P(E, E_{\gamma}) = 1.
\label{eq:matrixnorm}
\end{equation}

\begin{figure*}[tb]
\centering
\includegraphics[clip,totalheight=7cm]{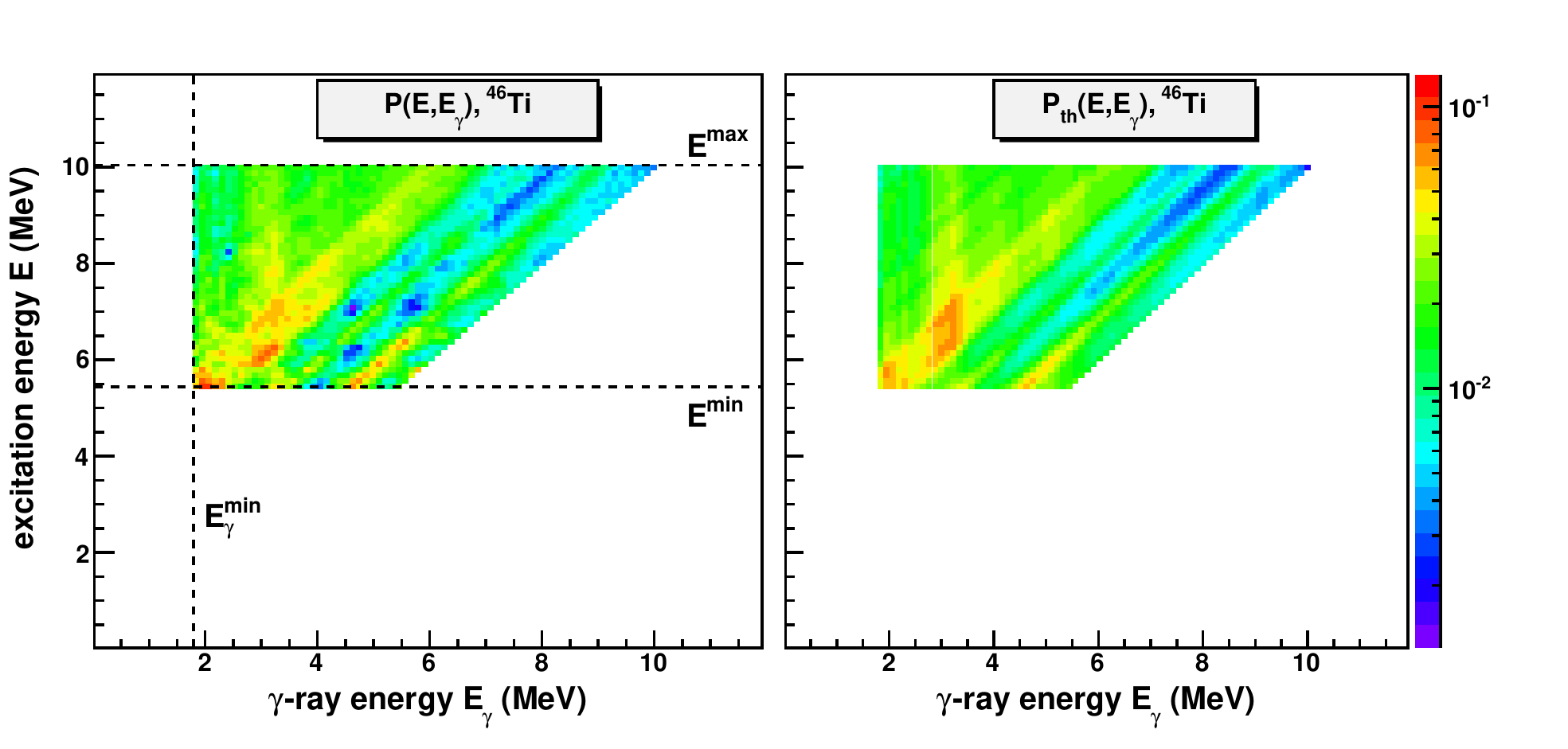}
\caption{(Color online). Experimental first-generation matrix $P(E, E_{\gamma})$  
(left) and the calculated $P_{\mathrm{th}}(E, E_{\gamma})$ (right) 
of $^{46}$Ti from the iteration procedure of A.~Schiller \textit{et al}.~\cite{schi0}. The dashed lines show the limits set 
in the experimental first-generation matrix for the fitting procedure. The data are taken from the experiment
 presented in Ref.~\cite{magne_46Ti}.}
\label{fig:rhosigdemo}
\end{figure*}

The experimental matrix $P(E, E_{\gamma})$ and the fitted matrix $P_{\mathrm{th}}(E, E_{\gamma})$ 
of $^{46}$Ti are displayed in Fig.~\ref{fig:rhosigdemo}. The energy limits set in the first-generation matrix for extraction 
are also shown. These limits ($E^{\mathrm{min}}$, $E^{\mathrm{max}}$, and $E_{\gamma}^{\mathrm{min}}$) are 
chosen to ensure that the data utilized are from the statistical excitation-energy region and that no 
$\gamma$-lines stemming from, e.g., yrast transitions, 
%which might not be subtracted correctly in the first-generation method, 
are used in the further analysis. Note that
the $\gamma$-ray energy bins are now re-binned to the same size (120 keV in this case) as the excitation-energy bins. 
%-----------------------------------------------------------------%
%-----------------------------------------------------------------%

\begin{figure*}[tb]
\centering
\includegraphics[clip,totalheight=9cm]{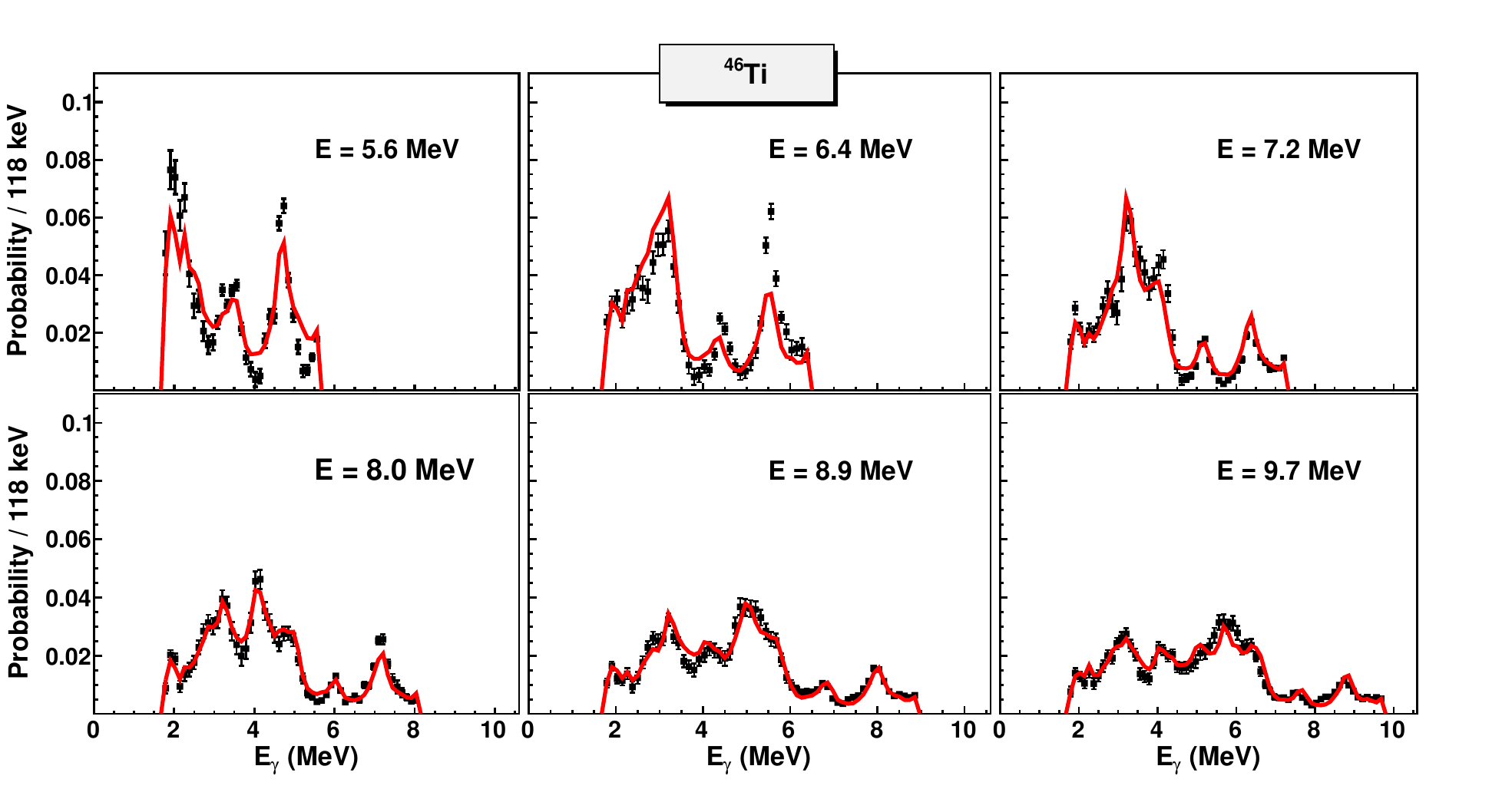}
\caption{(Color online). Experimental first-generation $\gamma$ spectra (data points with error bars) 
at six different initial excitation energies (indicated in the figure) compared to the 
$\chi^2$ fit (solid lines) for $^{46}$Ti. The fit is performed simultaneously on the entire 
first-generation matrix of which the six displayed spectra are a fraction. The data are taken from the experiment
 presented in Ref.~\cite{magne_46Ti}.}
\label{fig:rhosig}
\end{figure*}

\begin{figure*}[tb]
\centering
\includegraphics[clip,totalheight=9cm]{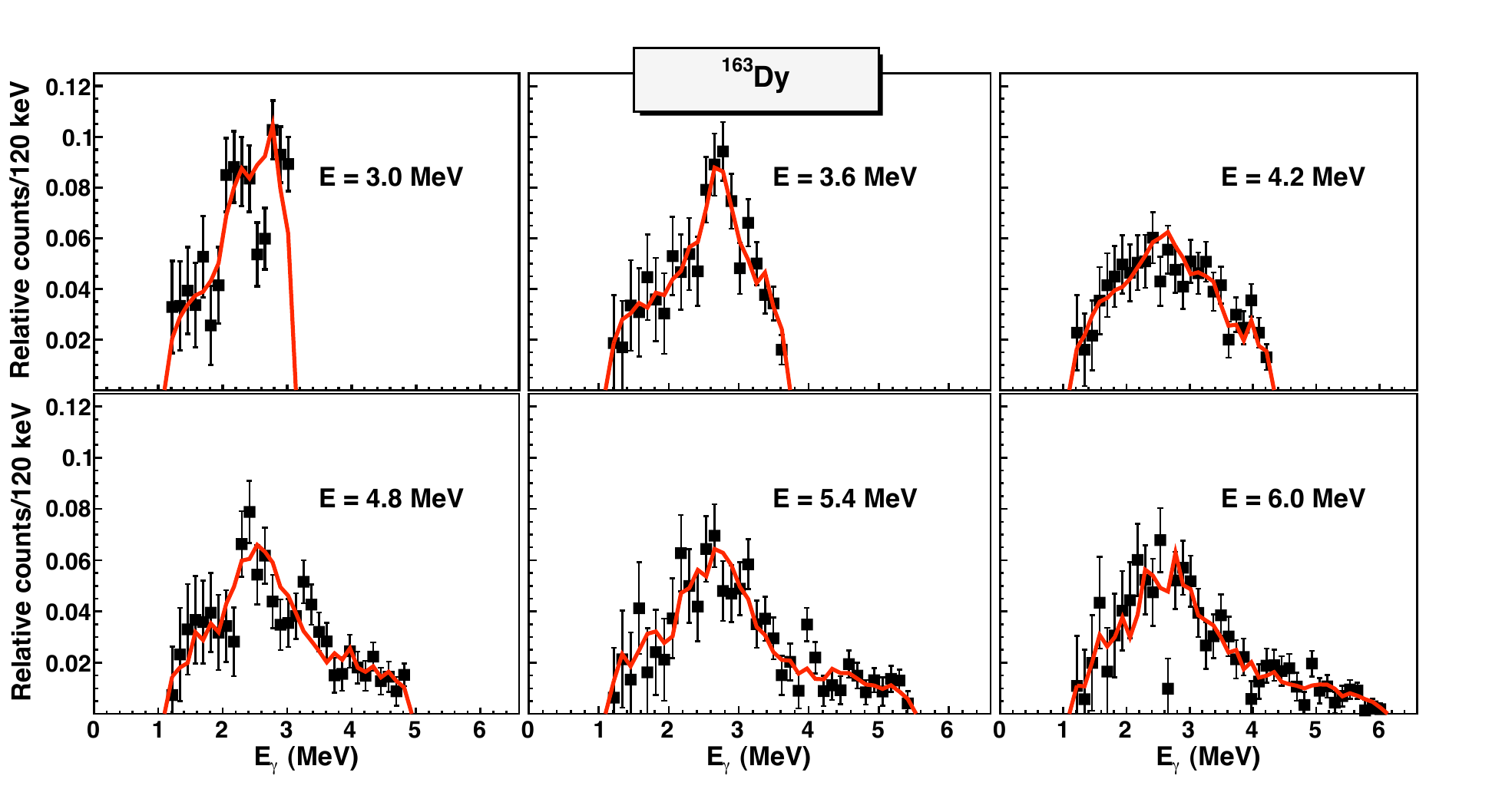}
\caption{(Color online). Same as Fig.~\ref{fig:rhosig} for $^{163}$Dy. The data are taken from the experiment
 presented in Ref.~\cite{hildes_DyRSF}.}
\label{fig:rhosig163Dy}
\end{figure*}

Each point of the $\rho$ and ${\mathcal{T}}$ functions is assumed to be an independent variable, 
so that the reduced $\chi^{2}$ of Eq.~(\ref{eq:chisquared}) is minimized for every argument $E-E_{\gamma}$ and $E$. 
The quality of the procedure when applied to $^{46}$Ti and $^{163}$Dy is shown in Figs.~\ref{fig:rhosig} and~\ref{fig:rhosig163Dy}, 
where the experimental first-generation spectra for various initial excitation energies are compared 
to the least-$\chi^{2}$ solution. In general, the agreement between the experimental data and the fit is very good. Note, however, 
that differences of several standard deviations do occur. In the case of $^{46}$Ti, this is particularly pronounced for 
the peaks at $E_{\gamma} \approx 4.7$ MeV for $E=5.6$ MeV and $E_{\gamma} \approx 5.5$ MeV for $E=6.4$ MeV. As these peaks 
correspond to the decay to the first excited state, one might expect large Porter-Thomas fluctuations~\cite{porter-thomas} in the  
strength of these transitions.
%-----------------------------------------------------------------%
%-----------------------------------------------------------------%
%-----------------------------------------------------------------%
%-----------------------------------------------------------------%

The global fitting to the data points only gives the functional form of $\rho$ and ${\mathcal{T}}$. 
In fact, it has been shown \cite{schi0} that if one solution for the multiplicative functions $\rho$ 
and ${\mathcal{T}} $ is known, one may construct an infinite number of other functions, which give 
identical fits to the $P(E, E_{\gamma})$ matrix by
\begin{eqnarray}
\tilde{\rho}(E-E_\gamma)&=&A\exp[\alpha(E-E_\gamma)]\,\rho(E-E_\gamma),
\label{eq:array1}\\
\tilde{{\mathcal{T}}}(E_\gamma)&=&B\exp(\alpha E_\gamma){\mathcal{T}} (E_\gamma).
\label{eq:array2}
\end{eqnarray}
Therefore the transformation parameters $\alpha$, $A$ and $B$, which correspond to the physical solution, must be
found from external data. 

\subsection{Normalization}
\label{subsec:norm}

\begin{figure}[tb]
\centering
\includegraphics[clip,width=\columnwidth]{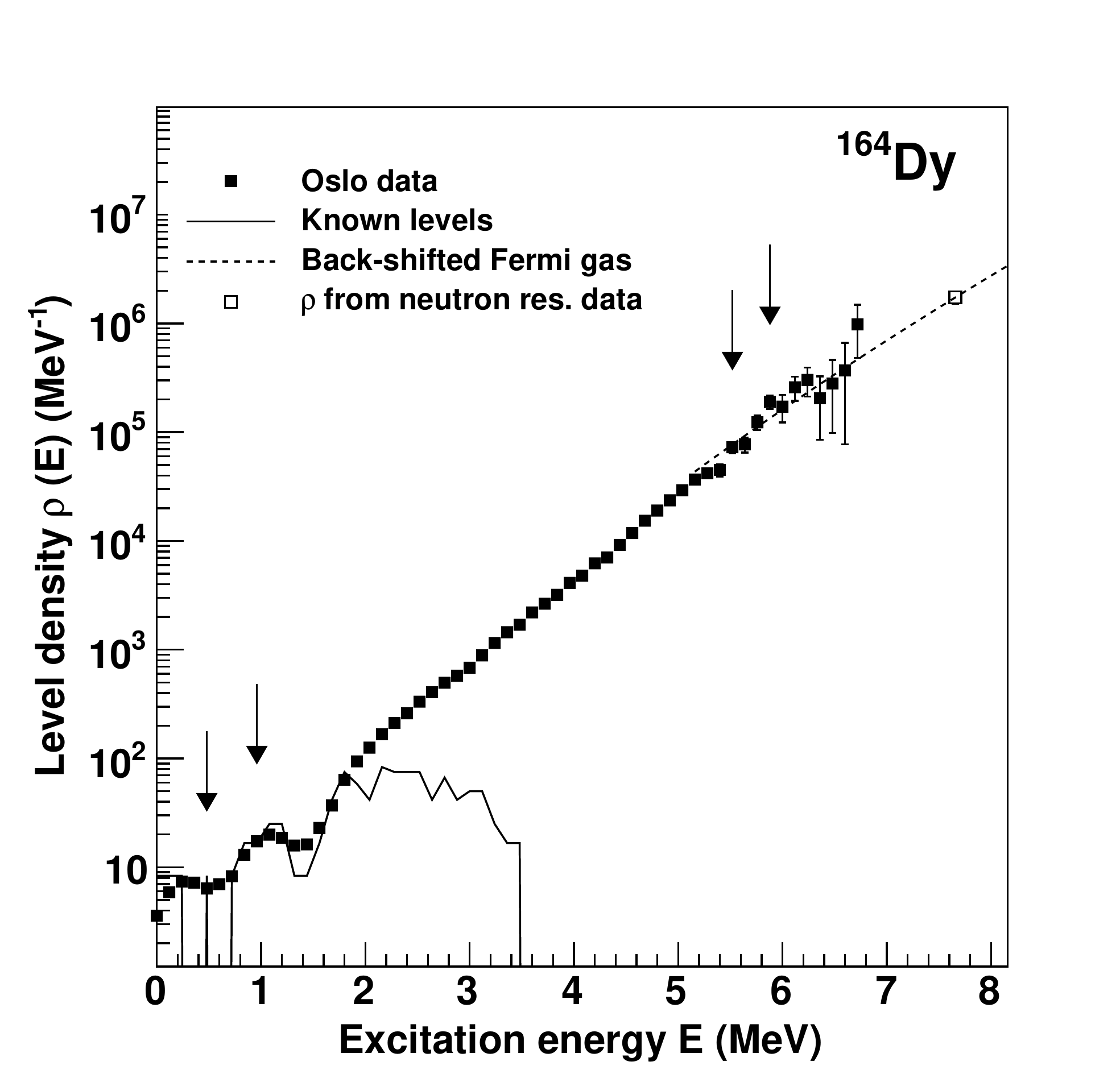}
\caption{Normalization procedure of the level density (data points) of $^{164}$Dy. 
The data points between the arrows are normalized to known levels at low excitation energy 
(solid line) and to the level density at the neutron separation energy (open square) using an 
interpolation with the Fermi-gas level density (dashed line). The data are taken from the experiment
 presented in Ref.~\cite{hildes_DyRSF}.}
\label{fig:normlev}
\end{figure}

In order to determine the correction $\alpha$ to the slope of the level density and the $\gamma$-ray transmission coefficient, 
and to determine the absolute value $A$ of the level density in Eq.~(\ref{eq:array1}), the $\rho$ function is adjusted to 
fit the number of known discrete levels at low excitation energy and neutron (or proton) resonance data at high excitation 
energy. This normalization is shown for $^{164}$Dy in Fig.~\ref{fig:normlev}. 
%-----------------------------------------------------------------%
%-----------------------------------------------------------------%
The data point at high excitation energy (open square in Fig.~\ref{fig:normlev}) is 
calculated in the following way according to Ref.~\cite{schi0}. The starting point is 
Eqs. (4) and (5) of Ref.~\cite{GC}:
\begin{equation}
\rho(U,J) = \frac {\sqrt{\pi}}{12} \frac{{\rm exp} (2\sqrt{aU})}{a^{1/4} U^{5/4}} 
\frac{(2J+1)\exp \left[-(J+1/2)^2/2\sigma^2\right]}{2\sqrt{2\pi}\sigma^3},
\label{eq:G&C_1}
\end{equation}
and
\begin{equation}
\rho(U) = \frac {\sqrt{\pi}}{12} \frac{{\rm exp} (2\sqrt{aU})}{a^{1/4} U^{5/4}} \frac{1}{\sqrt{2\pi}\sigma},
\label{eq:G&C_2}
\end{equation}
where $\rho(U,J)$ is the level density for a given spin $J$, and $\rho(U)$ is the level density 
for all spins. The intrinsic excitation energy $U$, the level-density parameter $a$, 
and the spin cutoff parameter $\sigma$ are normally taken from Ref.~\cite{egidy1}
in previous works, or from Ref.~\cite{egidy2} in recent works. 

Now, let $I_t$ be the spin of the target nucleus in a neutron 
resonance experiment. The average neutron resonance spacing $D_{\ell = 0}$ for 
s-wave neutrons can be written as 
\begin{equation}
\frac{1}{D_0} = \frac{1}{2}\left[ \rho(S_n,J=I_t+1/2) + \rho(S_n,J=I_t-1/2)\right],
\label{eq:D}
\end{equation}
because all levels with spin $J=I_t\pm 1/2$ are accessible in neutron resonance experiments, 
and because it is assumed that both parities contribute equally to the level density at the neutron 
separation energy $S_n$. Combining Eqs.~(\ref{eq:G&C_1})--(\ref{eq:D}) with $U = S_n$, 
one finds the total level density at the neutron separation energy to be
\begin{equation}
\rho(S_n) = \frac{2\sigma^2}{D_0} \cdot \frac{1}{(I_t+1)\exp\left[-(I_t+1)^2/2\sigma^2\right] + I_t\exp\left[-I_t^2/2\sigma^2\right]}.
\label{eq:oldD}
\end{equation} 
Note also that the resonance spacing between $p$-waves, $D_1$, could also be used for calculating $\rho(S_n)$, 
see Ref.~\cite{Pb}. 

Since our experimental data only reach up to excitation energies around $S_n-E_{\gamma}^{\mathrm{min}}$, an interpolation 
has been made between the Oslo data and $\rho(S_n)$ using the back-shifted Fermi gas model of Refs.~\cite{egidy1,egidy2}, 
as shown in Fig.~\ref{fig:normlev}. It should be noted that in most cases the gap between the data and $\rho(S_n)$ 
is small, so that the normalization is not very sensitive to the interpolation; a pure exponential function 
of the type $\rho(E) = C_0 \exp(C_1 E)$, where $C_0$ and $C_1$ are fitting parameters, gives an interpolation of equally good agreement
(see Ref.~\cite{Heidi_Sn}).
 
The slope of the $\gamma$-ray transmission coefficient ${\mathcal{T}} (E_{\gamma})$ has already been 
determined through the normalization of the level density, as explained above. The remaining constant 
$B$ in Eq.~(\ref{eq:array2}) gives the absolute normalization of ${\mathcal{T}}$, and it is determined 
using information from neutron resonance decay on the average total radiative width $\langle\Gamma_\gamma\rangle$ 
at $S_n$ according to Ref.~\cite{voin1}.

The starting point is Eq.~(3.1) of Ref.~\cite{Kopecky&Uhl_2},
\begin{align}
\langle\Gamma_{\gamma}(E,J,\pi)\rangle= &\frac{1}{2 \pi \rho(E, J, \pi)} \sum _{XL}\sum_{J_{\mathrm{f}},\pi_{\mathrm{f}}}
\int_{E_{\gamma}=0}^{E}{\mathrm{d}}E_{\gamma} {\mathcal{T}}_{XL} (E_{\gamma})
\nonumber\\
& \times \rho(E-E_{\gamma},J_{\mathrm{f}},\pi_{\mathrm{f}}),
\label{eq:longGamma}
\end{align}
where $\langle \Gamma_{\gamma}(E,J,\pi)\rangle$ is the average total radiative width of levels with energy $E$, 
spin $J$ and parity $\pi$. The summation and integration are going over all final levels with spin $J_\mathrm{f}$ 
and parity $\pi_{\mathrm{f}}$ that are accessible through $\gamma$ transitions with energy $E_{\gamma}$, 
electromagnetic character $X$ and multipolarity $L$. Assuming that the main contribution to the experimental 
${\mathcal T}$ is from dipole radiation ($L = 1$), we get
\begin{equation}
B {\mathcal T}(E_{\gamma}) = 
B \sum_{XL} {\mathcal T}_{XL}(E_{\gamma}) 
\approx B \left[{\mathcal T}_{\mathrm{E1}}(E_{\gamma}) + {\mathcal T}_{\mathrm{M1}}(E_{\gamma})\right],
\label{eq:Bnorm}
\end{equation}
from which the total, experimental $\gamma$-ray strength function can easily be calculated:
\begin{equation}
f(E_{\gamma}) = \frac{1}{2\pi E_{\gamma}^3} B {\mathcal T}(E_{\gamma}),
\label{eq:RSF}
\end{equation}
from the relation between $\gamma$-ray strength function and $\gamma$-ray transmission
coefficient~\cite{RIPL}:
\begin{equation}
{\mathcal T}_{XL}(E_{\gamma}) = 2\pi E_{\gamma}^{(2L+1)} f_{XL}(E_{\gamma})\:.
\label{eq:Ttof}
\end{equation}

Further, we also assume that there are equally many accessible levels with positive and negative parity 
for any excitation energy and spin, so that the level density is given by
\begin{equation}
\rho(E-E_{\gamma}, J_{\mathrm{f}},\pm\pi_{\mathrm{f}})=\frac{1}{2}\rho(E-E_{\gamma},J_{\mathrm{f}}).
\label{eq:rhopar}
\end{equation}
Now, by combining Eqs.~(\ref{eq:longGamma}), (\ref{eq:Bnorm}) and (\ref{eq:rhopar}), 
the average total radiative width of neutron s-wave capture resonances with spins 
$I_t \pm 1/2$ expressed in terms of the experimental ${\mathcal T}$ is obtained:
\begin{align}
\langle \Gamma_{\gamma}(S_n,I_t\pm 1/2,&\pi_t)\rangle =
 \frac{B}{4\pi\rho(S_n,I_t\pm 1/2,\pi_t)}\int_{E_{\gamma}=0}^{S_n}\mathrm{d}E_{\gamma}\mathcal{T}(E_{\gamma}) \nonumber \\ 
 &\times \rho(S_n-E_{\gamma}) \sum_{J= -1}^{1} g(S_{n}-E_{\gamma},I_{t}\pm 1/2+J),
\label{eq:width}
\end{align}
where $I_t$ and $\pi_t$ are the spin and parity of the target nucleus in the $(n,\gamma)$ reaction, 
and $\rho(S_n-E_{\gamma})$ is the experimental level density. Note that the factor 
$1/\rho(S_n,I_t\pm 1/2,\pi_t)$ equals the neutron resonance spacing $D_0$. 
The spin distribution of the level density is assumed to be given by~\cite{GC}:
\begin{equation}
g(E,J) \simeq \frac{2J+1}{2\sigma^2}\exp\left[-(J+1/2)^2/2\sigma^2\right].
\label{eq:spindist}
\end{equation}
The spin distribution is normalized so that $\sum_J g(E,J) \approx 1$. The experimental value of  
$\langle\Gamma_\gamma\rangle$ at $S_n$ is then the weighted sum of the level widths of states with 
$I_t\pm 1/2$ according to Eq.~(\ref{eq:width}). From this expression the normalization constant $B$ 
can be determined as described in Ref.~\cite{voin1}. However, some considerations must be done before 
normalizing according to Eq.~(\ref{eq:width}). 

\begin{figure}[htb]
\centering
\includegraphics[clip,width=\columnwidth]{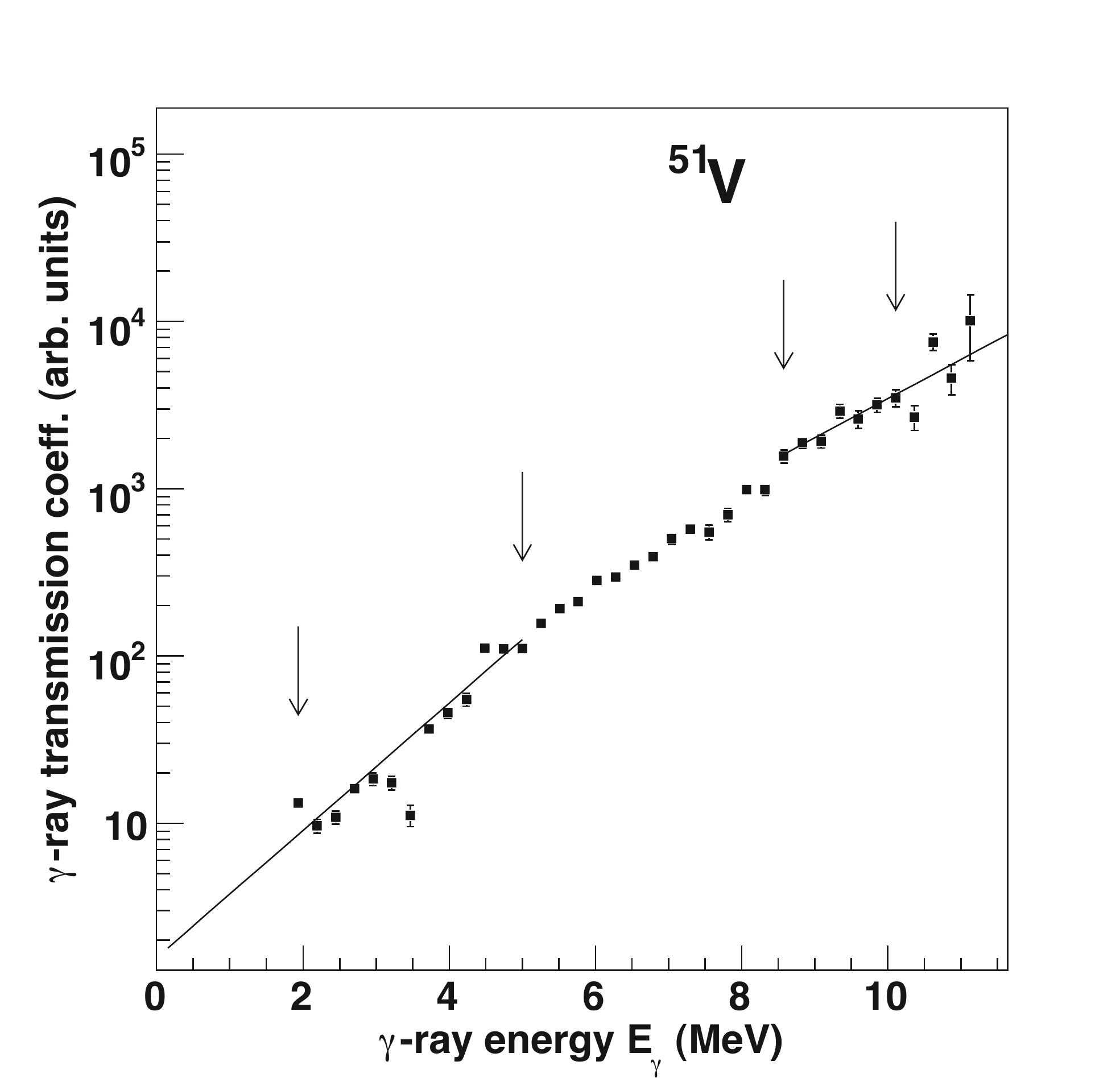}
\caption{\small Extrapolation of the $\gamma$-ray transmission coefficient of $^{51}$V. 
The data points between the arrows in the low and high $\gamma$-energy regions are utilized 
to fit the exponential functions to the data. The data are taken from the experiment
 presented in Ref.~\cite{V}.}
\label{fig:sigext}
\end{figure}

Methodical difficulties in the primary $\gamma$-ray extraction prevent determination of the function 
${\mathcal{T}}(E_{\gamma})$ for $E_\gamma<E_{\gamma}^{\mathrm{min}}$ as discussed previously. 
In addition, the data at the highest $\gamma$-energies suffer from poor statistics. 
Therefore, ${\mathcal{T}}$ is extrapolated with an exponential function, as demonstrated for $^{51}$V 
in Fig.~\ref{fig:sigext}. The contribution of the extrapolation to the total radiative width given by 
Eq.~(\ref{eq:width}) does not normally exceed $15$\%, thus the errors due to a possibly poor extrapolation 
are of minor importance~\cite{voin1}. 
%-----------------------------------------------------------------%
%-----------------------------------------------------------------%

\section{Uncertainties and possible systematic errors}
\label{sec:err}
In the following, we will go through the Oslo method step by step with a close look at the 
uncertainties and possible systematic errors connected to it. 

\subsection{Unfolding of $\gamma$-ray spectra}
\label{sec:unf}
The unfolding method is described in great detail in Ref.~\cite{Gut96}. The method is based on a successive 
subtraction-iteration technique in combination with a special treatment of the Compton background. 
The stability of the method has been extensively tested in previous works and has proven to be 
very robust and reliable (see, e.g., Refs.~\cite{V,Sc,Pb}). To some extent also the impact of 
slightly erroneous response functions has been investigated in Ref.~\cite{hend1}. As this is the 
part of the method that has the largest potential of influencing the final results, it is further 
addressed here. 

\begin{figure}[tb]
\centering
\includegraphics[clip,width=\columnwidth]{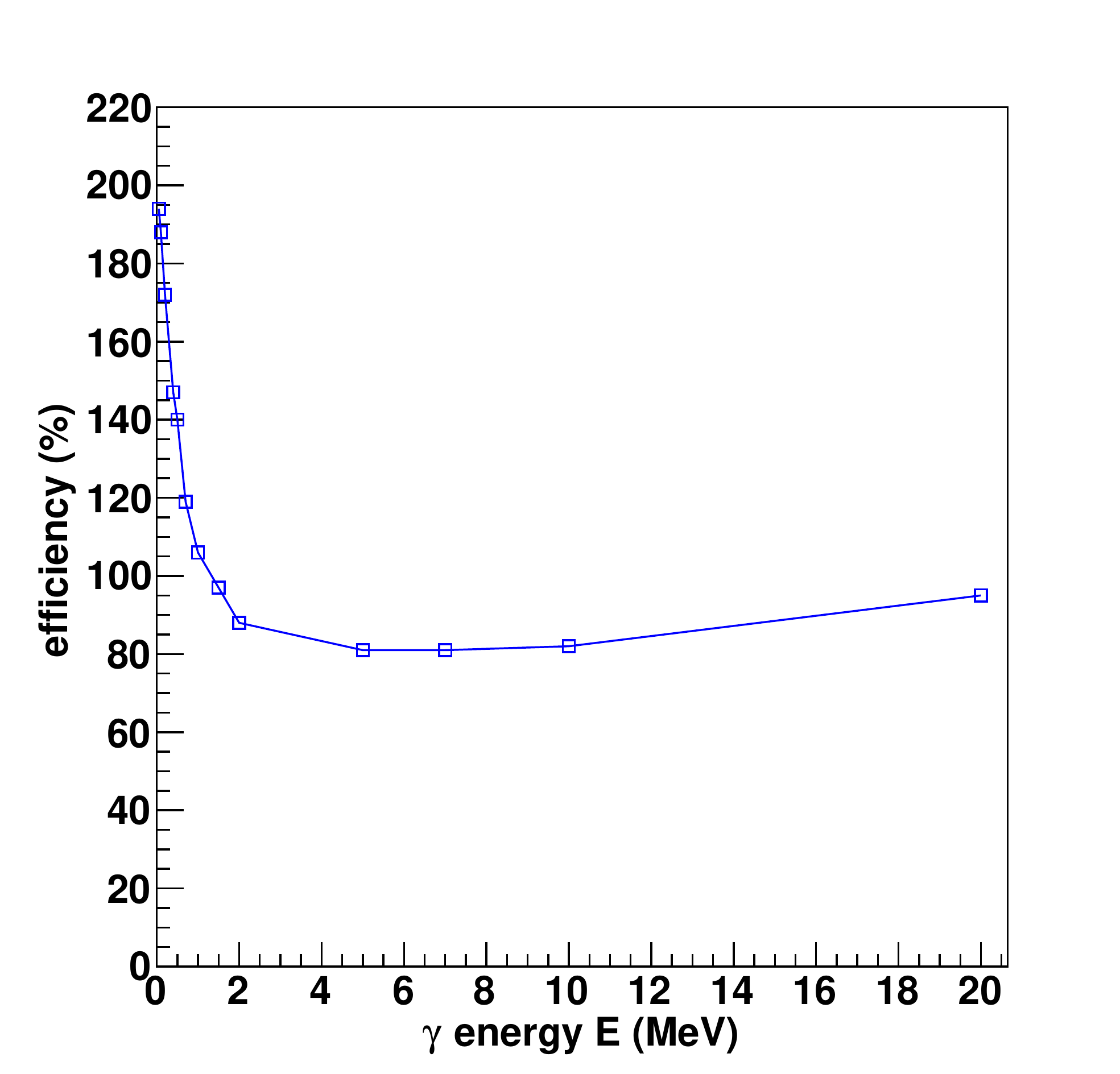}
\caption{(Color online). Efficiencies relative to the 1.33-MeV efficiency used in the unfolding method.}
\label{fig:eff}
\end{figure}

In the unfolding method, the $\gamma$-ray spectra are corrected for the total absorption efficiency of the 
NaI crystals for a given $\gamma$ energy. The applied efficiencies (normalized to the 
efficiency at 1.33 MeV) are shown in Fig.~\ref{fig:eff}. 
%-----------------------------------------------------------------%
%-----------------------------------------------------------------%

We have tested the effect of reducing these efficiencies by up to $\approx 20$\% for $\gamma$ energies 
above 1 MeV, using simulated particle-$\gamma$ coincidences generated with the DICEBOX code~\cite{DICEBOX}.
In the DICEBOX algorithm, a complete decay scheme of an artificial
nucleus is generated. In this case we have considered an artificial nucleus resembling $^{57}$Fe. 
Below an excitation energy of about 2.2 MeV,
all information from the known decay scheme is used; above this energy the levels and decay properties are generated
from a chosen model of the level density and $\gamma$-ray strength function. 
The code allows to take into account Porter-Thomas fluctuations of the  
individual transition intensities as well as assumed fluctuations in the actual density of levels. 
Each particular set of the level scheme and the decay intensities is called a {\em nuclear realization}. 
For more details see Ref.~\cite{DICEBOX}. 
A restriction on the 
spin distribution of the initial excitation-energy levels of $1/2 \leq J \leq 13/2$ was applied, and the chosen 
bin size was 120 keV. In these simulations, each level in a bin was populated with the same probability independently of 
its spin and parity. This means that Porter-Thomas fluctuations were not considered in the population of 
levels via the direct population, but only in the $\gamma$ decay.

\begin{figure}[tb]
\centering
\includegraphics[clip,width=\columnwidth]{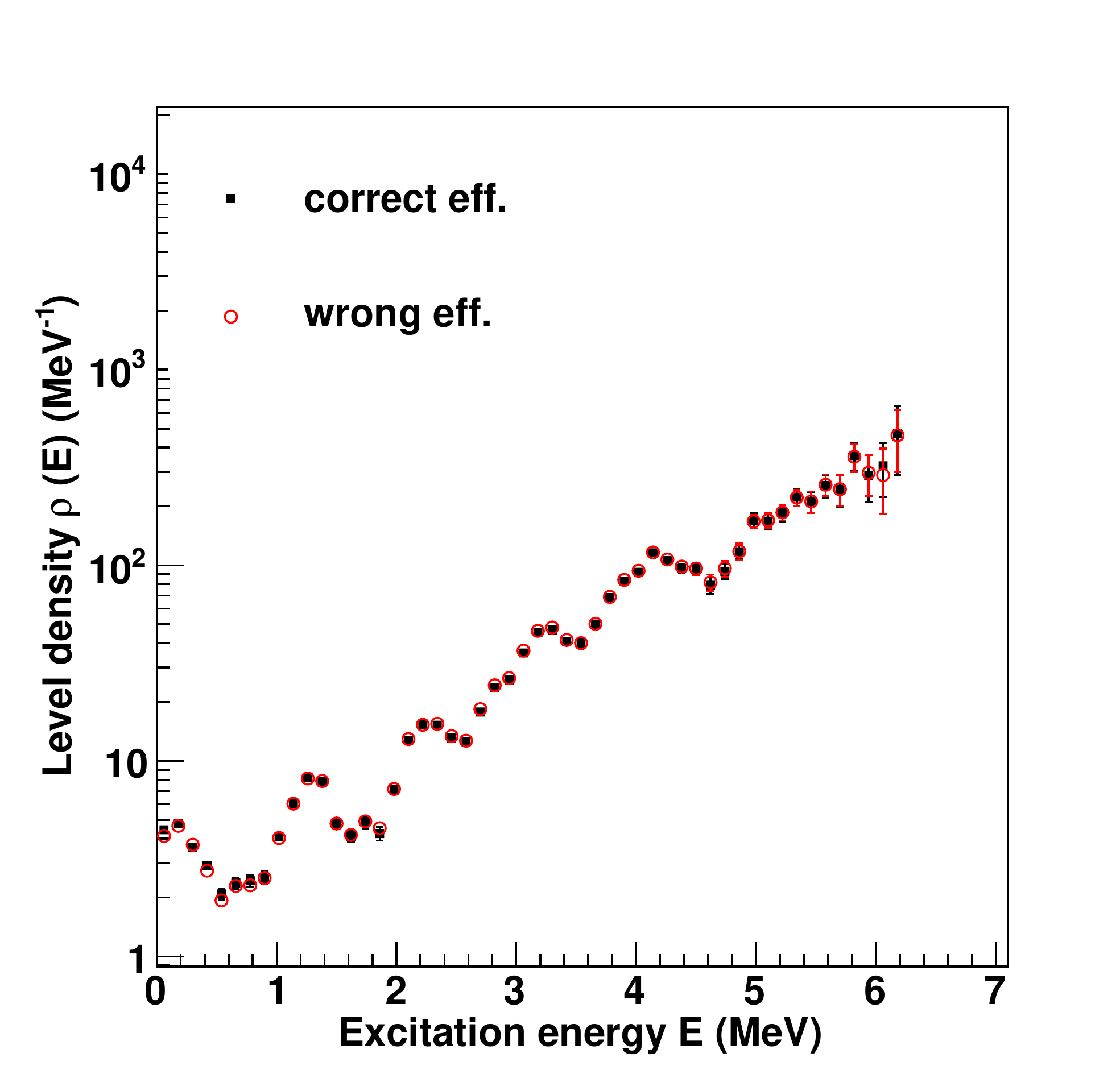}
\includegraphics[clip,width=\columnwidth]{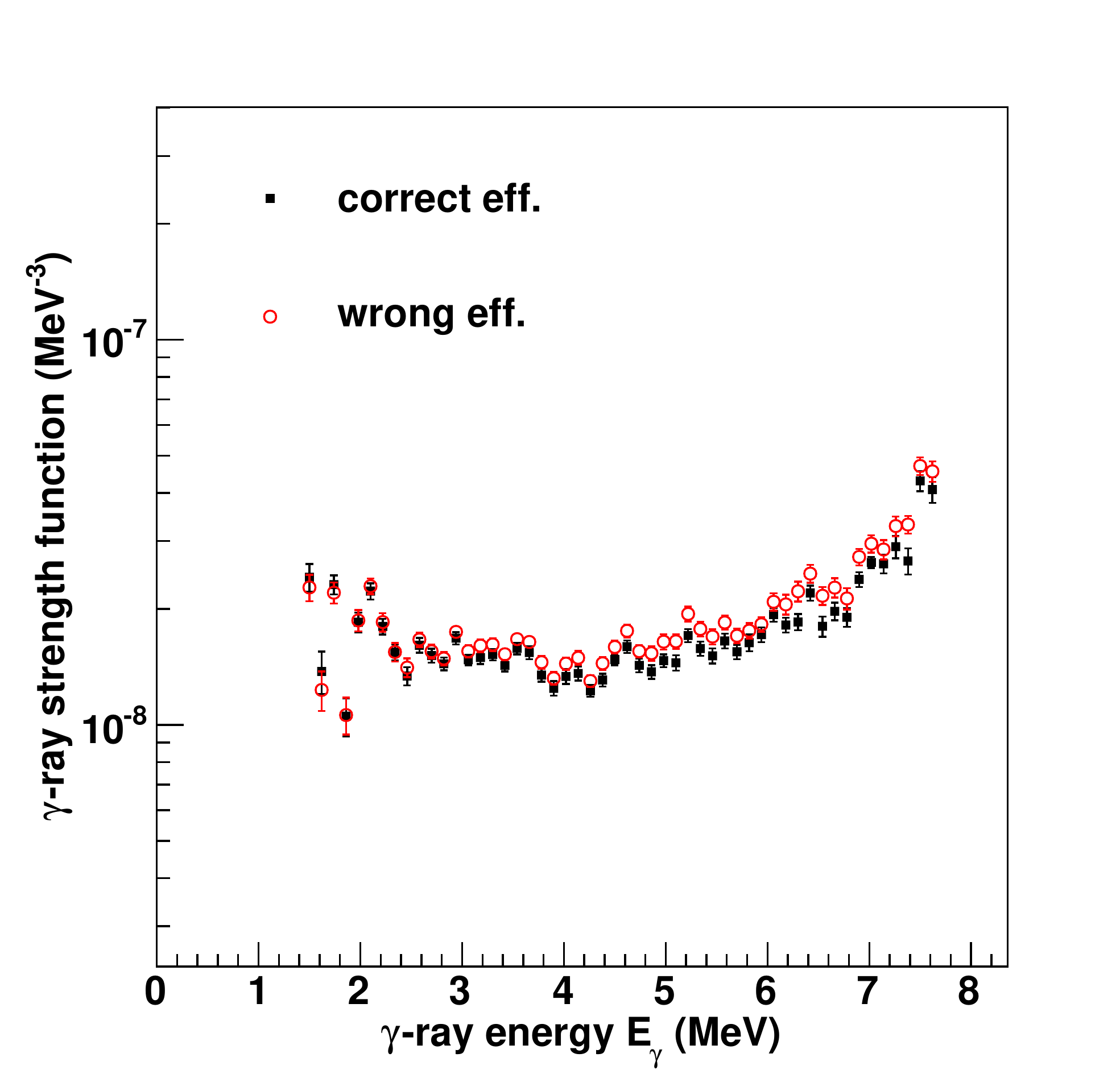}
\caption{(Color online). Test of sensitivity on the total $\gamma$ efficiency 
on the level density (top) and $\gamma$-ray strength function (bottom) extracted from simulated data. }
\label{fig:efftest}
\end{figure}

The results on the extracted level density and $\gamma$-ray strength function are shown in Fig.~\ref{fig:efftest}. 
%-----------------------------------------------------------------%
%-----------------------------------------------------------------%
It is seen that the level density is not very sensitive to the total efficiency, 
but the slope of the $\gamma$-ray strength function is increased when using a too
low efficiency for the high-energy $\gamma$ rays, thus leading to a too large correction in the unfolding procedure. 
However, it is very gratifying that the overall shape is indeed preserved, and the deviation in absolute value 
of the two $\gamma$-ray strength functions does not exceed 20\%, corresponding to 
the maximum change in the absolute efficiency.

\subsection{The first-generation method}
\label{subsec:fgerr}

The first-generation method, which is applied to extract the distribution of primary $\gamma$ rays
from each excitation energy, is a sequential subtraction technique where an iterative procedure 
is applied to determine the weighting coefficients $w_{ij}$, which correspond to the primary $\gamma$-ray spectrum
as described in Sec.~\ref{subsec:firstgen}.

The main assumption of the first-generation method is that the $\gamma$ decay from any excitation-energy 
bin is independent on how the nucleus was excited to this bin. In other words, the decay routes are 
the same whether they were initiated directly by the nuclear reaction or by $\gamma$ decay from 
higher-lying states, giving rise to the same shape of the $\gamma$ spectra. 
This assumption is automatically fulfilled when states have the same cross 
section to be populated by the two processes, since $\gamma$ branching ratios are properties of 
the levels themselves. 

\begin{figure}[tb]
\centering
\includegraphics[clip,width=\columnwidth]{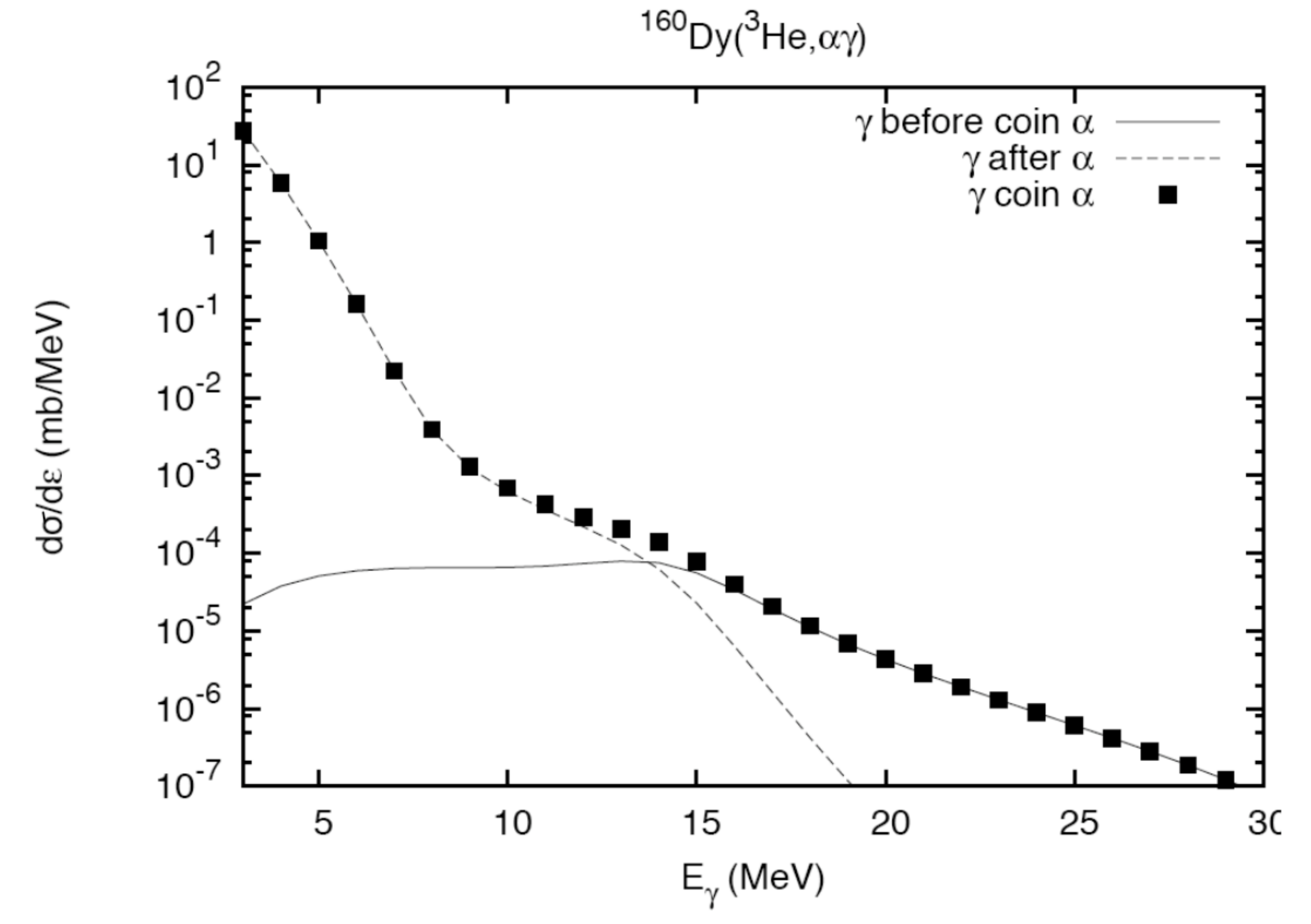}
\caption{Calculated $\gamma$-ray spectra from the $^{160}$Dy($^3$He,$\alpha \gamma$) 
reaction at 45 MeV~\cite{betac}. The solid line shows the pre-equilibrium
component of the total $\gamma$-ray spectrum ($\gamma$ before $\alpha$),
the dashed line represents the equilibrium part ($\alpha$ before $\gamma$),
and the filled squares give the total spectrum.}
\label{fig:betak}
\end{figure}

\begin{figure}[tb]
\centering
\includegraphics[clip,width=\columnwidth]{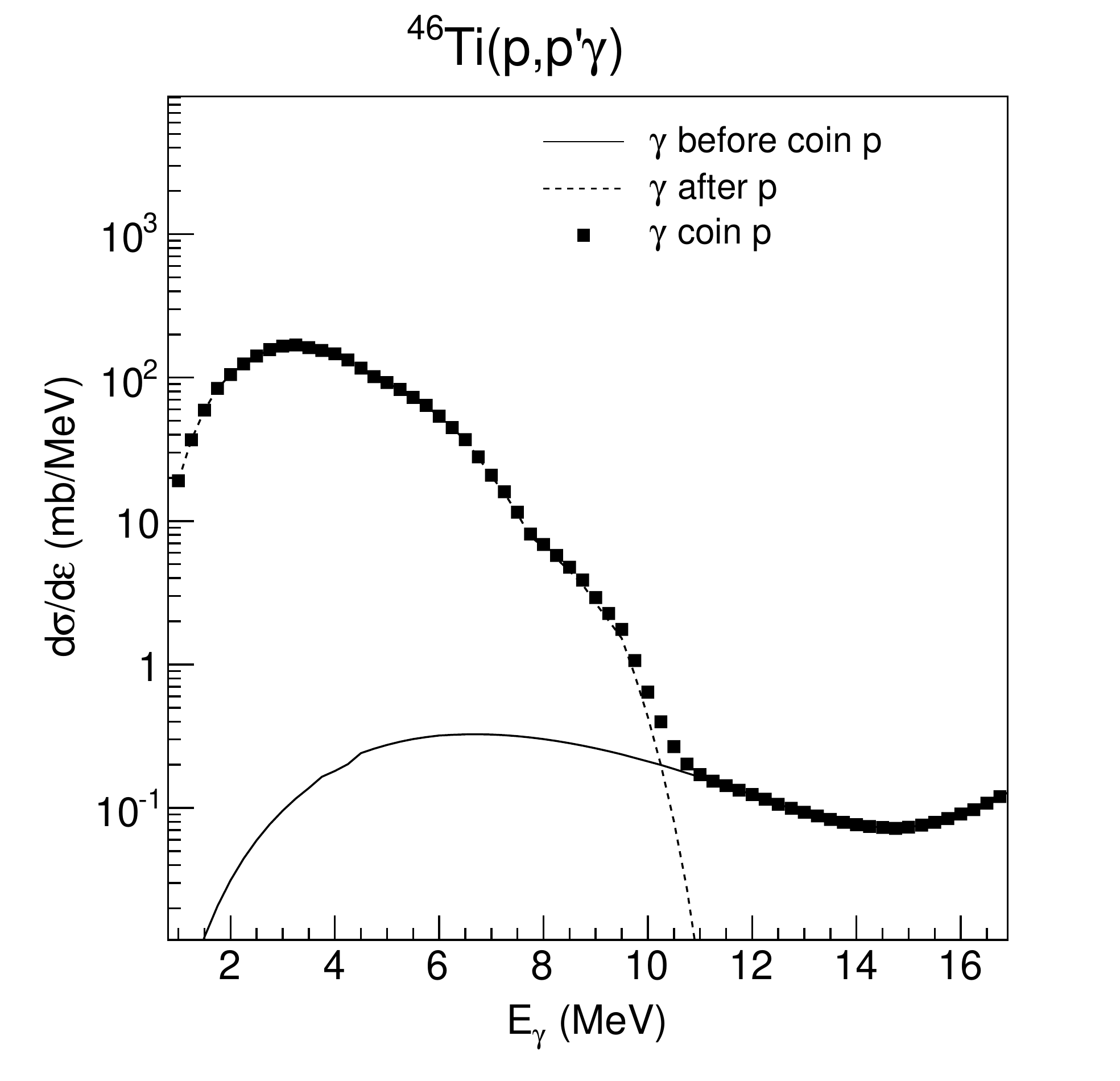}
\caption{Same as Fig.~\ref{fig:betak}, but for the inelastic scattering
$^{46}$Ti($p,p' \gamma$) at $E_p = 15$ MeV.}
\label{fig:betak2}
\end{figure}

In the region of high level density, the nucleus seems to attain a 
compound-like system before emitting $\gamma$ rays even though direct reactions are utilized. 
This is due to two factors. First, a significant configuration mixing of the levels will appear 
when the level spacing is comparable to the residual interaction. Second, the reaction time, 
and thus the time it takes to create a complete compound state, 
is $\approx 10^{-18}$s, while the typical life time of states in the quasi-continuum is $\approx 10^{-15}$s. 
Therefore, it is reasonable to assume that the nucleus has thermalized prior to $\gamma$ decay.
This is supported by recent calculations~\cite{betac} based on the Iwamoto-Harada-Bispinghoff model, showing that for the
 $^{160}$Dy($^3$He,$\alpha \gamma$) reaction with a 45-MeV $^{3}$He beam, the pre-equilibrium
("direct") component of the $\gamma$-ray spectra is very small for $\gamma$ energies below $\approx 10$ MeV 
(see Fig.~\ref{fig:betak}). The same is true for the $^{46}$Ti($p,p'\gamma$) reaction with proton beam energy $E_p = 15$ MeV, 
see Fig.~\ref{fig:betak2}. Note that the "direct" component is calculated using the pre-equilibrium 
(i.e. statistical) formalism. 
%-----------------------------------------------------------------%
%-----------------------------------------------------------------%
%-----------------------------------------------------------------%
%-----------------------------------------------------------------%

Experimentally, the independence of the reaction mechanism has been tested by creating the same compound
nucleus with the two different reactions ($^3$He,$\alpha$) and ($^3$He,$^3$He'). This has been done for, e.g., 
$^{96,97}$Mo~\cite{Mo_RSF}, $^{161,162}$Dy~\cite{bagheri}, and $^{171,172}$Yb~\cite{undraa}. One observes an 
excellent agreement with the level density and $\gamma$-ray strength function 
resulting from the two reactions within the experimental error bars. 
However, at very low excitation energies, there is a significant difference in the obtained level density: the 
inelastic scattering gives consistently a higher level density close to the ground state than does the pick-up
reaction. This could be a sign that the inelastic scattering populates states with wave functions having a large
overlap with the ground state and the low-lying excited states. Thus, the decay to the ground state and low-lying
states will be very fast, and cannot be characterized as compound decay. 

The region at low excitation energies can be tricky also in other aspects. Vertical ridges and/or valleys can occur in 
the primary $\gamma$-ray 
matrix as a consequence of differences in feeding of these discrete states, giving significantly different shapes
of the $\gamma$ spectra at low $E$ compared to higher excitation energies. The direct reaction cross section depends 
strongly on the intrinsic wave functions of low lying states, as seen in the particle spectra in 
Figs.~\ref{fig:alphaSn}--\ref{fig:pTi}. One can encounter the situation where some of these
states are very weakly populated in the reaction, but strongly fed through decay from above-lying states. 
This means that some higher-order $\gamma$ rays are not fully subtracted in the first-generation method, giving an erroneous
primary $\gamma$-ray spectrum for low $E_\gamma$. On the other hand, the reaction might populate very strongly some
low-lying states that are more moderately populated by decay from above. One can then subtract too much of the
$\gamma$ rays from these states. 

The latter case is demonstrated for $^{50}$V in the right panel of Fig.~\ref{fig:matrix50V}. 
A state at excitation energy 910 keV with spin/parity 7$^{+}$ decaying 100\% to the ground state,
is strongly populated in the neutron pick-up reaction, which favors high-$\ell$ transfer 
(here $\ell = 3$, see Ref.~\cite{V50_spec}). However, 
it is not so strongly populated by $\gamma$ decay from above, and the result is that there is a vertical valley 
with zero counts in the 
primary $\gamma$-ray matrix at this $\gamma$ energy.

Furthermore, it is important that the populated spin distribution is (at least approximately) independent
on the excitation energy. Else, the bins at high excitation energy will contain decay from states with 
higher spin than the bins at lower excitation energy, again disturbing the low-energy part of the primary
$\gamma$-ray spectra. For the $^{163}$Dy($^3$He, $\alpha$)$^{162}$Dy reaction,
the spin population has been extensively studied in  Ref.~\cite{hend1} and references therein. In this specific case,
the spin distribution turned out to be approximately constant in the excitation-energy region investigated. Also, 
for the lighter nuclei it is observed that indeed, the spins are populated with the same relative intensities within the
error bars. This is seen for $^{50}$V in the left panel of Fig.~\ref{fig:matrix50V}: the relative feeding to the low-lying
states (vertical lines) is approximately the same for the whole quasi-continuum region. This is also the 
case for $^{46}$Ti~\cite{magne_46Ti}.

Another potential problem could arise from the finite detector resolution. To illustrate this, consider a 
first-generation $\gamma$-ray of 8 MeV, decaying from $E_i=10$ MeV to $E_f = 2$ MeV. This $\gamma$-ray would
typically have a resolution of $\approx 250$ keV, while the particle resolution could be $\approx 150$ keV
(for 15-MeV protons). This means that the weighting function (see Sec.~\ref{subsec:firstgen}) 
for $E_f = 2$ MeV is about 100 keV broader than
the excitation-energy resolution at this point. If the situation is that there is only one level within $E_f$,
one could then "miss" some of the weighting function because
it is broader than the particle peak. It could also be that the opposite situation applies: low-energy $\gamma$ rays
with, say, 50-keV resolution might decay to excitation energies with resolution ranging from $150-300$ keV, leading to
too narrow weighting functions. 

\begin{figure}[bt]
\begin{center}
\includegraphics[clip,width=0.8\columnwidth]{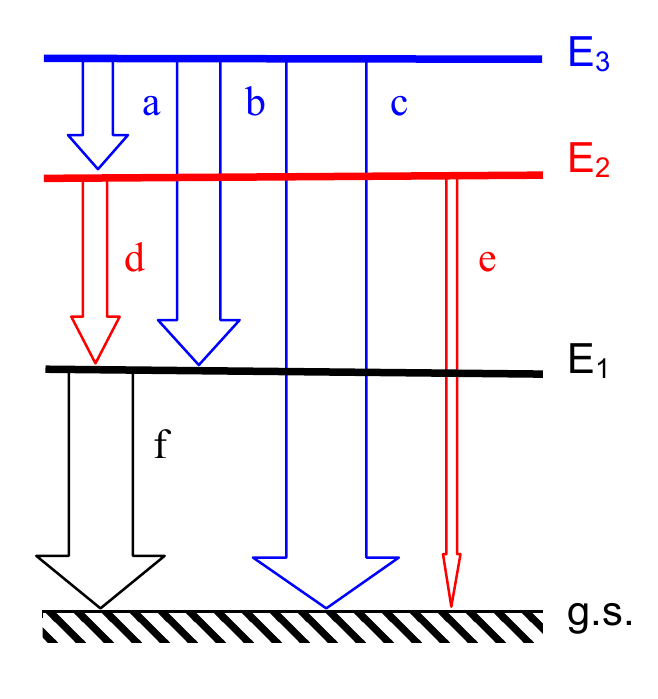}
\caption{(Color online). Hypothetical decay scheme of an artificial nucleus (see text). }
\label{fig:decay}
\end{center}
\end{figure}

We have tested the effects of different resolutions by employing a very simple, artificial decay scheme, 
see Fig.~\ref{fig:decay}. A hypothetical
nucleus with three excited states at $E_1 = 3.5$ MeV, $E_2 = 6$ MeV, and $E_3 = 8$ MeV, was assumed to have the following
decay scheme: 
\begin{itemize}
\item{from $E_3$: 30\% $\gamma_a$, 
20\% $\gamma_b$, and 50\% $\gamma_c$.}
\item{from $E_2$: 67\% $\gamma_d$, and 33\% $\gamma_e$.} 
\item{from $E_1$: 100\% $\gamma_f$.} 
\end{itemize}
The $\gamma$-ray energies involved are: $\gamma_a = 2.0$ MeV, $\gamma_b = 4.5$ MeV, $\gamma_c = 8.0$ MeV,
$\gamma_d = 2.5$ MeV, $\gamma_e = 6.0$ MeV, and $\gamma_f = 3.5$ MeV. The first-generation $\gamma$ rays from 
$E_3$ are then $\gamma_a$, $\gamma_b$, and $\gamma_c$, from $E_2$ $\gamma_d$ and $\gamma_e$, and from
$E_1$ $\gamma_f$. 
%--------------------------------------------------------------------------------------%
%--------------------------------------------------------------------------------------%

\begin{figure}[bt]
\begin{center}
\includegraphics[clip,width=\columnwidth]{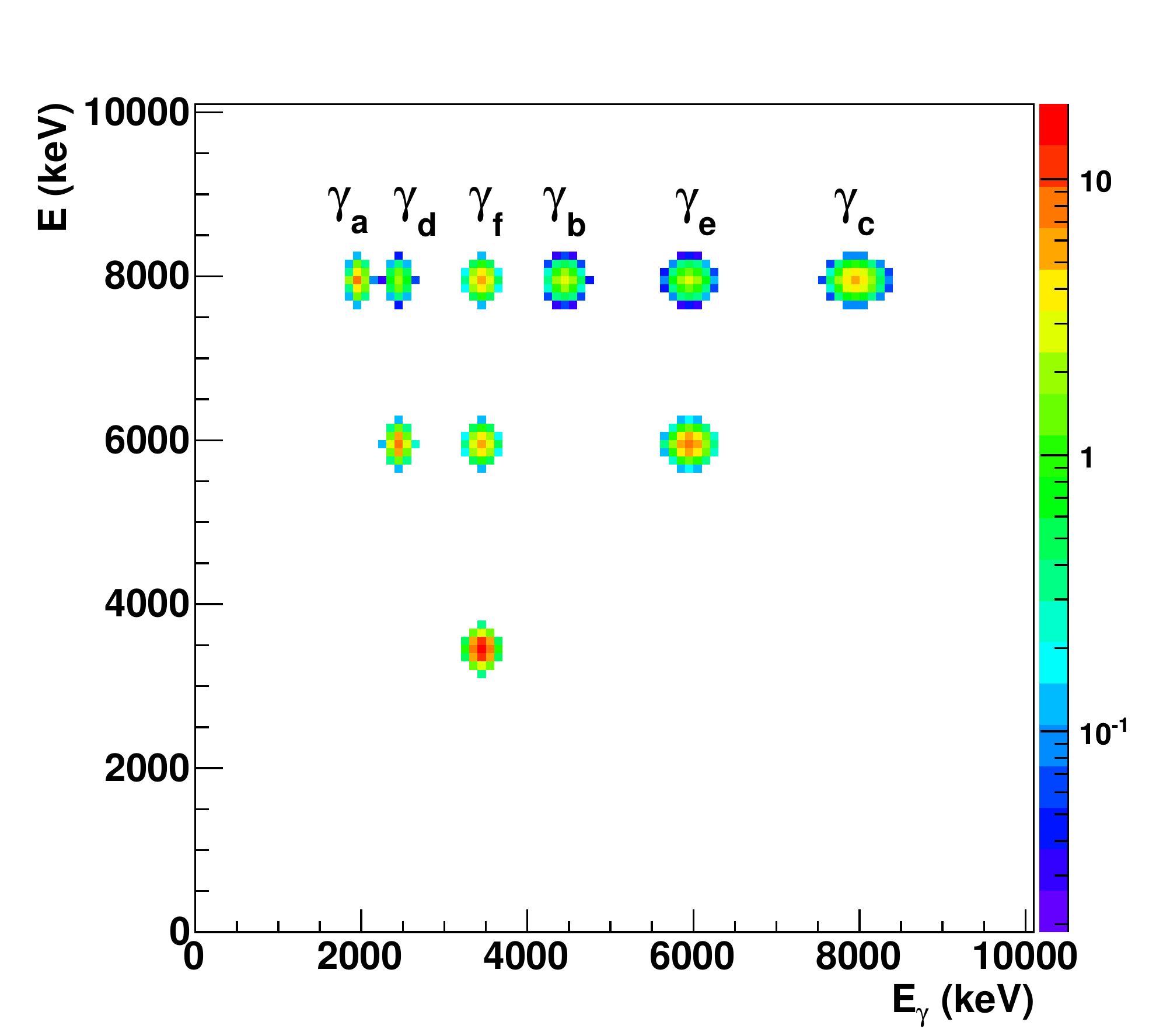}
\includegraphics[clip,width=\columnwidth]{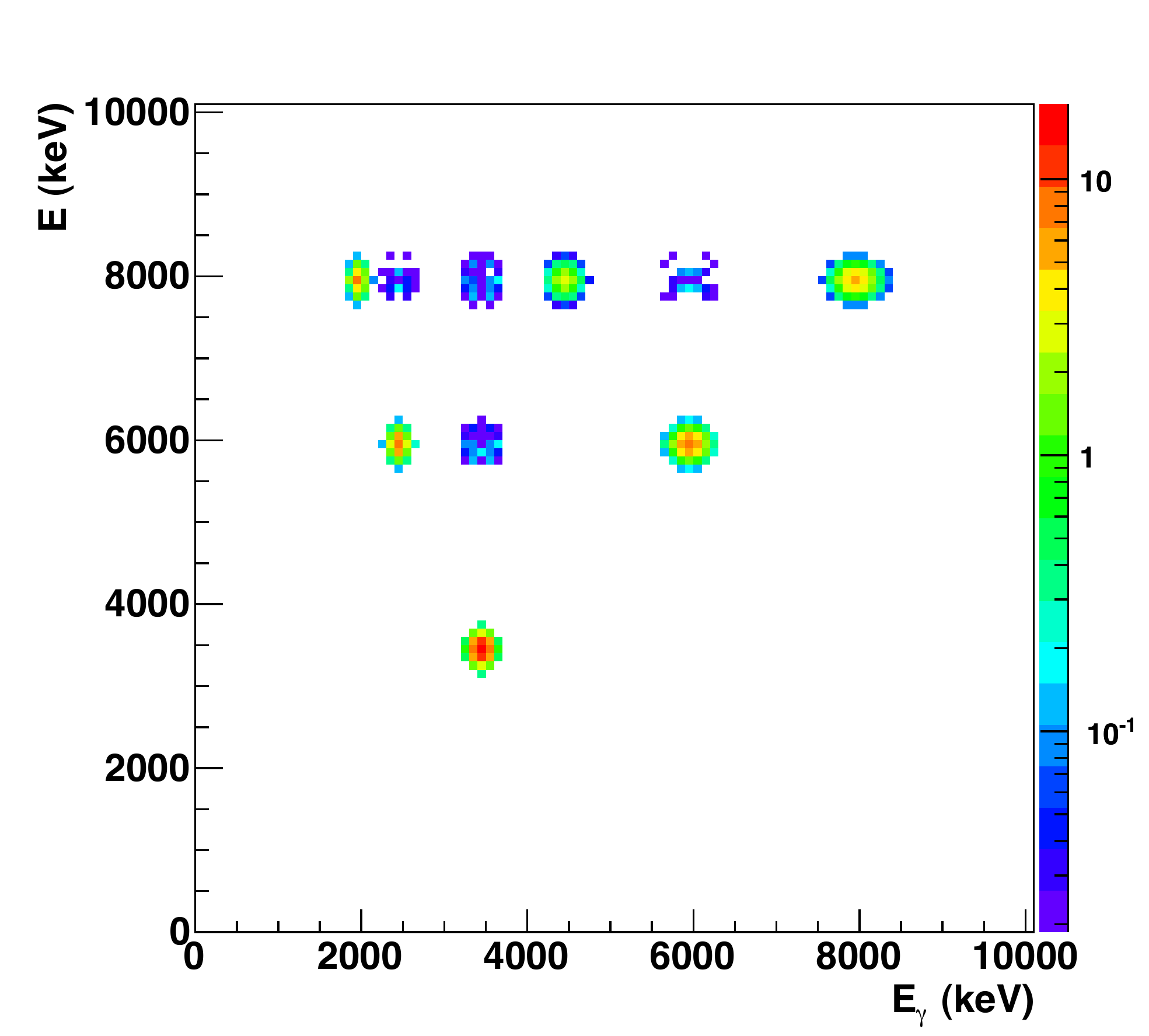}
\caption{(Color online). Matrix of energy levels and $\gamma$ transitions in an artificial nucleus. Top: 
	original matrix; bottom: result after the first-generation method is applied. }
\label{fig:artmatrix}
\end{center}
\end{figure}

Applying no smoothing for all excitation energies, i.e., the $\gamma$-ray peaks are $\delta$ functions,
the exact result is obtained from the first-generation method. 
Then, we assume 200-keV resolution for all excitation energies, but with an energy-dependent resolution of 
the $\gamma$-ray spectrum with 50-keV resolution for 1-MeV $\gamma$ rays and
300-keV resolution for 9-MeV $\gamma$ rays, similar to the experimental conditions. This constructed matrix is
shown in Fig.~\ref{fig:artmatrix}. 
%--------------------------------------------------------------------------------------%
%--------------------------------------------------------------------------------------%
When applying the first-generation method on this smoothed matrix, we get the result shown in the lower
part of Fig.~\ref{fig:artmatrix}. It is seen that it is not exact any more, but the differences are 
small. For example, for $E_3 = 8$ MeV, $\gamma_d$ is not a primary transition and 
should have been completely gone in the first-generation spectrum,
but still about 5\% of the counts in the original peak is  present. For the $\gamma_e$ and $\gamma_f$ peaks,
the situation is the same; also so for $\gamma_f$ at $E_2 = 6$ MeV. This means that one might expect leftovers
of higher-generation $\gamma$ rays of the order of 5\% in the primary spectra. Compared to 
values of experimental errors, which are typically within $5-30$\%, 
this is a relatively small effect (the error propagation is discussed in detail in Ref.~\cite{schi0}). 

To check more thoroughly what effect possible errors in the first-generation method might have on the final
results, namely the extracted level density and strength function, we have performed simulations 
with the generalized version of DICEBOX~\cite{DICEBOX}, as already discussed biefly in Sec.~\ref{sec:unf}. Again,
we have considered an artificial nucleus resembling $^{57}$Fe, with a 
spin distribution of the initial excitation-energy levels of $1/2 \leq J \leq 13/2$, and  
bin size of 120 keV. Note also that equally many negative- and positive-parity states are assumed above the region
of known, discrete levels.

\begin{figure}[!hbt]
\begin{center}
\includegraphics[clip,width=\columnwidth]{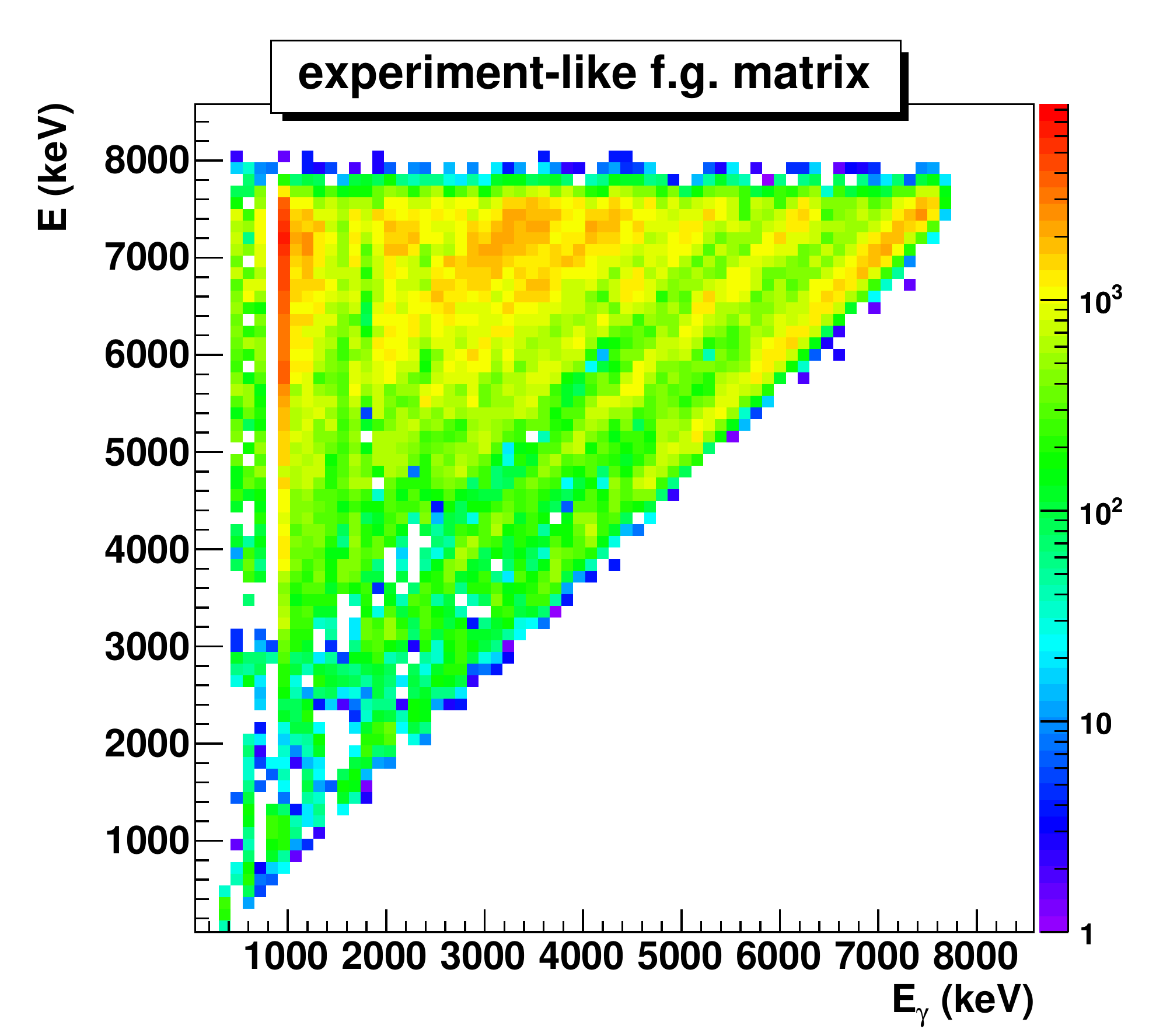}
\includegraphics[clip,width=\columnwidth]{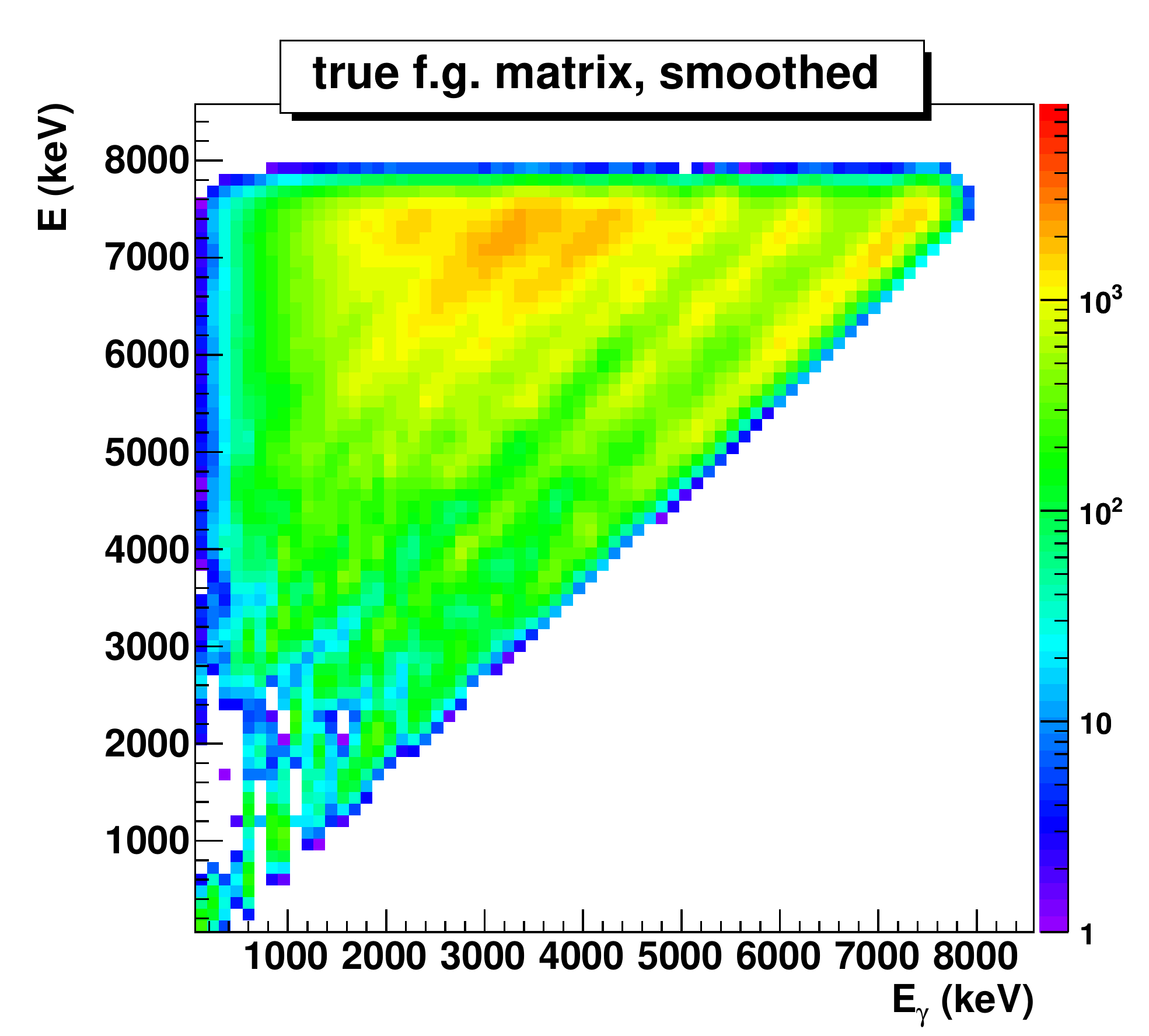}
\caption{(Color online). Simulated first-generation matrix for $^{57}$Fe. Top: spectra from folded data set. 
	Bottom: true spectra with smoothing similar to the experimental resolution.}
\label{fig:testfg}
\end{center}
\end{figure}

First, the simulated spectra were 
folded with the CACTUS response functions, and also a Gaussian smoothing was applied giving a 
full width at half maximum (FWHM) of 250 keV for all excitation energies. These spectra were thus made to
be as similar as possible to experimental spectra. Then, we applied the unfolding technique and the first-generation
method to obtain the first-generation spectra. We then compared the extracted first-generation matrix with the
true first-generation matrix from the simulations, after smoothing the true 
first-generation spectra with a resolution similar to the experimental one. 
Examples of two such matrices for one nuclear realization are shown in Fig.~\ref{fig:testfg}. 
%--------------------------------------------------------------------------------------%
%--------------------------------------------------------------------------------------%
The overall good similarity between the two matrices is gratifying. However, in the low-$E_{\gamma}$ region, 
there are significant differences 
between the results extracted from the experiment-like matrix and the true, smoothed
matrix. In particular, one can see that there are some vertical lines in the experiment-like first-generation matrix, 
e.g., for $E_{\gamma} \approx $1020 keV, that are not present in the true matrix. 
This particular vertical ridge originates from the 7/2$^{-}$ state at 1007 keV, which for this 
nuclear realization is strongly populated in the decay cascades at high excitation energy. However, at low excitation
energies, which corresponds to the population from the direct reaction, this state is only moderately populated. Thus, its
$\gamma$ decay is not
correctly subtracted in the first-generation procedure. It is therefore important to exclude such leftovers
from higher-generation $\gamma$ rays in the further analysis, as mentioned in Sec.~\ref{subsec:nld_rsf}.

\begin{figure}[!hbt]
\begin{center}
\includegraphics[clip,width=.94\columnwidth]{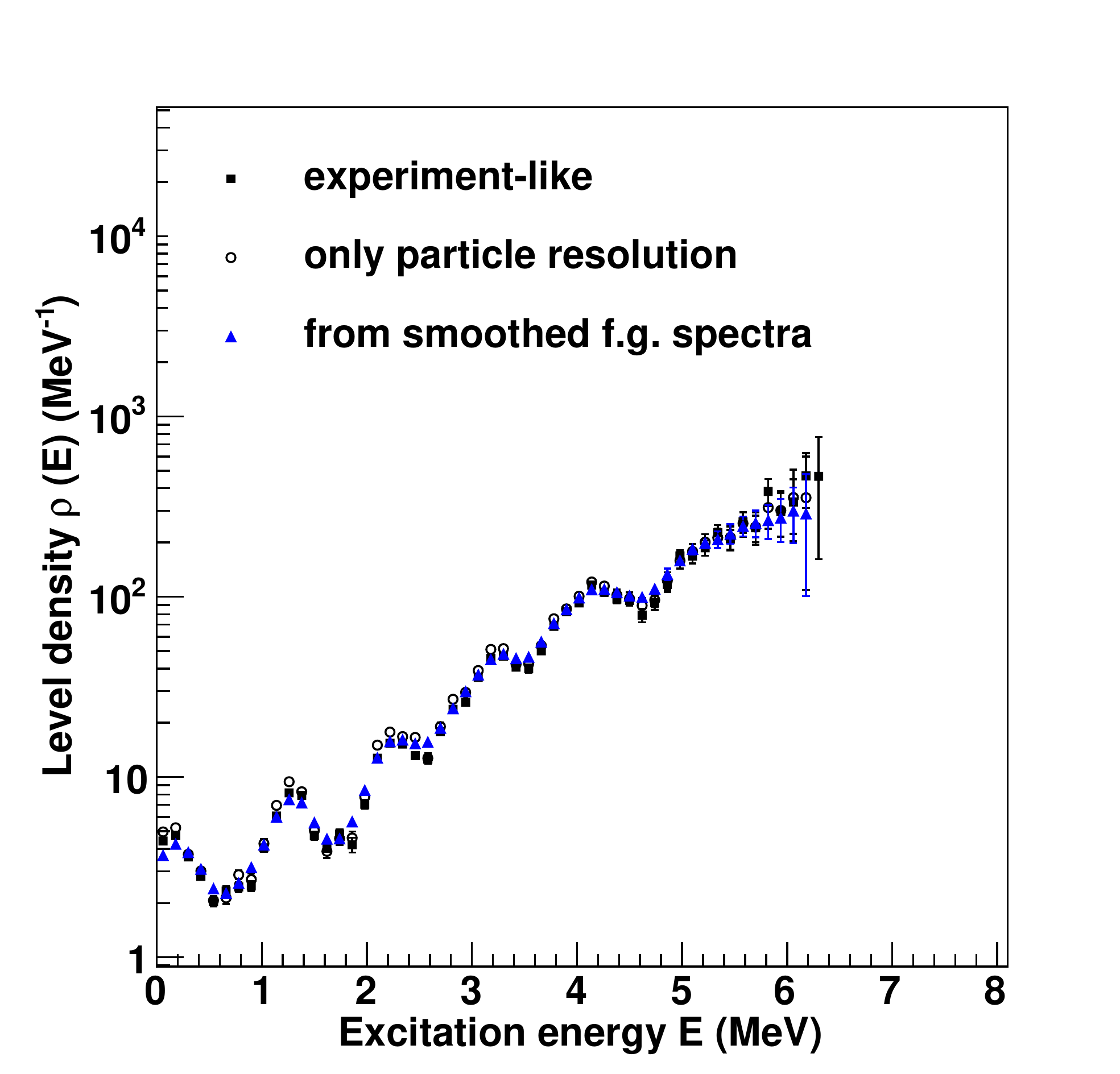}
\includegraphics[clip,width=.94\columnwidth]{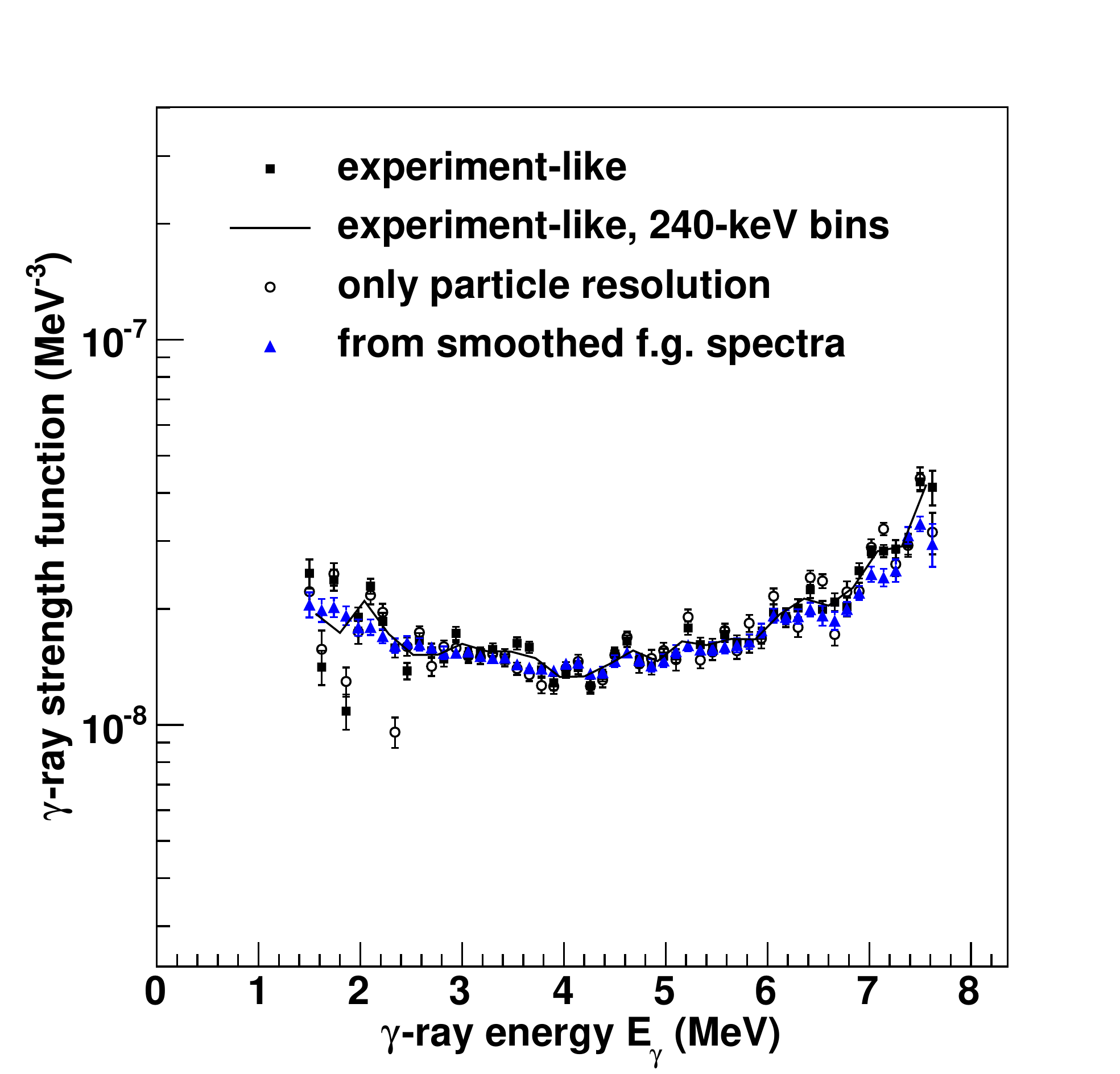}
\caption{(Color online). Results from simulated spectra on $^{57}$Fe. Top: level density. 
	Bottom: $\gamma$-ray strength function.}
\label{fig:testrho_and_rsf}
\end{center}
\end{figure}

We also tested the case with 
a Gaussian smoothing on the particle resolution, 
and the ideal response on the $\gamma$-detection side. The extracted 
level densities and strength functions for all three cases are displayed in Fig.~\ref{fig:testrho_and_rsf}.  
%--------------------------------------------------------------------------------------%
%--------------------------------------------------------------------------------------%
In general, the results agree very well. We note that there are larger fluctuations
in the strength function extracted from the experiment-like matrix as well as the case where only the 
particle resolution is applied, especially in the region below $E_{\gamma} \approx 2.5$ MeV. These fluctuations
are related to uncertainties in the first-generation subtraction procedure and could be due to small variations in the shape
of the $\gamma$ spectra. However, by compressing the experiment-like $\gamma$ spectra by a factor of two (bin size of 240 keV), 
we obtain practically the same shape of the $\gamma$-ray strength function 
as from the true first-generation spectra 
(see Fig.~\ref{fig:testrho_and_rsf}).  
 
\begin{figure}[!hbt]
\begin{center}
\includegraphics[clip,width=.94\columnwidth]{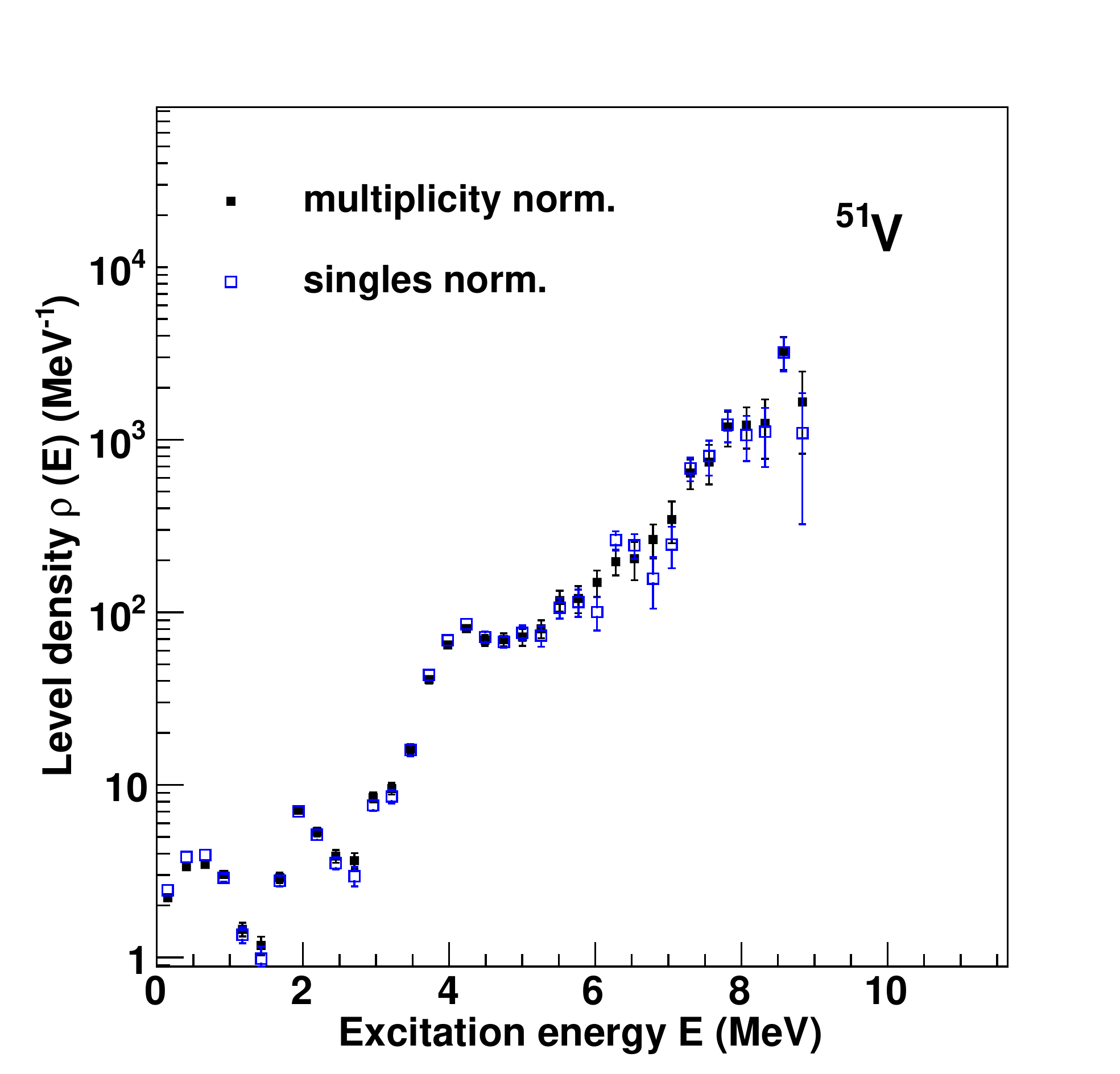}
\includegraphics[clip,width=.94\columnwidth]{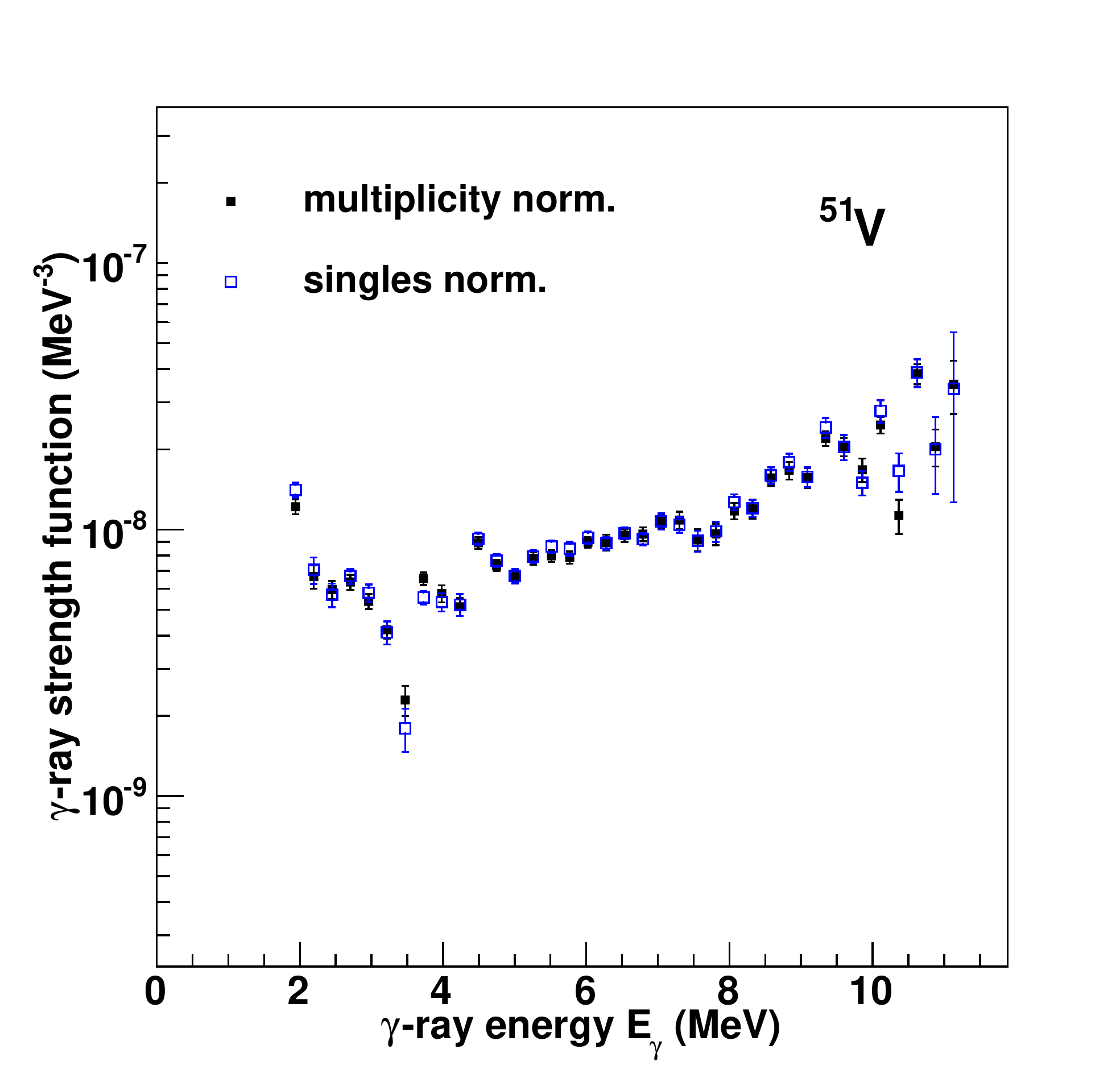}
\caption{(Color online). Test of the two normalization options in the first-generation method for $^{51}$V. 
	Top: level density. 
	Bottom: $\gamma$-ray strength function. The data are taken from the experiment
 presented in Ref.~\cite{V}.}
\label{fig:Vtest}
\end{center}
\end{figure}

Finally, we have tested the two normalization options (singles or multiplicity) of the first-generation method on experimental
spectra to investigate the effect on the extracted level density and $\gamma$-ray strength function. 
The result for $^{51}$V is shown in Fig.~\ref{fig:Vtest}. 
%--------------------------------------------------------------------------------------%
%--------------------------------------------------------------------------------------%
We observe that the two options give very similar results, only a very few data points are outside the experimental error bars. 
It is clear that the two methods do not give any difference in the overall shape of neither the level density nor the strength function.

\subsection{The Brink hypothesis}
\label{subsec:brink}

The $\gamma$-ray transmission coefficient $\mathcal{T}(E_{\gamma})$ in Eq.~(\ref{eq:brink}) is assumed to be 
independent of excitation energy
(and thus nuclear temperature) according to the generalized Brink hypothesis~\cite{brink}, as discussed 
in Sec.~\ref{subsec:nld_rsf}. This hypothesis is 
violated when high temperatures and/or spins are involved in the nuclear reactions, as shown for giant dipole 
excitations (see Ref.~\cite{Andreas&Thoennessen} and references therein). However, since both the temperature reached and the spins 
populated are rather low for the Oslo experiments, these dependencies are usually assumed to be of minor importance in the 
relatively low excitation-energy region considered here. 

The effect of the Brink hypothesis has been tested by analyzing simulated spectra on an artificial nucleus resembling $^{163}$Dy. 
The simulations were again performed  
with the DICEBOX code~\cite{DICEBOX} for a specific spin range on the initial excited levels, 
$1/2 \leq J \leq 13/2$. 

\begin{figure}[bt]
\begin{center}
\includegraphics[clip,width=\columnwidth]{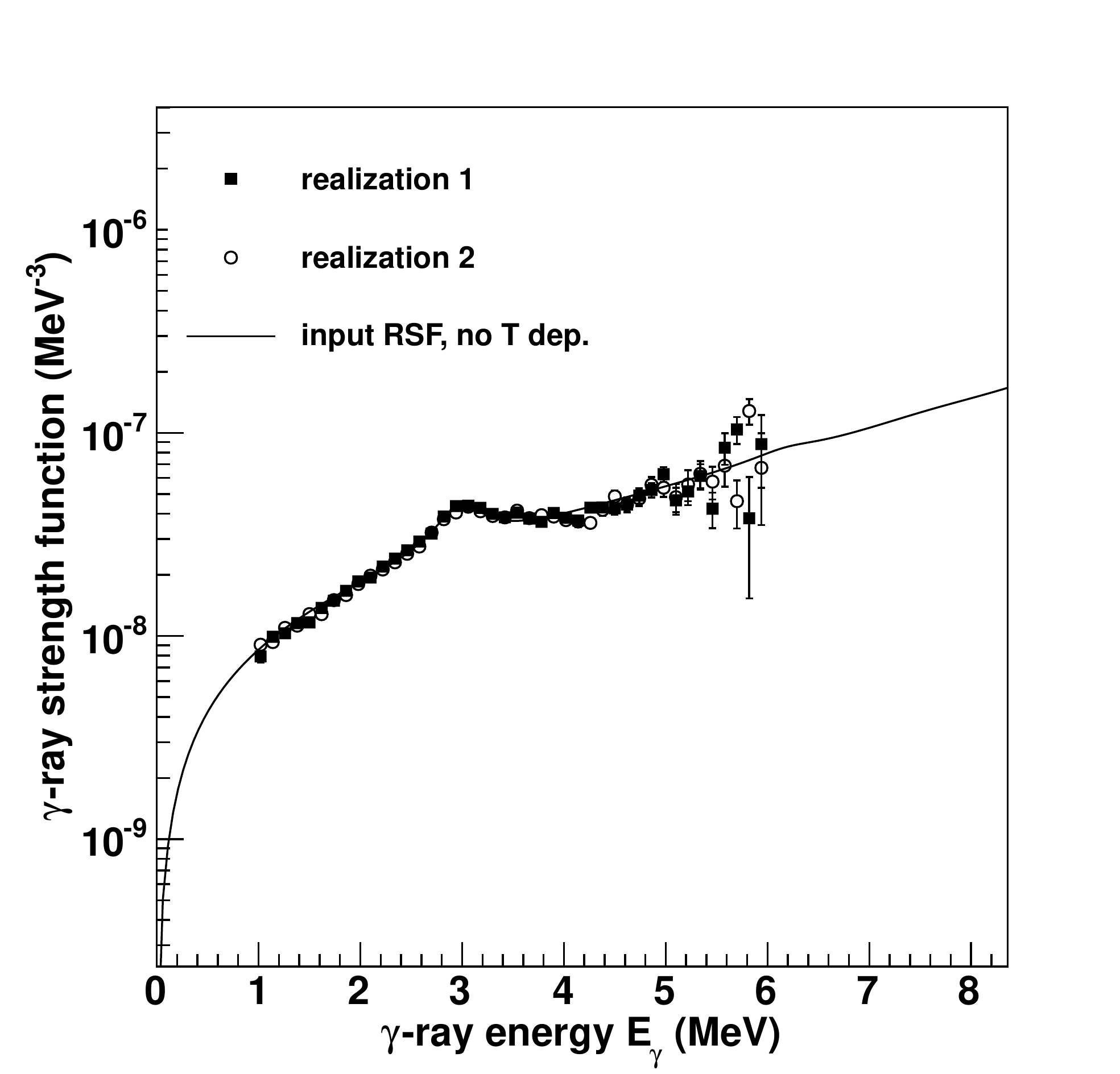}
\caption{Results from simulated spectra of the artificial nucleus $^{163}$Dy.  The extracted 
	$\gamma$-ray strength functions %normalized to the experimental $\left< \Gamma_\gamma \right>$
	are shown as black points and open circles for two different nuclear realizations, while the 
	input $\gamma$-ray strength model is shown as a solid line.}
\label{fig:testDy_BA}
\end{center}
\end{figure}

As a first step, a temperature-independent model for the $\gamma$-ray strength function was used as input for the simulations. 
The extracted and input $\gamma$-ray strength function are shown in Fig.~\ref{fig:testDy_BA}.
As expected, the Oslo method works very well in this case. 
%--------------------------------------------------------------------------------------%
%--------------------------------------------------------------------------------------%

In the next test, a temperature-dependent input $\gamma$-ray strength function was used, with temperature $T \propto \sqrt{E_{\mathrm{f}}}$.
%(the Kadmenski{\u{\i}},
%Markushev and Furman (KMF) model \cite{ka83}). 
In principle, it is 
not possible to disentangle the input level density and $\gamma$-ray strength function anymore, since now we have
\begin{equation}
P(E, E_{\gamma}) \propto  \rho (E_{\mathrm{f}}) {\mathcal{T}}  (E_{\mathrm{f}},E_{\gamma}).
\end{equation}
This we keep in mind when we use the procedure of Ref.~\cite{schi0} in order to extract the level density and $\gamma$-ray strength function. 

\begin{figure}[ht]
\begin{center}
\includegraphics[clip,width=\columnwidth]{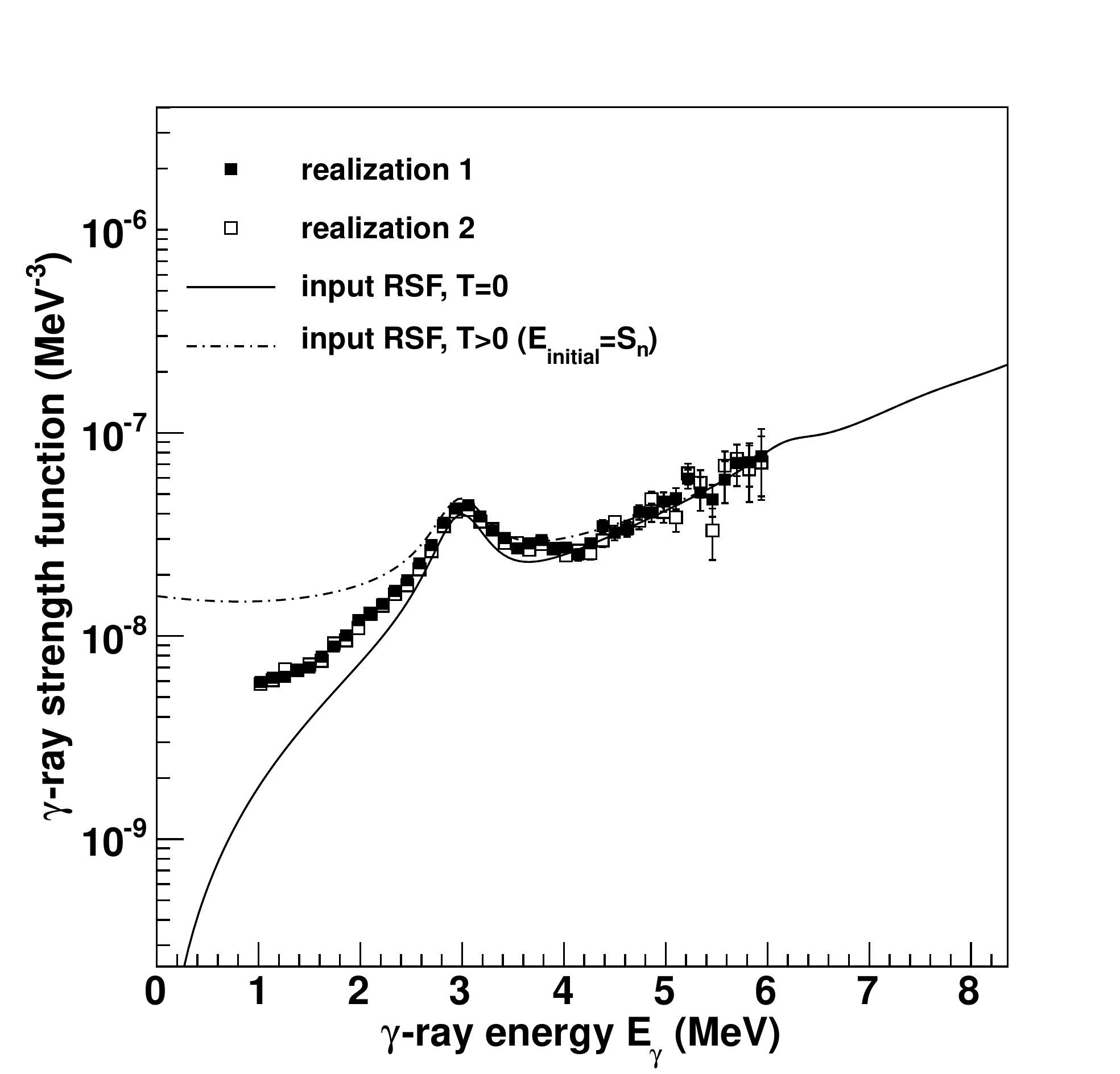}
\caption{Results from simulated spectra of $^{163}$Dy. The extracted 
	$\gamma$-ray strength functions are displayed as black and open squares for two different 
	nuclear realizations. The input $\gamma$-ray strength function models for $T_f=0$ (solid line) and 
	$T_f \propto \sqrt{(S_n - E_\gamma)}$ (dashed-dotted line) are also shown. }
\label{fig:testDy_KMF}
\end{center}
\end{figure}

The extracted $\gamma$-ray strength functions for two different nuclear realizations are shown in Fig.~\ref{fig:testDy_KMF}, within the excitation-energy
range $2.1 < E < 6.2$ MeV.
%--------------------------------------------------------------------------------------%
%--------------------------------------------------------------------------------------%
It is seen that the extracted $\gamma$-ray strength function lies in between the two extremes of the temperature-dependent
input model, and thus an average strength function for the excitation energy region under study is
found. The shape of the extracted $\gamma$-ray strength function is therefore quite reasonable, although it is clear that the low-energy
part with $1 \leq E_{\gamma} \leq 2.5$ MeV must necessarily be quite different from the two extremes of the input.

\begin{figure}[ht]
\begin{center}
\includegraphics[clip,width=\columnwidth]{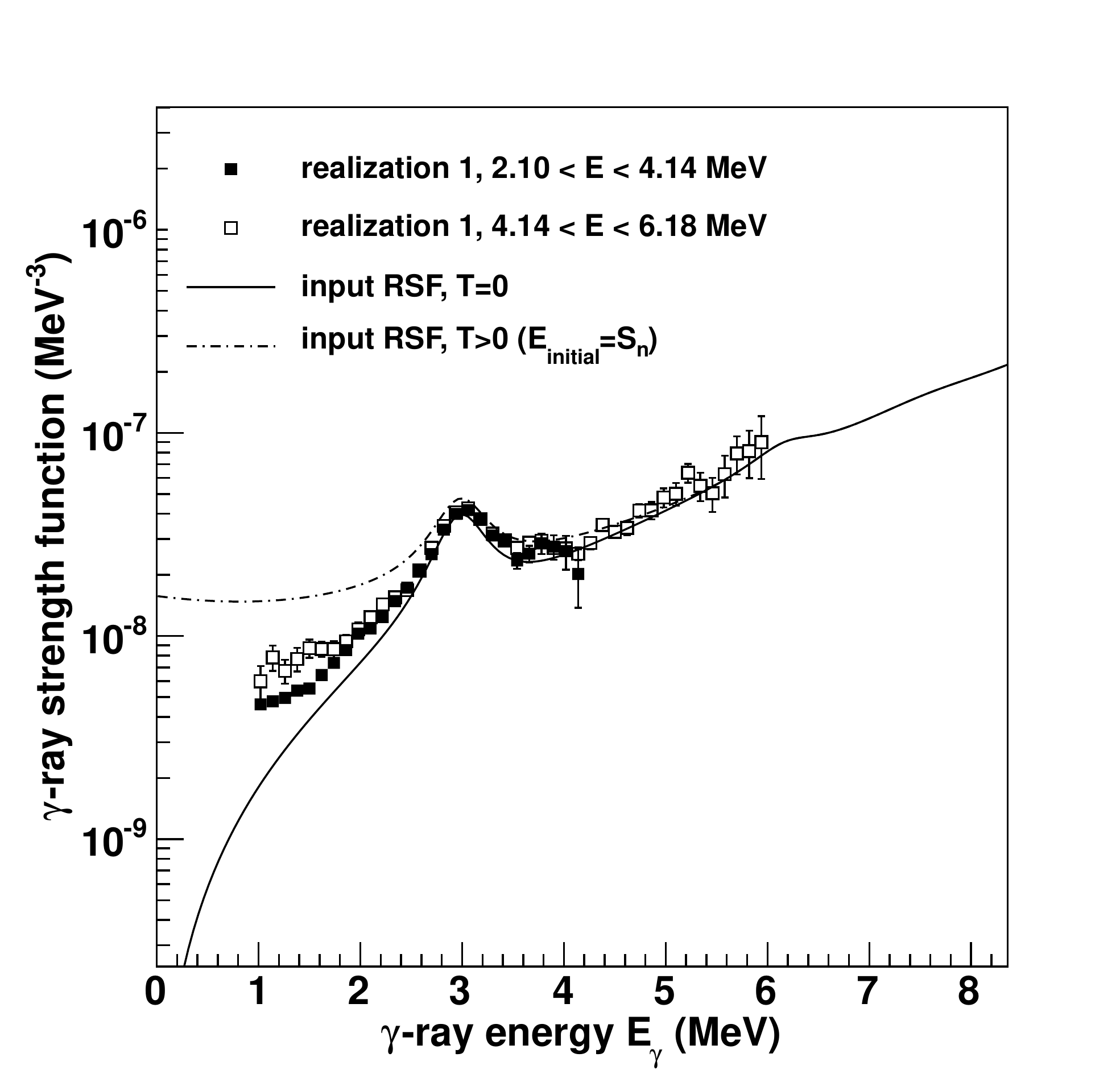}
\caption{Results from simulated spectra of the artificial nucleus $^{163}$Dy. The extracted 
	$\gamma$-ray strength functions for two excitation-energy regions are displayed as black and open squares. 
	The input $\gamma$-ray strength function models for $T_f=0$ (solid line) and 
	$T_f \propto \sqrt{(S_n - E_\gamma)}$ (dashed-dotted line) are also shown. }
\label{fig:testDy_lowhigh}
\end{center}
\end{figure}

This is further illustrated in Fig.~\ref{fig:testDy_lowhigh}, where we have extracted the $\gamma$-ray strength function 
for two separate 
excitation-energy regions: $2.1 < E < 4.1$ MeV and $4.1 < E < 6.2$ MeV. 
%--------------------------------------------------------------------------------------%
%--------------------------------------------------------------------------------------%
It is easily seen that the $\gamma$-ray strength function for $\gamma$ energies between $\approx 1-2$ MeV is different in the two cases;
the higher excitation energies lead to a higher temperature of the final states and thus a higher strength function. 

\begin{figure}[tb]
\centering
\includegraphics[clip,width=\columnwidth]{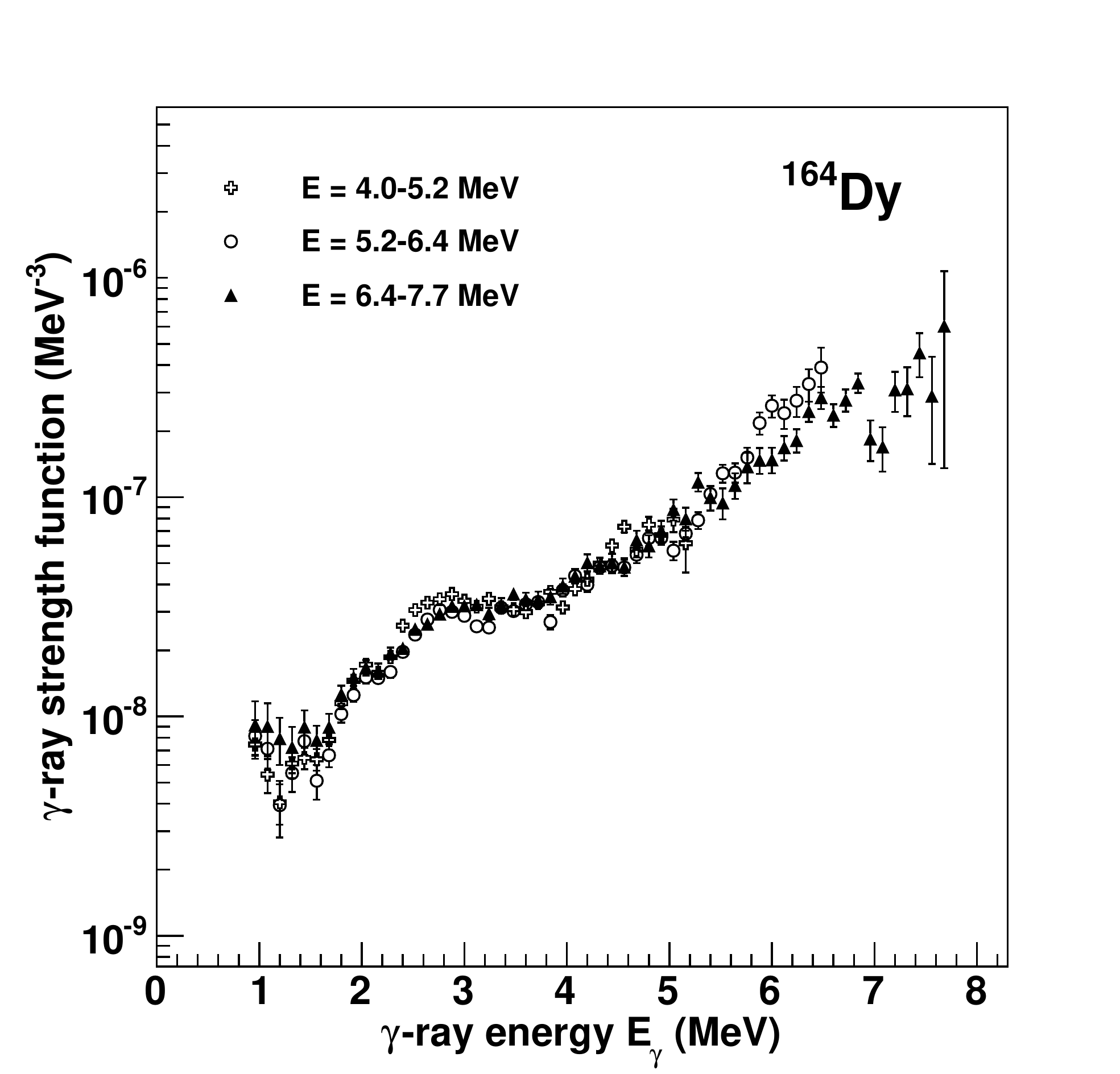}
\caption{Experimental $\gamma$-ray strength function of $^{164}$Dy for three 
	sets of initial excitation energies. The data are taken from the experiment
 presented in Ref.~\cite{hildes_DyRSF}.}
\label{fig:Dy164rsf}
\end{figure}

However, one should keep in mind that the experimental $\gamma$-ray strength functions have been tested 
against the assumption of temperature dependence for many nuclei, e.g., $^{45}$Sc~\cite{Sc}, $^{56,57}$Fe \cite{Fe_Alex},
$^{96,98}$Mo \cite{Mo_RSF}, and $^{117}$Sn~\cite{Sn_RSF}. This is also shown for $^{164}$Dy~\cite{hildes_DyRSF} in Fig.~\ref{fig:Dy164rsf}, 
where the $\gamma$-ray strength function has been extracted for three sets of initial excitation energies. 
As can be seen from the figure, the similarity of the three $\gamma$-ray strength functions is striking. 
%-----------------------------------------------------------------%
%-----------------------------------------------------------------%
There is, therefore, no experimental evidence in this excitation-energy region for a strong temperature dependence in the 
strength function. Hence, 
the Brink hypothesis seems to be valid here.

\subsection{The parity distribution}
\label{subsec:parity}

As mentioned previously, for both the normalization of the level density and the $\gamma$-ray transmission coefficient, 
the assumption of equally many levels with positive and negative parity is used. We will in the following investigate this 
assumption in detail.

Using $\rho_+$ and $\rho_-$ to denote the level density with positive and negative parity levels, 
the parity asymmetry $\alpha$ is defined as \cite{gary}
\begin{equation}
\alpha=\frac{\rho_+-\rho_-}{\rho_++\rho_-},
\label{eq:asym}
\end{equation}
which gives $-1$ and $1$ for only negative and positive parities, respectively, and 0 when both parities 
are equally represented.  

Another expression widely used in literature is the ratio $\rho_{-}/\rho_{+}$, which relates to $\alpha$ by
\begin{equation}
\frac{\rho_{-}}{\rho_{+}} = \frac{1-\alpha}{1+\alpha}.
\end{equation}

\begin{figure}[tb]
\centering
\includegraphics[clip,width=\columnwidth]{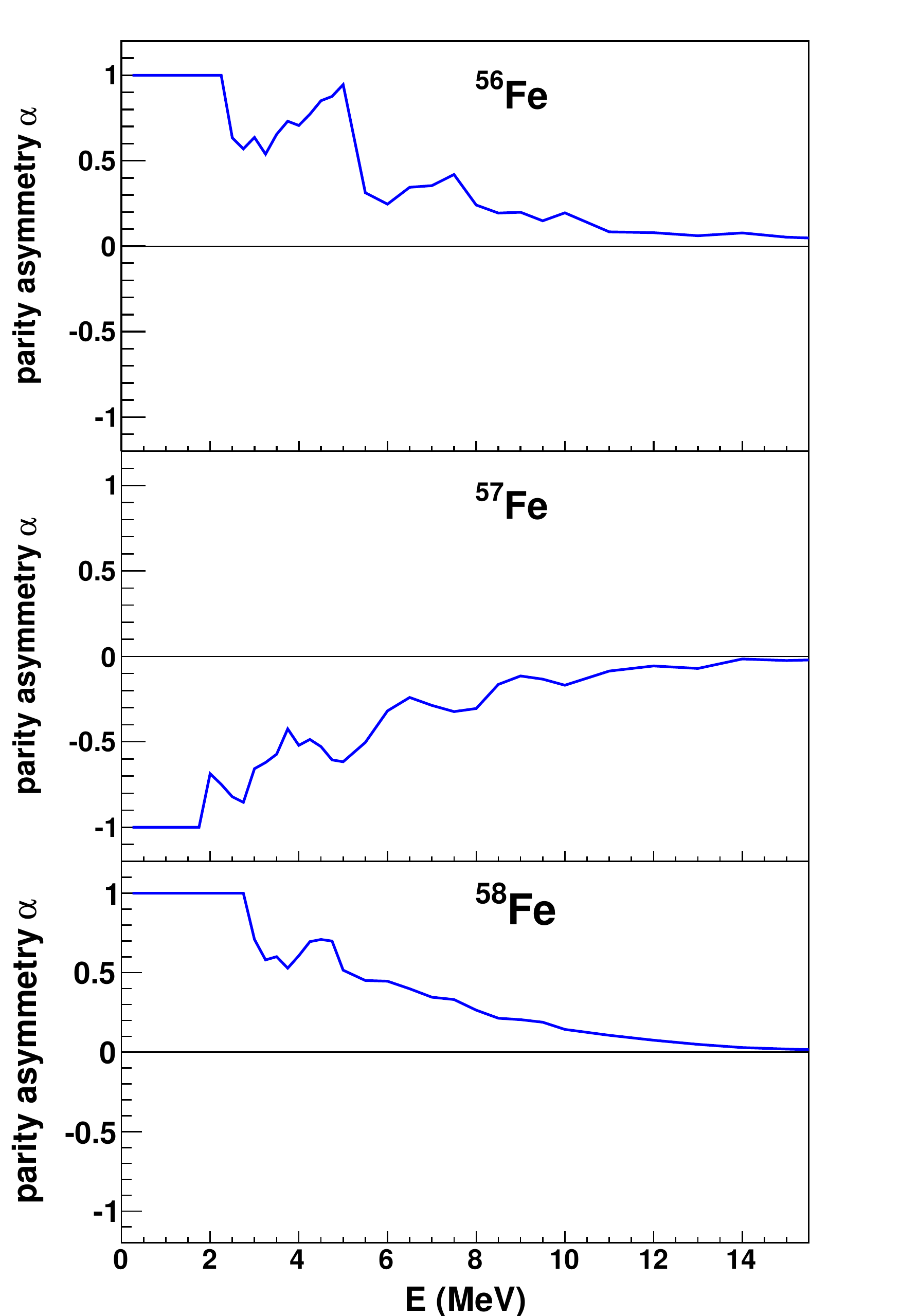}
\caption{(Color online). Calculated parity distributions as a function of 
	excitation energy calculated for $^{56-58}$Fe, from Ref.~\cite{go08}.}
\label{fig:parFe}
\end{figure}

\begin{figure}[tb]
\centering
\includegraphics[clip,width=\columnwidth]{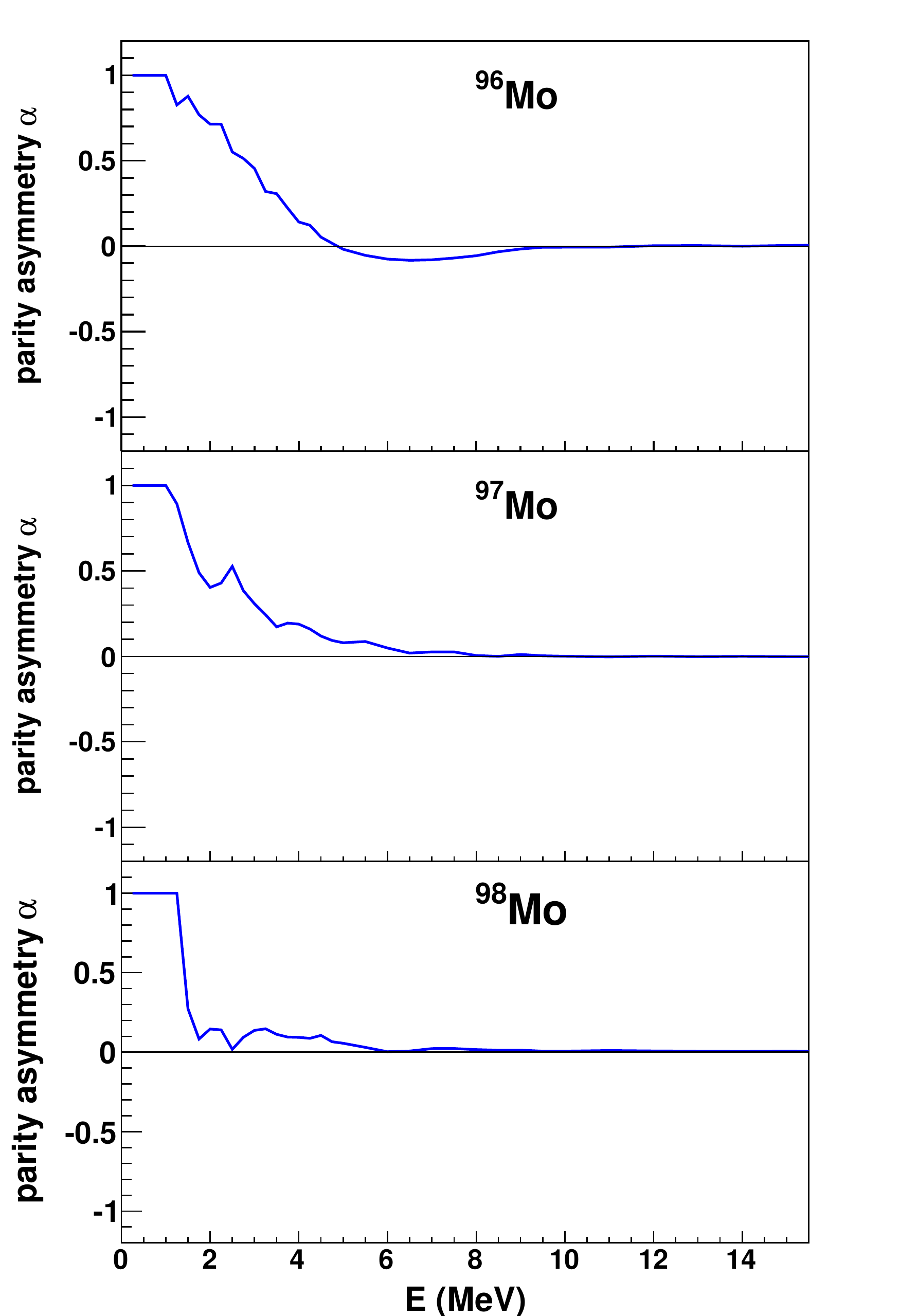}
\caption{(Color online). Same as Fig.~\ref{fig:parFe} for $^{96-98}$Mo.}
\label{fig:parMo}
\end{figure}

\begin{figure}[tb]
\centering
\includegraphics[clip,width=\columnwidth]{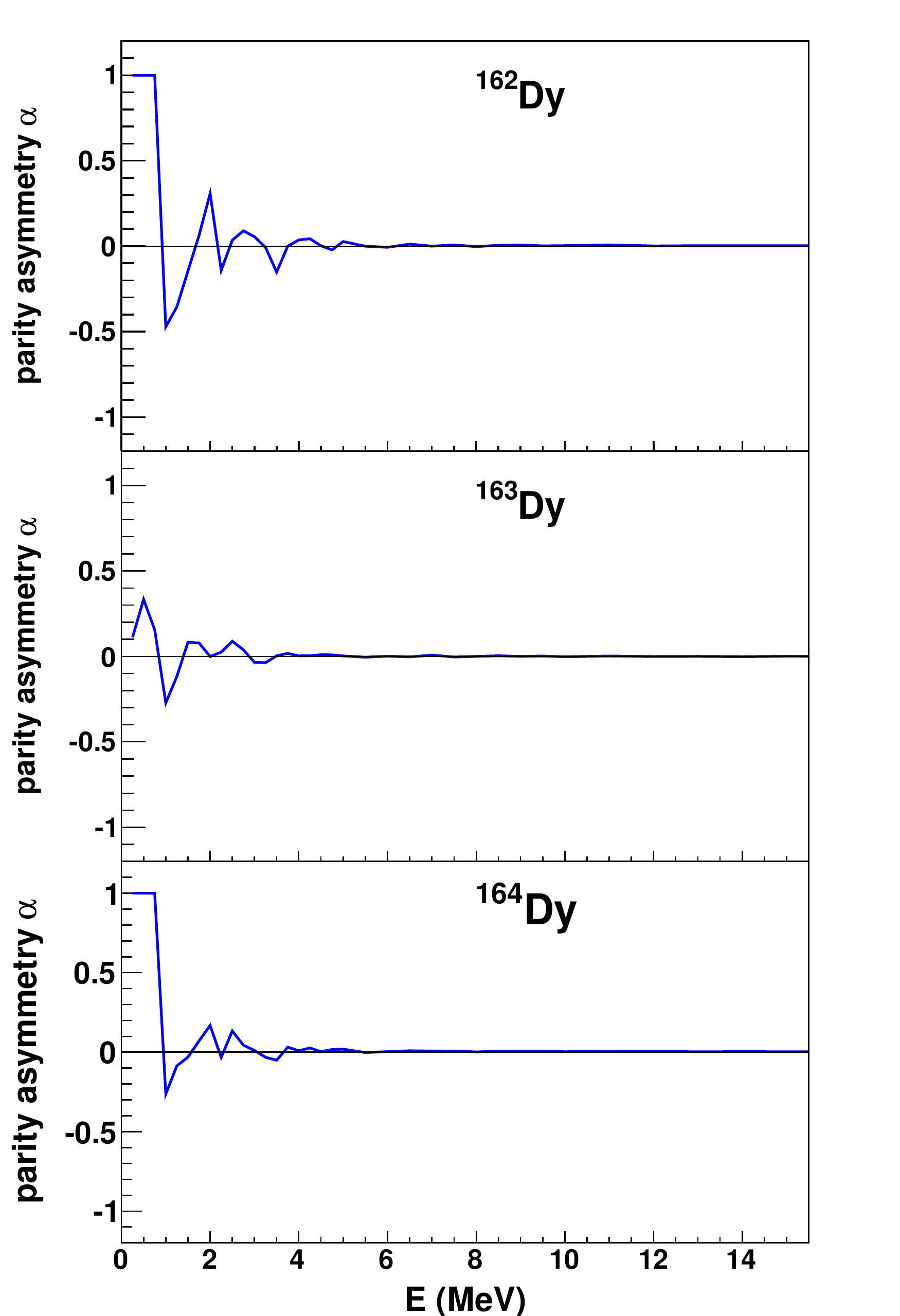}
\caption{(Color online). Same as Fig.~\ref{fig:parFe} for $^{162-164}$Dy.}
\label{fig:parDy}
\end{figure}

We have considered theoretical parity distributions from combinatorial plus Hartree-Fock-Bogoliubov calculations
of spin- and parity-dependent level densities~\cite{go08}. Applying the definition in Eq.~(\ref{eq:asym}), we
find the calculated parity distributions for several Fe, Mo, and Dy isotopes as shown in Figs.~\ref{fig:parFe}--\ref{fig:parDy}.
As one might expect from the fact that more orbits are accessible at increasing excitation energy, the 
parity distributions are seen to approach zero as the excitation energy increases. However, it is clear that for the lighter
nuclei, in particular the Fe isotopes, the assumption of zero parity asymmetry is not fulfilled in the calculations for 
excitation energies below 
$\approx 10$ MeV. 
%-----------------------------------------------------------------%
%-----------------------------------------------------------------%
%-----------------------------------------------------------------%
%-----------------------------------------------------------------%
%-----------------------------------------------------------------%
%-----------------------------------------------------------------%

We have also looked at other theoretical work such as shell-model Monte Carlo calculations~\cite{Alhassid} and 
macroscopic-microscopic calculations~\cite{Rauscher}. In Fig.~2 of Ref.~\cite{Rauscher}, 
the ratio $\rho_{-}/\rho_+$ is shown for $^{56}$Fe, indicating a value of $\rho_{-}/\rho_+ \simeq 0.1$ 
at 10 MeV excitation energy. From Fig.~4 in Ref.~\cite{Alhassid}, the ratio $\rho_{-}/\rho_+ \simeq 0.2$ for $E=10$ MeV.
In contrast to this, the combinatorial calculations of Ref.~\cite{go08} give $\rho_{-}/\rho_+ \simeq 0.7$ at $E=10$ MeV, 
which is also in accordance with other microscopic calculations based on the Nilsson model and BCS quasi-particles~\cite{Fe_Emel},
where $\rho_{-}/\rho_+ \simeq 0.5$. These results indicate considerably more negative-parity states in $^{56}$Fe 
than found in Refs.~\cite{Alhassid,Rauscher}. In this specific case, the amount of positive-parity states is 
very sensitive to the position of the $g_{9/2}$ orbital relative to the Fermi level. 

To our understanding, there are currently no experimental data on the parity distribution in $^{56}$Fe. 
However, recent measurements on level
densities of $J^{\pi} = 2^{+}$ and $2^-$ states in $^{58}$Ni and $^{90}$Zr~\cite{vonNeumann-Cosel} show no indication 
of a significantly larger amount of states  with one of the parities in none of the nuclei under study at $E \approx 10$ MeV. 
Also, from the study of proton resonances in $^{45}$Sc~\cite{gary}, equally many $1/2^{+}$ and $1/2^{-}$ states were found, 
again at $E \approx 10$ MeV.
Thus it seems reasonable to assume that the parity asymmetry is at least very small for these excitation energies, in support
of the assumption of equal parity as described in Sec.~\ref{subsec:norm}. 

We would nevertheless like to investigate the impact of the assumption of parity symmetry on the calculations of $\rho(S_n)$. 
Let us assume that the spin- and parity-projected level density 
$\rho(E,J,\pi)$ can be described by~\cite{Rauscher}
\begin{equation}
\rho(E,J,\pi) = \rho(E) \cdot g(E,J) \cdot \mathcal{P}(E,\pi),
\label{eq:newrhoBn}
\end{equation}
where $\rho(E)$ is the total level density at excitation energy $E$, $g(E,J)$ 
is the spin distribution given by Eq.~(\ref{eq:spindist}), and $\mathcal{P}(E,\pi)$ is 
the parity projection factor. According to Eq.~(\ref{eq:D}), we get
\begin{equation}
\frac{1}{D_0} = \rho(S_n) \cdot g(S_n,J=I_t\pm 1/2) \cdot \mathcal{P}(S_n,\pi_t)
\end{equation}
for the neutron resonance spacing at $S_n$ reaching states with parity 
$\pi_t \cdot (-1)^{\ell} = \pi_t$ for s-wave neutrons having $\ell = 0$. 
Now, we define the parity projection factor for positive and negative parities as 
\begin{equation}
\mathcal{P}_+  \equiv \mathcal{P}(E,\pi=\pi_+) =  \frac{\rho_+}{\rho} = \frac{1+\alpha}{2}, \\
\label{eq:parity_g}
\end{equation}
and
\begin{equation}
\mathcal{P}_-  \equiv \mathcal{P}(E,\pi=\pi_-) =  \frac{\rho_-}{\rho} = \frac{1-\alpha}{2},
\label{eq:parity_s}
\end{equation}
using
\begin{equation}
\mathcal{P}_+ + \mathcal{P}_- = 1.
\label{eq:parunity}
\end{equation}
Further,
\begin{align}
\frac{1}{D_0} & = \rho(S_n) \left[ g(S_n,J=I_t+1/2) + g(S_n,J=I_t-1/2)\right]  \mathcal{P} (S_n,\pi) \\
	& = \rho(S_n) \left[ g(S_n,J=I_t+1/2) + g(S_n,J=I_t-1/2)\right] \frac{1\pm\alpha}{2},
\end{align}
which gives
\begin{equation}
\rho(S_n) = \frac{\sigma^2}{D_0} \frac{(1\pm\alpha)/2}{(I_t+1)\exp\left[-(I_t+1)^2/2\sigma^2\right] + I_t\exp\left[-I_t^2/2\sigma^2\right]},
\label{eq:newD}
\end{equation}
using Eq.~(\ref{eq:spindist}). If the target nucleus in the neutron-capture experiment has positive parity in the ground state, the 
factor $(1+\alpha)/2$ is used, and for negative ground-state parity we use $(1-\alpha)/2$.

For several
key cases, we have applied Eq.~(\ref{eq:newD}) for calculating $\rho(S_n)$ and compared to the result using Eq.~(\ref{eq:oldD}). 
For example, for $^{58}$Fe with neutron resonance spacing $D_0 = 6.5$ keV at $S_n = 10.044$ MeV~\cite{RIPL}, 
and using the spin cutoff parameter $\sigma(S_n) = 3.93$ from the prescription of Ref.~\cite{egidy2}, we obtain $\rho(S_n) = 2518$ MeV$^{-1}$
if the assumption of equal parity is used, and $\rho(S_n) = 2939$ MeV$^{-1}$ if we correct for the parity asymmetry $\alpha = 0.14$
predicted by the combinatorial model of \cite{go08}. Thus, including the parity asymmetry gives about 17\% higher level
density at $S_n$. For $^{96}$Mo, with $D_0 = 105$ eV~\cite{RIPL}, $\sigma(S_n) = 5.15$, and $\alpha = -0.017$~\cite{go08}, 
we get $\rho(S_n) = 1.01\cdot 10^5$ MeV$^{-1}$ when no parity asymmetry is taken into account, and $\rho(S_n) = 1.03\cdot 10^5$ MeV$^{-1}$
when the parity asymmetry is considered; only a change of $\approx 2$\%.

To check the effect of the parity asymmetry on the normalization, we have renormalized the data on $^{96}$Mo using
the above-mentioned values for $\rho(S_n)$ with and without parity correction. Since the estimated $\alpha$ was so small
for this case, we have also assumed a much larger parity asymmetry of $\alpha=-0.5$ leading to $\rho(S_n) = 2.02\cdot 10^5$ MeV$^{-1}$,
a factor of 2 larger level density than the one assuming equal parity. Assuming that  $\alpha=0.5$ gives 
$\rho(S_n) = 6.74\cdot 10^4$ MeV$^{-1}$, roughly a factor of 2/3 reduction compared to the parity-symmetry case.

\begin{figure}[tb]
\centering
\includegraphics[clip,width=\columnwidth]{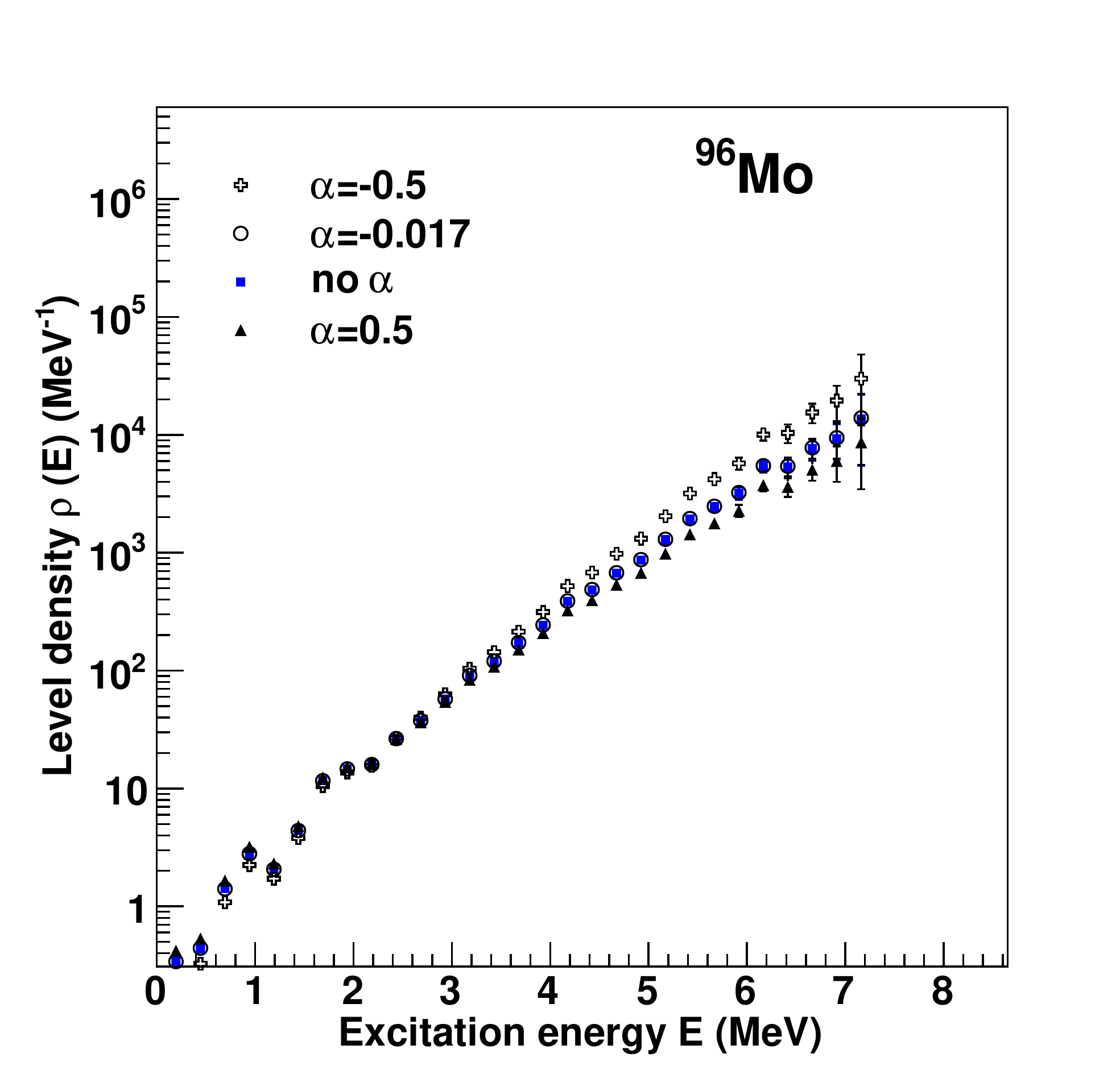}
\includegraphics[clip,width=\columnwidth]{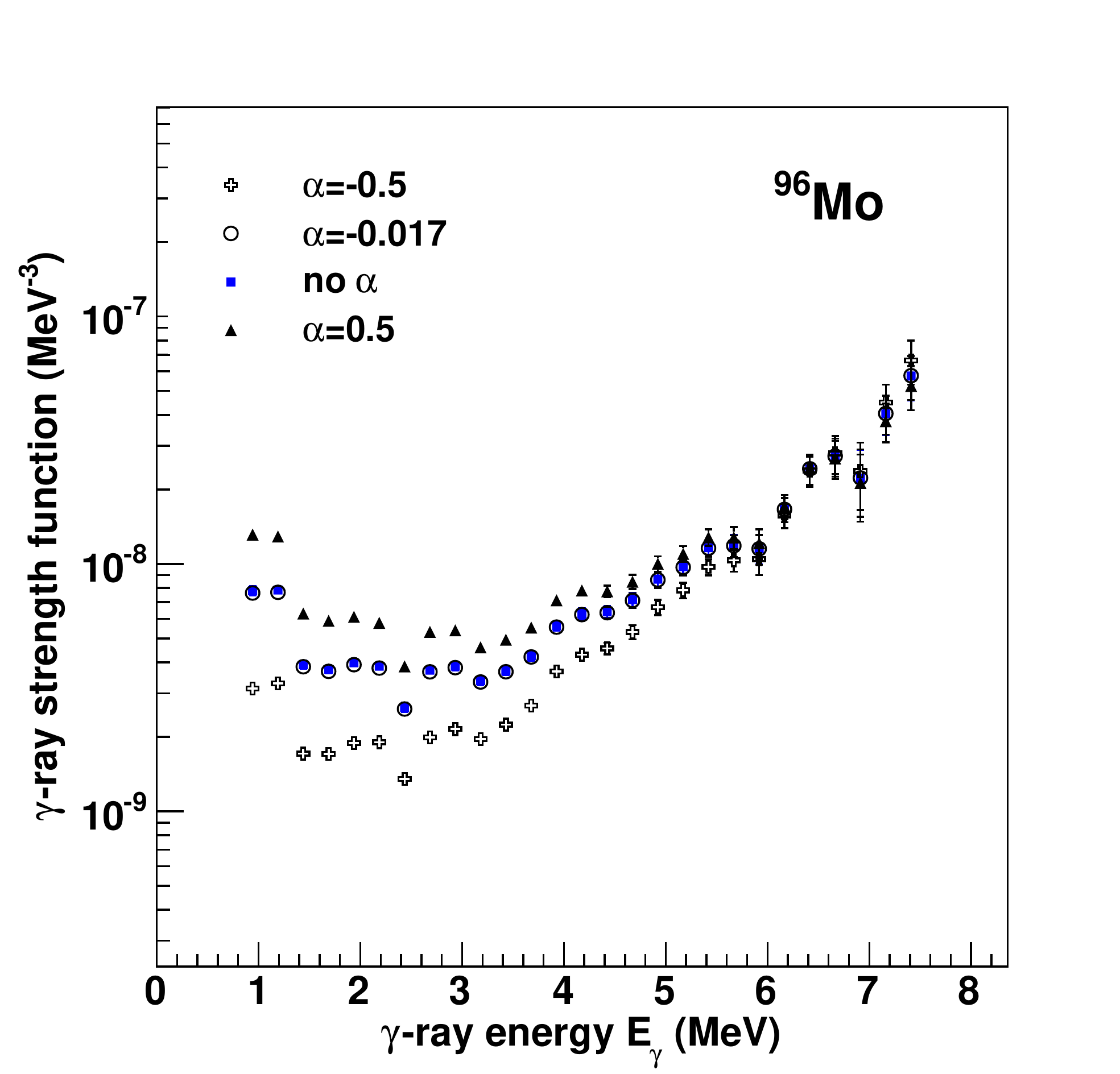}
\caption{(Color online). Effect of parity asymmetry on the experimental level density (upper panel) 
	and strength function (lower panel) 
	of $^{96}$Mo. The data are taken from the experiment
 presented in Ref.~\cite{Mo_RSF}.}
\label{fig:Mopartest}
\end{figure}

The resulting level density and strength function are shown in Figs.~\ref{fig:Mopartest}.
%-----------------------------------------------------------------%
%-----------------------------------------------------------------%
Here, it is easily seen that a small parity asymmetry does not give any significant changes of the normalization
in neither the level density nor the strength function. However, for $\alpha=-0.5$ corresponding to a factor of 2 
larger $\rho(S_n)$ gives an overall larger level density for excitation energies larger than $\approx 3$ MeV, 
and the steeper slope is reflected also in the strength function. In addition, one sees a suppression in the 
$\gamma$-ray strength function for $E_\gamma \lesssim 6$ MeV, a direct consequence of the changes in the level density. The opposite is 
true for $\alpha=0.5$; here, $\rho(S_n)$ and thus the slope of the level density is reduced. Consequently,
the slope is reduced also in the $\gamma$-ray strength function, but the absolute value for $E_\gamma \lesssim 6$ MeV is in this case
increased with respect to the parity-symmetry case. Note that such large values of $\alpha$ represent extreme cases;
both from experimental data~\cite{gary,vonNeumann-Cosel} and several theoretical estimates (e.g, \cite{go08,Fe_Emel})
the asymmetry around $S_n$ should be smaller than typically $\pm0.2$. 

The assumption of parity symmetry influences also the 
normalization of the $\gamma$-ray transmission coefficient. To take into account the parity distribution, 
one can modify Eq.~(\ref{eq:rhopar}) according to Eq.~(\ref{eq:newrhoBn}) so that
\begin{equation}
\rho(E-E_{\gamma}, J_{\mathrm{f}},\pi_{\mathrm{f}}) = \rho(E-E_{\gamma}) 
\cdot \mathcal{P}(E-E_{\gamma},\pi_{\mathrm{f}})\cdot g(E-E_{\gamma},J_{\mathrm{f}}) . 
\end{equation}
We will restrict ourselves to consider $\ell=1$ radiation only ($E1$ and $M1$), since in general, this multipolarity 
is expected to give by far the largest contribution to the strength function in the quasicontinuum region (see, e.g., studies
by Kopecky and Uhl~\cite{Kopecky&Uhl_2}). 

For $E1$ radiation, the parity of the final state is opposite to the initial state. In s-wave neutron capture experiments,
the parity of the target nucleus' ground state is equal to the parity of the neutron resonances of the created compound nucleus.
Therefore, the accessible parity of the final states must be the opposite of the initial state. 
For $M1$ radiation, the accessible 
final states must have the same parity as the initial state. 

Based on Eq.~(\ref{eq:longGamma}) and taking the parity distribution into account, one finds
\begin{align}
\langle \Gamma_{\gamma}(S_n,I_t\pm & 1/2,\pi_t)\rangle = \frac{B}{2\pi\rho(S_n,I_t\pm 1/2,\pi_t)} \nonumber \\
& \times \int_{E_{\gamma}=0}^{S_n}\mathrm{d}E_{\gamma}\left[\mathcal{T}_{E1}(E_{\gamma})\mathcal{P}_{-/+} + \mathcal{T}_{M1}(E_{\gamma})\mathcal{P}_{+/-}\right] \nonumber \\ 
& \times  \rho(S_n-E_{\gamma}) \sum_{J= -1}^{1} g(S_{n}-E_{\gamma},I_{t}\pm 1/2+J). 
\label{eq:newwidth}
\end{align}
for target nuclei with positive/negative ground-state parity. 
It is easily seen that if $\mathcal{P}_+ = \mathcal{P}_- = 1/2$, Eq.~(\ref{eq:width}) is restored. 

The Oslo method does not enable the separation of  $\mathcal{T}(E_\gamma)$ into its $E1$ and $M1$ components. We have therefore 
applied models for the level density and the $E1$ and $M1$ strength functions in order to investigate the influence 
of including parity on the integral in Eq.~(\ref{eq:newwidth}). For the level density, we have used the results of~\cite{go08}.
For the $E1$ strength function, we have used the the Kadmenski{\u{\i}},
Markushev and Furman (KMF) model \cite{ka83} with a constant temperature $T_f=0.3$ MeV, 
while for the $M1$ component (the spin-flip resonance~\cite{BM}) we
have applied a Lorentzian shape (see Ref.~\cite{RIPL}). We use again $^{96}$Mo as a test case, with
experimental GDR parameters taken from Ref.~\cite{RIPL} and $M1$ Lorentzian parameters from systematics~\cite{RIPL}. The target nucleus
$^{95}$Mo in the $(n,\gamma)^{96}$Mo reaction has spin/parity $5/2^{+}$. 

First, we take the parity distribution from Ref.~\cite{go08} as shown in Fig.~\ref{fig:parMo} and apply in the integral of Eq.~(\ref{eq:newwidth}).
The resulting value is only about 1\% smaller than the value obtained assuming $\mathcal{P}_+ = \mathcal{P}_- = 1/2$ for all excitation
energies, implying that the effect of parity is negligible in this case. As a further test we used the parity distribution of $^{56}$Fe
shown in Fig.~\ref{fig:parFe} on $^{96}$Mo; then, we found a $\approx 47$\% reduction using Eq.~(\ref{eq:newwidth}) 
compared to the parity-symmetry
case. For the very extreme cases assuming that $\alpha=1$ for all $E$ (allowing only for positive-parity states and $M1$ transitions), we get 
about a factor of 3 reduction, and using $\alpha=-1$ for all $E$ (allowing only for negative-parity states and $E1$ transitions), we obtain
about 67\% increase of the normalization integral in Eq.~(\ref{eq:newwidth}) relative to the case where no parity distribution is considered.

To summarize, for nuclei in the Mo mass region and above,  
we find only small corrections of the order of a few percent to the estimation 
of $\rho(S_n)$ and also for the absolute normalization of the strength function when using 
realistic parity distributions (compared to the extreme cases). 
However, for light nuclei such as Fe, effects of the 
parity distribution could be significant on both the level-density and strength-function normalization, of the order of 30--50\%. 

The parity distribution could also potentially influence the $\gamma$-ray strength function in other ways than just the normalization. 
In the quasi-continuum region, one expects
that $E1$ transitions largely dominate as long as the parity asymmetry is close to zero. However, if the parity asymmetry is large,
$M1$ transitions will be favored over those of $E1$ type. We have investigated this using DICEBOX~\cite{DICEBOX} 
to generate simulated data on an artificial 
nucleus resembling $^{57}$Fe, as already described in Sec.~\ref{subsec:fgerr}. We have applied 
a symmetric parity distribution in the first case, and an asymmetric parity distribution
in the second case. Note that the parity distribution is implemented directly in the level 
density only; the input $\gamma$-ray strength function is
kept fixed.

\begin{figure}[tb]
\centering
\includegraphics[clip,width=\columnwidth]{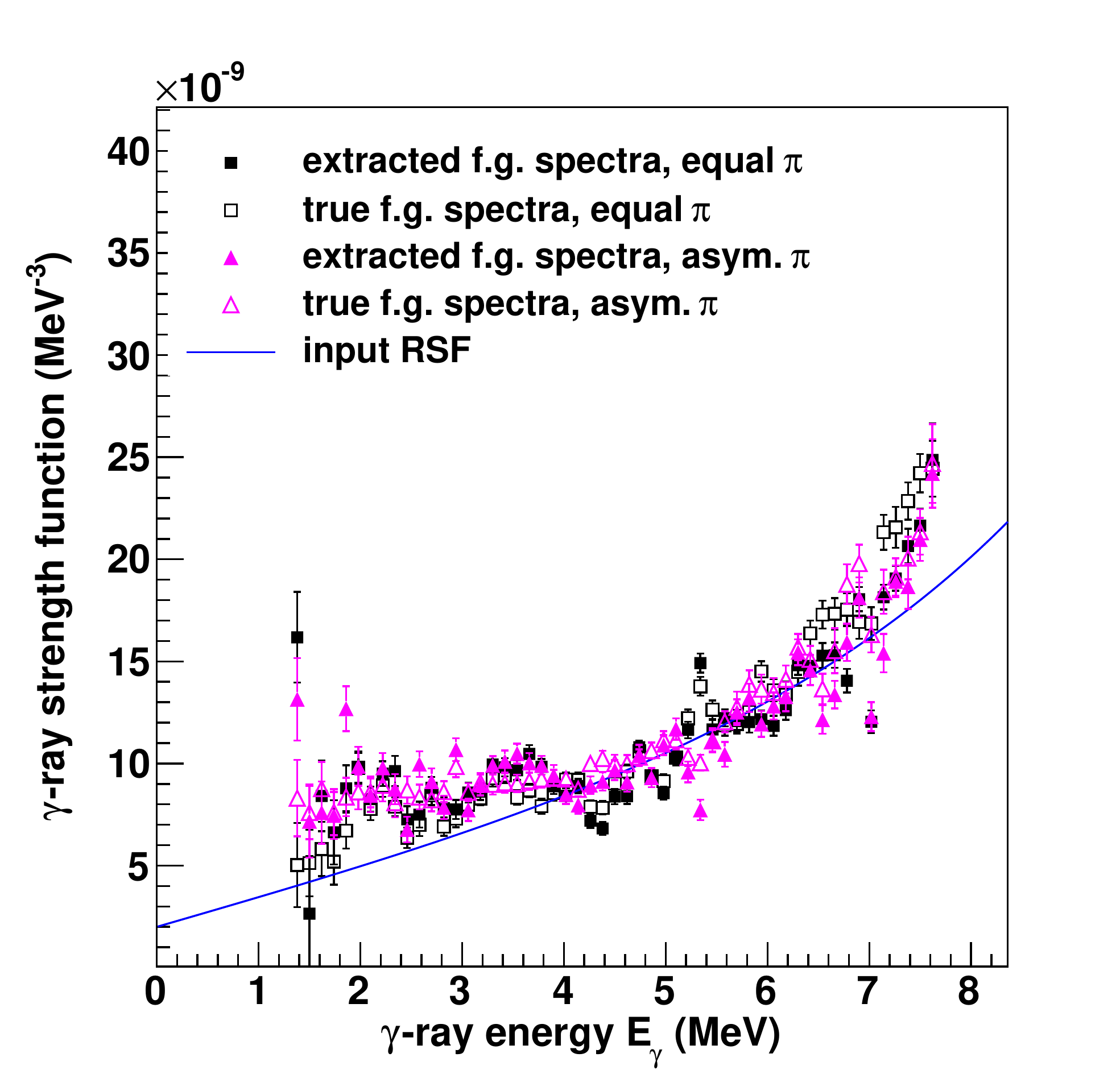}
\caption{(Color online). Extracted strength functions from simulated data for an artificial $^{57}$ Fe nucleus, with and without 
	parity asymmetry in the level density.}
\label{fig:rsfpartestCT}
\end{figure}

The resulting $\gamma$-ray strength functions are shown in Fig.~\ref{fig:rsfpartestCT}.
%-----------------------------------------------------------------%
%-----------------------------------------------------------------%
Visually, it is hard to see any difference between the $\gamma$-ray strength functions with and without parity asymmetry. 
By taking the average $\gamma$-ray strength function
for $\gamma$ energies between $1.4-3$ MeV, we find an average increase of about 20\% for the case with 
parity asymmetry relative to
the equal parity case, using the true first-generation spectra. The statistical fluctuations here are typically less than 
10\% for $E_\gamma = 2-3$ MeV, and up to 40\% for $\gamma$ rays below 2 MeV. If we use the extracted first-generation spectra
from the folded data, the difference between the two cases is approximately 6--7\%, less than the statistical fluctuations 
which are larger than when using the true first-generation spectra. For higher $E_\gamma$, there is no effect of parity 
within the fluctuations. We therefore conclude that it could be an effect of 
$\approx 20$\% on the low-energy part of the $\gamma$-ray strength function. Hence, we find it reasonable to believe that 
even a considerable parity asymmetry on the level density
as in the $^{57}$Fe case will not drastically change the shape of the $\gamma$-ray strength function.

In Fig.~\ref{fig:rsfpartestCT}, we note that there is an enhanced strength 
for $\gamma$-ray energies below $\approx 3.5$ MeV compared to the input $\gamma$-ray strength function, 
regardless of the parity distribution.
This is in fact due to the spin range of the initial levels, which is further
investigated in the following section. %Sec.~\ref{subsec:spin}. 

\subsection{The spin distribution}
\label{subsec:spin}

The uncertainty in the spin distribution may lead to errors in the Oslo method in three ways: 
(\textit{i}) the extraction of first-generation $\gamma$ rays and subsequent effects on the extracted level density and $\gamma$-ray strength function,
(\textit{ii}) the estimation of $\rho(S_n)$,
and (\textit{iii}) the spin range accessed experimentally (typically $0-8\hbar$ depending on the reaction and target spin)
compared to the true, total spin distribution. These issues will be discussed in the following.

We have once more relied on simulated data applying the DICEBOX code~\cite{DICEBOX} 
in order to test the sensitivity on the final results with respect to 
the spin range of the initial levels. 
We simulated again a light nucleus that resembles $^{57}$Fe, as we expect that light nuclei would be most sensitive 
to the initial spin population. This is because the higher spin levels are missing at low excitations. 
Again, the critical energy was set to 2.2 MeV, as there are no available data in the literature 
about levels with higher spins above this energy. 
As before, all levels had the same probability of population.
The input $\gamma$-ray strength function model was independent of excitation energy.

\begin{figure}[tb]
\centering
\includegraphics[clip,width=\columnwidth]{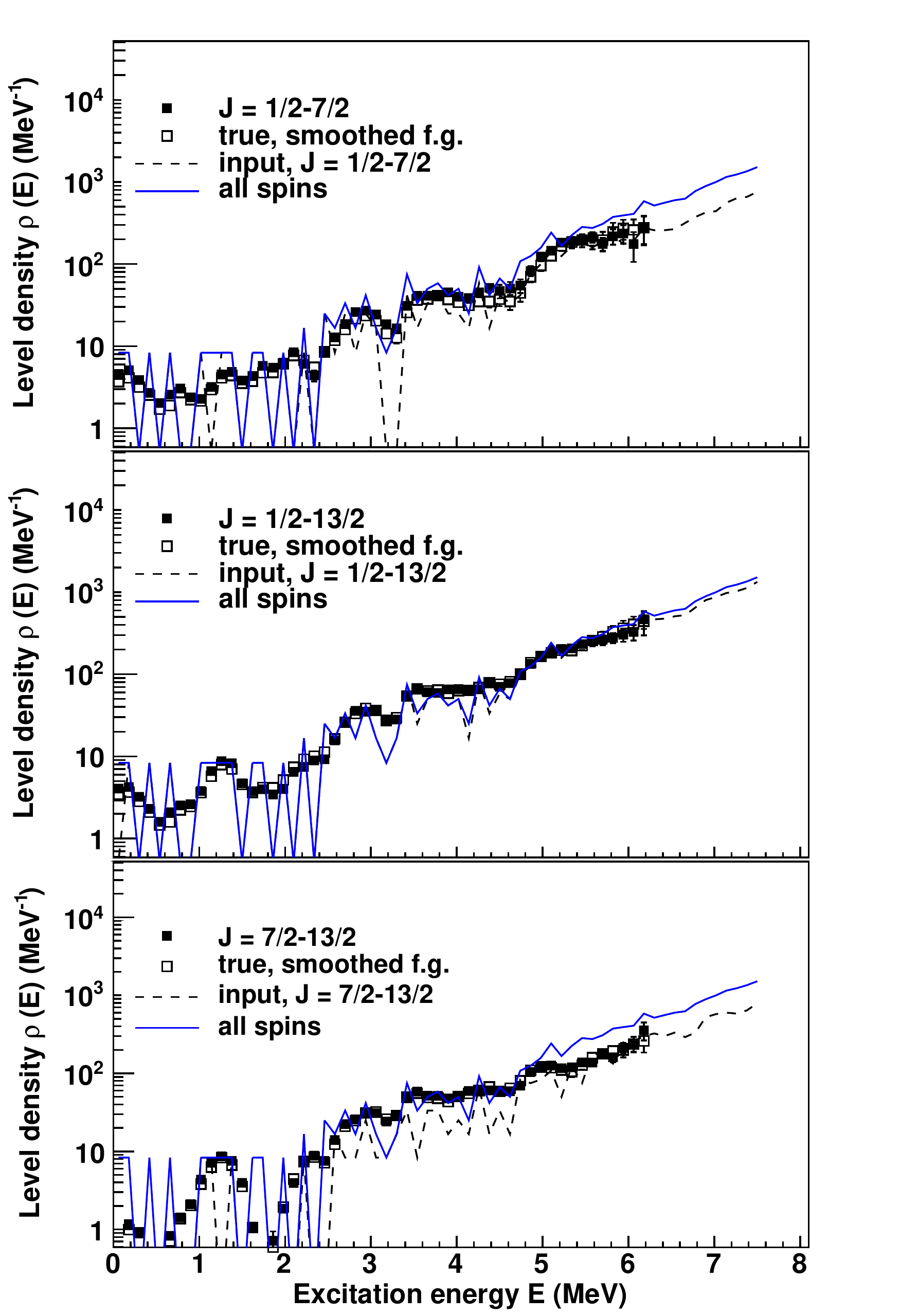}
\caption{(Color online). Extracted level densities from simulated data for initial spin range $1/2 \leq J \leq 7/2$ (top panel),
	$1/2 \leq J \leq 13/2$ (middle panel), and $7/2 \leq J \leq 13/2$ (bottom panel) on the initial levels.}
\label{fig:nldspintest}
\end{figure}

\begin{figure}[tb]
\centering
\includegraphics[clip,width=\columnwidth]{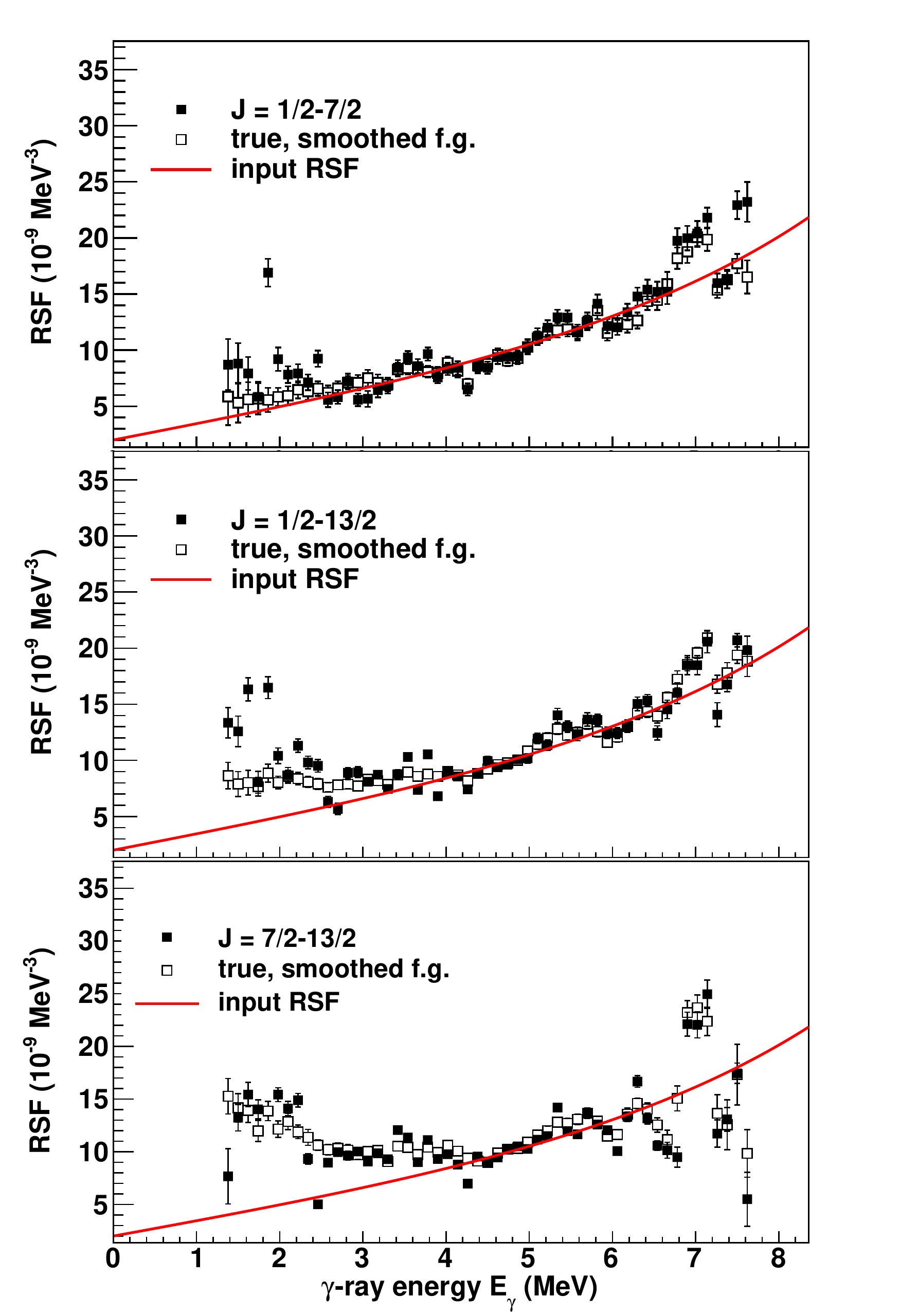}
\caption{(Color online). Extracted strength functions for spin range $1/2 \leq J \leq 7/2$ (top),
	$1/2 \leq J \leq 13/2$ (middle), and $7/2 \leq J \leq 13/2$ (bottom) on the initial levels.}
\label{fig:rsfspintest}
\end{figure}

Three spin ranges were used: $1/2 \leq J \leq 7/2$, $1/2 \leq J \leq 13/2$,
and $7/2 \leq J \leq 13/2$. This means that the highest possible spin reached at the final excitation energy 
was $9/2$ or $15/2$ depending on the maximum allowed initial spin. The weighting of the allowed spins followed the 
spin distribution of the input level density, which in this case was from the microscopic calculations of Ref.~\cite{go08}. 
For all initial spin ranges we applied the full Oslo method,
and in parallel we extracted the level density and $\gamma$-ray strength function from the true first-generation spectra.  
The resulting level densities and $\gamma$-ray strength functions are displayed in Figs.~\ref{fig:nldspintest} and \ref{fig:rsfspintest}, 
respectively.
%-----------------------------------------------------------------%
%-----------------------------------------------------------------%
%-----------------------------------------------------------------%
%-----------------------------------------------------------------%

As can be seen from Fig.~\ref{fig:rsfspintest}, there is an enhanced strength at low $\gamma$ energies 
for the spin windows including the highest spins. Also, by looking at the
average multiplicity $\langle M \rangle$ at 7.0 MeV of excitation energy, we get 1.8 for $1/2 \leq J \leq 7/2$,
2.2 for $1/2 \leq J \leq 13/2$, and 2.5 for $7/2 \leq J \leq 13/2$. At $E_{\gamma} = 2.1$ MeV, we find that
the $\gamma$-ray strength function is increased with a factor of $\approx 1.7$ for the spin range $1/2 \leq J \leq 13/2$, and a factor of  
$\approx 2.6$ for the spin range $7/2 \leq J \leq 13/2$ with respect to the input $\gamma$-ray strength function. This is not an artifact
from the unfolding or first-generation method, as it is also seen in the extracted $\gamma$-ray strength functions from the true
first-generation spectra. 

The explanation for the observed behavior is probably connected to three issues: (\textit{i}) the dominance of 
dipole radiation, which carries $L=1$; (\textit{ii}) the applied spin restriction on the initial levels;
(\textit{iii}) the low level density in this mass region at low excitation energies, especially for high spins. 
Let us take, as an example, 
the population of a 13/2 level at high excitation energy. If it is to decay to a low-lying level, involving
a high-energy $\gamma$ transition, a level with appropriate spin must be present at this excitation energy. This is not
necessarily the case as there are only about $2-10$ levels per 120 keV for excitation energies below 4.7 MeV. Among these
few levels, there are probably no high-spin levels at all.
This leads to a higher average multiplicity and an enhanced probability for decaying with low-energy $\gamma$ rays.

From Fig.~\ref{fig:rsfspintest}, it is clear that the enhancement in the extracted $\gamma$-ray strength function is 
not present in the input $\gamma$-ray strength function, 
but is an effect of the three factors mentioned above. This means that the simple factorization in Eq.~(\ref{eq:brink})
of the first-generation spectra is not able to reproduce the input $\gamma$-ray strength function. The extracted $\gamma$-ray strength function, which 
could be considered as an "effective" $\gamma$-ray strength function, is however fully capable of reproducing the true primary $\gamma$-ray spectra,
and it is seen from Fig.~\ref{fig:nldspintest} that the extracted level density is very reasonable indeed. 

However, it is not obvious whether the spin range of the inital states 
is the full explanation of the observed low-energy enhancement in light and medium-mass nuclei. 
%As discussed in Sec.~\ref{subsec:fgerr}, the experimentally populated spin range varies very little with excitation energy.
Further investigations are therefore needed to clarify this issue. It should also be noted that similar tests have
been performed for an artificial nucleus resembling $^{163}$Dy. No such enhancement was seen in this case, in 
agreement with experimental findings in this mass region. This is not surprising since in these nuclei, the level 
density is much higher and relatively high spins are available already at low excitation energies.

The quantity $\rho(S_n)$ is calculated assuming a bell-like spin distribution according to Ref.~\cite{GC} 
given by Eq.~(\ref{eq:spindist}) and using a model for the spin cutoff parameter $\sigma$ usually taken from 
Ref.~\cite{GC} or from Ref.~\cite{egidy2}. Both these assumptions could in principle be a source of uncertainty, 
as it is hard or even impossible to measure experimentally the total spin distribution at high excitation energy. 
Thus, the theoretical spin distributions are seldom constrained to data for excitation energies higher than typically
$2-3$ MeV. 

\begin{figure}[tb]
\centering
\includegraphics[clip,width=\columnwidth]{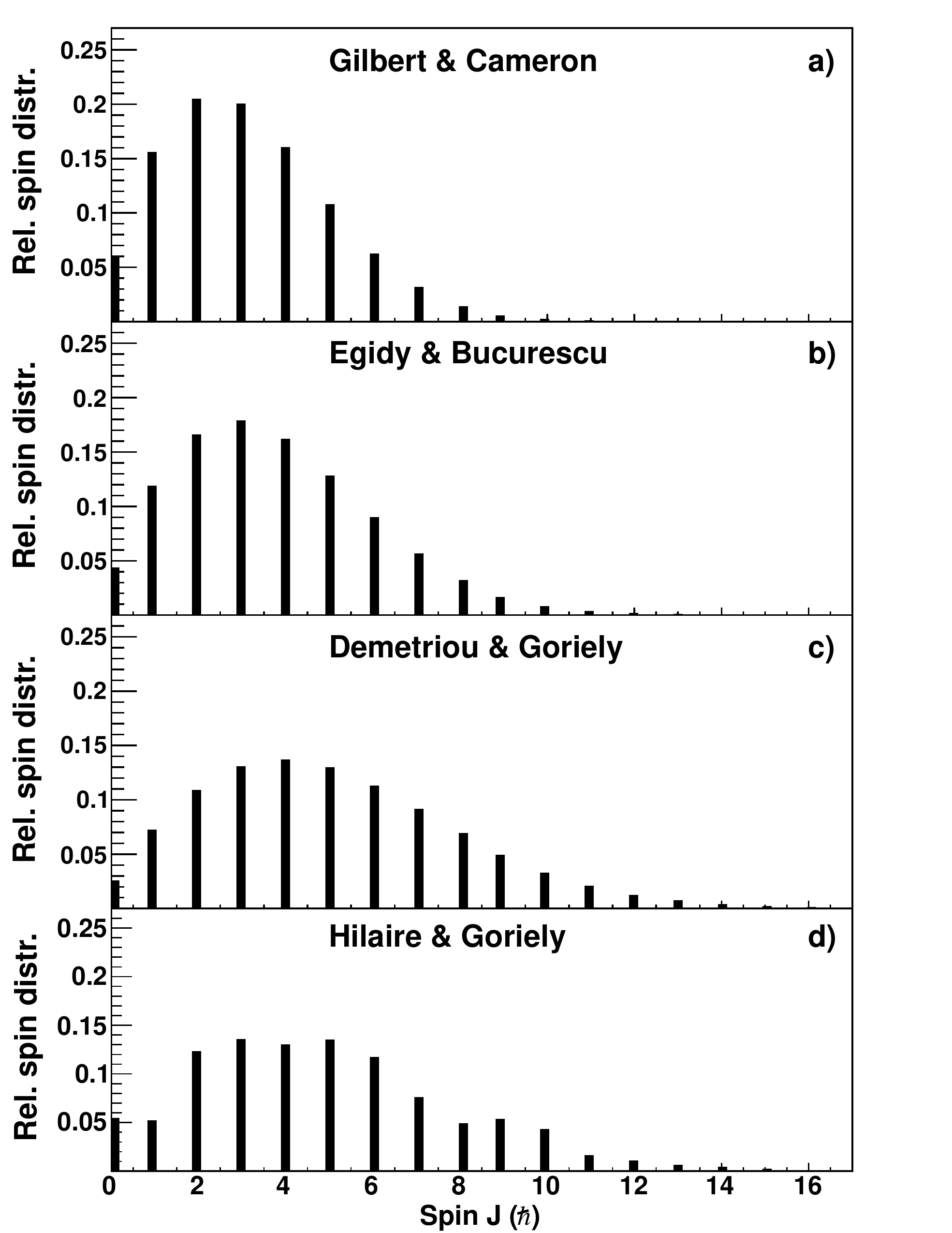}
\caption{Relative spin distributions calculated for $E=8.0$ MeV of $^{44}$Sc (see text).}
\label{fig:spindist}
\end{figure}

In Fig.~\ref{fig:spindist} various spin distributions for $^{44}$Sc are shown, calculated at an excitation energy of 8.0 MeV. 
In the two upper panels, the spin distribution given in Eq.~(\ref{eq:spindist}) has been used, but with the 
expression for the spin cutoff 
parameter of Refs.~\cite{GC,egidy1} in panel a) and the formalism of Ref.~\cite{egidy2} in panel b). 
In panel c) the spin distribution of the spin-dependent level densities of Ref.~\cite{Goriely1} are shown. Here, the authors 
also have assumed a bell-shaped spin distribution according to Eqs.~(7) and (8) in Ref.~\cite{Goriely1}. 
It is clear from the figure that the spin distributions in panel b) and c) are broader with  
centroids shifted to higher spins compared to the one in panel a). 

In panel d), the spin distribution of the calculated spin- and parity-dependent level density of Ref.~\cite{Goriely2} 
is shown. There are no underlying assumptions for the spin distribution in these calculations. It is seen from 
this distribution that there is a significant difference in the relative numbers of states with spin 0 and 1.  
The normalization procedure for the level density described above is especially sensitive to such variations at 
low spin if the neutron resonance spacing $D_0$ is measured from a neutron capture reaction where the target 
nucleus is even-even, that is, with zero ground-state spin. Then the states reached in neutron capture can only 
have spin $1/2^+$, and the number of all other states must be estimated using a certain spin cutoff parameter, 
introducing a larger uncertainty in the calculated $\rho(S_n)$. Therefore it is preferred to calculate 
$\rho(S_n)$ from both $D_0$ and $D_1$ resonance spacings if possible, since in the latter, also states with $3/2^-$ 
are reached for target nuclei with $I_t^{\pi} = 0^+$, and will therefore decrease this uncertainty.
%-----------------------------------------------------------------%
%-----------------------------------------------------------------%

\begin{figure}[tb]
\centering
\includegraphics[clip,width=\columnwidth]{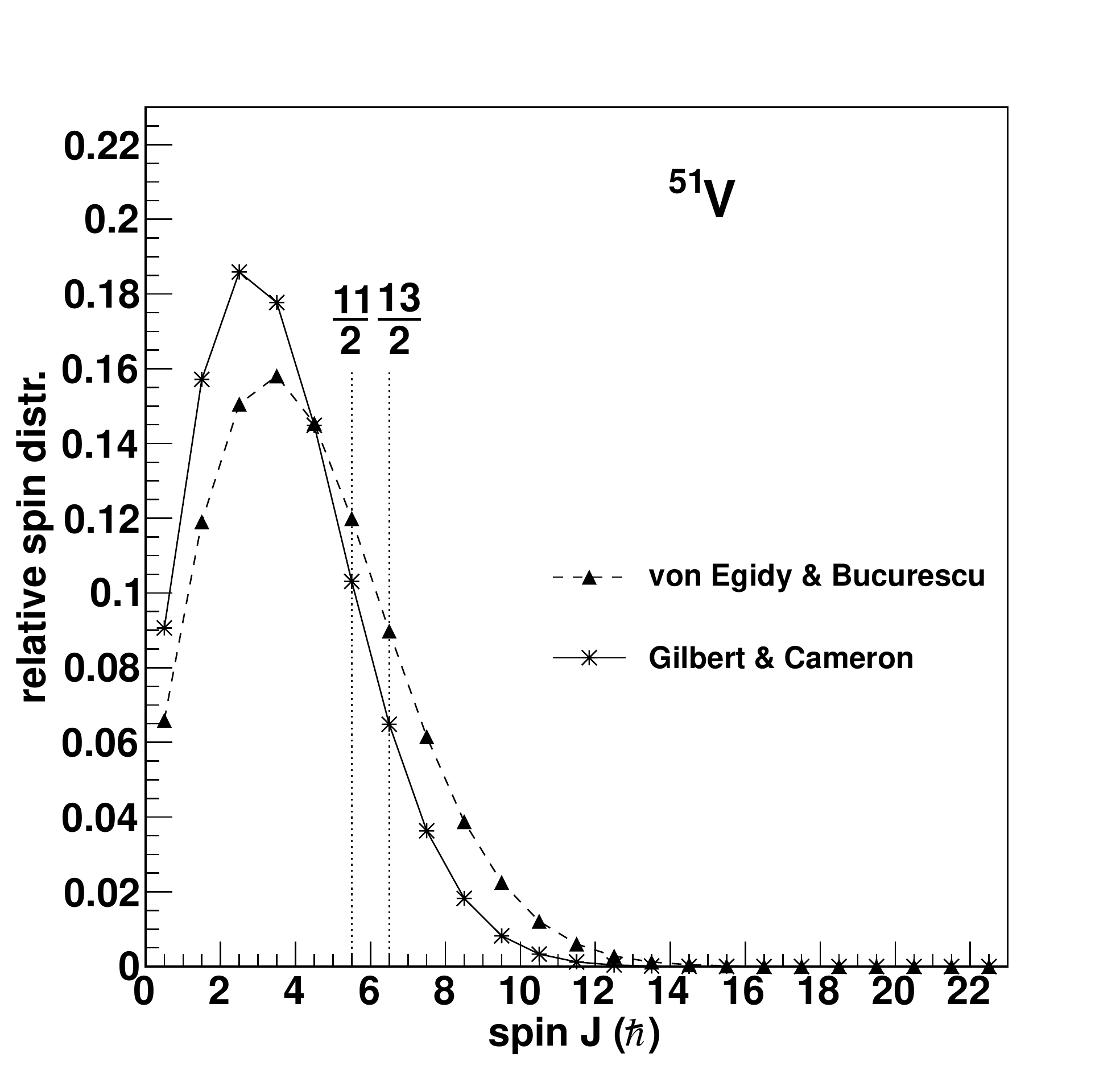}
\caption{Relative spin distributions calculated at $S_n = 11.051$ MeV of $^{51}$V (see text).}
\label{fig:spindist_V}
\end{figure}

We have investigated the relative difference in the value of $\rho(S_n)$ for several nuclei using the spin-cutoff parameter
of \cite{GC} with global parameterization of \cite{egidy1} (referred to as Gilbert \& Cameron), and the prescription of \cite{egidy2} 
(hereafter called von Egidy \& Bucurescu),
see Tab.~\ref{tab:parameters}. For the cases investigated here, the general trend is that the Gilbert \& Cameron approach leads to 
a lower value for $\rho(S_n)$ than the von Egidy \& Bucurescu parameterization. However, this depends on the spins reached in the 
$(n,\gamma)$ reaction. For example, for $^{51}$V, we note that the obtained $\rho(S_n)$ using the spin cutoff of von Egidy \& Bucurescu 
is lower than that of Gilbert \& Cameron.
This is easily understood  from the fact that in this case, relatively high spins are populated in the ($n,\gamma$) reaction. 
We see from Fig.~\ref{fig:spindist_V} that the Gilbert \& Cameron 
parameterization gives lower relative values for states with spin 11/2 and 13/2 compared to the von Egidy \& Bucurescu distribution. Thus, 
one divides by a smaller number in Eq.~(\ref{eq:oldD}) and $\rho(S_n)$ gets larger. 
%In the case of $^{74}$Ge, the two approaches predict
%very similar intensities for the populated spins, giving a relative difference of $\approx 8$\% only. 
\begin{table*}[htb]
\caption{Variation of the calculated $\rho(S_n)$ using different spin cutoff parameters. The target spin in the $(n,\gamma)$ reaction is 
denoted by $I_t^{\pi}$, and $E_1$ is the total backshift for the back-shifted Fermi-gas model while $a$ is the level-density parameter. 
All level spacings ($D_0$) are taken from
\cite{RIPL}. The parameter $\eta$ is the ratio  
$\rho_{EB}(S_n)/\rho_{GC}(S_n)$.} 
\begin{tabular}{lccc|cccc|cccc|c}
\hline
\hline
	&	&	&	 & \multicolumn{4}{|c|}{Gilbert \& Cameron} & \multicolumn{4}{|c|}{von Egidy \& Bucurescu} &  \\
Nucleus    & $I_t^\pi$ & $S_n$ & $D_0$ & $a$        & $E_{1}$ & $\sigma$ & $\rho_{GC}(Sn)$ & $a$   &  $E_{1}$ & $\sigma$ & $\rho_{EB}(Sn)$ & $\eta$ \\
           &             & (MeV) & (keV) & (MeV$^{-1}$) & (MeV) & &   (MeV$^{-1}$)    & (MeV$^{-1}$)  &  (MeV) & & (MeV$^{-1}$) & \\
\hline
$^{51}$V   & 6$^+$ & 11.05 & 2.3(6) & 6.42 & $-0.511$ & 3.24 & $5.18\times10^{3}$ & 6.17 & $-0.153$ & 3.83 & $4.15\times10^{3}$ & 0.80\\
$^{57}$Fe  & 0$^+$ & 7.646 & 25.4(22) & 7.08 & $-0.910$ & 3.20 & $8.46\times10^{2}$ & 6.58 & $-0.523$ & 3.83 & $1.20\times10^{3}$ & 1.41 \\
%$^{74}$Ge  & 9/2$^+$ & 10.20 & 0.062(15) & 8.88 & 1.674 & 3.69 & $1.11\times10^{5}$ & 10.03 & 1.021 & 4.38 & $1.20\times10^{5}$ & 1.08 \\
$^{96}$Mo  & 5/2$^+$ & 9.154 & 0.105(10) & 11.14 & 1.016 & 4.21 & $7.38\times10^{4}$ & 11.39 & 0.779 & 5.15 & $1.01\times10^{5}$ & 1.37\\
$^{117}$Sn  & 0$^+$ & 6.944 & 0.380(130) & 13.23 & 0.197 & 4.48 & $1.08\times10^{5}$ & 13.62 & $-0.210$ & 5.58 & $1.67\times10^{5}$ & 1.54\\
$^{164}$Dy  & 5/2$^+$ & 7.658 & 0.0068(6) & 17.75 & 0.416 & 5.49 & $1.74\times10^{6}$ & 18.12 & 0.310 & 6.91 & $2.59\times10^{6}$ & 1.49 \\
\hline
\hline
\end{tabular}
\\
\label{tab:parameters}
\end{table*}
%-----------------------------------------------------------------%
%-----------------------------------------------------------------%

One can conclude that the two approaches studied here give a $\approx 10 - 50$\% change in the resulting $\rho(S_n)$, 
while other calculations~\cite{go08} indicate even larger variations. Using this HFB-plus-combinatorial approach gives 
typically a factor of two change compared to $\rho_{GC}(S_n)$ for most cases and up to 
a factor of 3.7 for the extreme case of $^{172}$Dy (see Tab.~II in Ref.~\cite{go08}). The effect of these differences
is of course an uncertain determination of $\rho(S_n)$ and thus of the slope of the level density and the $\gamma$-ray strength function, in a similar 
fashion as for the parity dependence (see Sec.~\ref{subsec:parity} and Fig.~\ref{fig:Mopartest}).  

%\begin{table*}
%\caption{Input parameters for the model calculations.} 
%\begin{tabular}{ccccccccccc}
%\hline
%\hline
%Nucleus   &$\epsilon_2$& $\kappa$ & $\mu$ &$\Delta_{\pi}$&$\Delta_{\nu}$& $A_{\rm rot}$ & $\hbar \omega _{0}$ & $\hbar \omega _{\rm vib}$ & $\lambda_{\pi}$ &  $\lambda_{\nu}$\\ 
%          &            &          &       & (MeV)   &     (MeV)    &      (MeV)    &  (MeV)  &       (MeV)  & (MeV) & (MeV)\\
%\hline
%&&&&&\\
%$^{56}$Fe &     0.240   &  0.0660 & 0.32 & 1.568       &  1.363       &    0.120      &   10.72 & 2.66; 3.07       & 45.89 & 48.23  \\
%$^{164}$Dy &     0.348   &  0.0637 & 0.42& 0.875       &  0.832       &    0.012      &   7.49 & 0.76; 1.04       & 43.03 & 48.77    \\
%\hline
%\hline
%\end{tabular}
%\\
%\label{tab:tab2}
%\end{table*}

There is also a question of how the spin range accessed experimentally relative to the true, total spin range might influence
structures in the level density. 
In the analysis, we assume that the excluded spins contribute on average with a scaling factor of the level density, which is 
corrected for by normalizing to $\rho(S_n)$, i.e. the total level density for all spins at this energy. However, this 
relies on the hypothesis that structures in the level density due to, e.g., nucleon pair breaking are not severely affected by the 
spin window. 

To test this,
we have performed simplistic calculations with the code COMBI~\cite{Ti45}, which is based on a microscopic, combinatorial model using
Nilsson single-particle energies and the concept of quasiparticles from the nuclear Bardeen-Cooper-Schrieffer (BCS) theory. 
The chosen spin windows were $0\leq J \leq 6$ and $0\leq J \leq 30$ (in units of $\hbar$).
Details of the model can be found in Ref.~\cite{Ti45}. 
%, and the applied input parameters for the test cases $^{56}$Fe and
%$^{164}$Dy can be found in Tab.~\ref{tab:tab2}. 

\begin{figure}[tb]
\centering
\includegraphics[clip,width=\columnwidth]{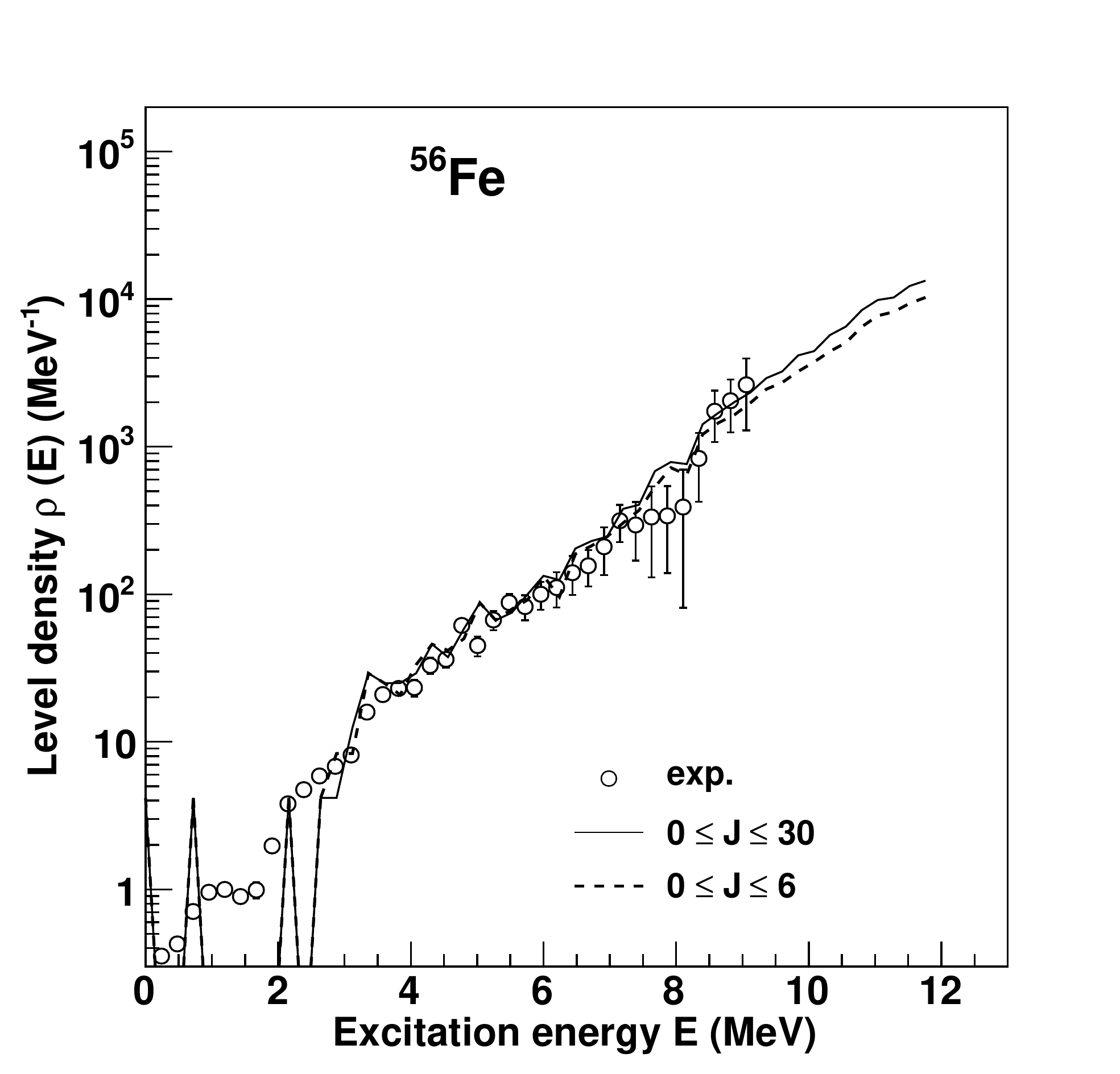}
\caption{Calculated level density of $^{56}$Fe for two spin ranges: $0 \leq J \leq 6$ (dashed line)
	and $0 \leq J \leq 30$ (solid line) compared to experimental data (open circles, from Ref.~\cite{Fe_Emel}) normalized to the 
	total level density at $S_n$.}
\label{fig:spinFe}
\end{figure}

\begin{figure}[tb]
\centering
\includegraphics[clip,width=\columnwidth]{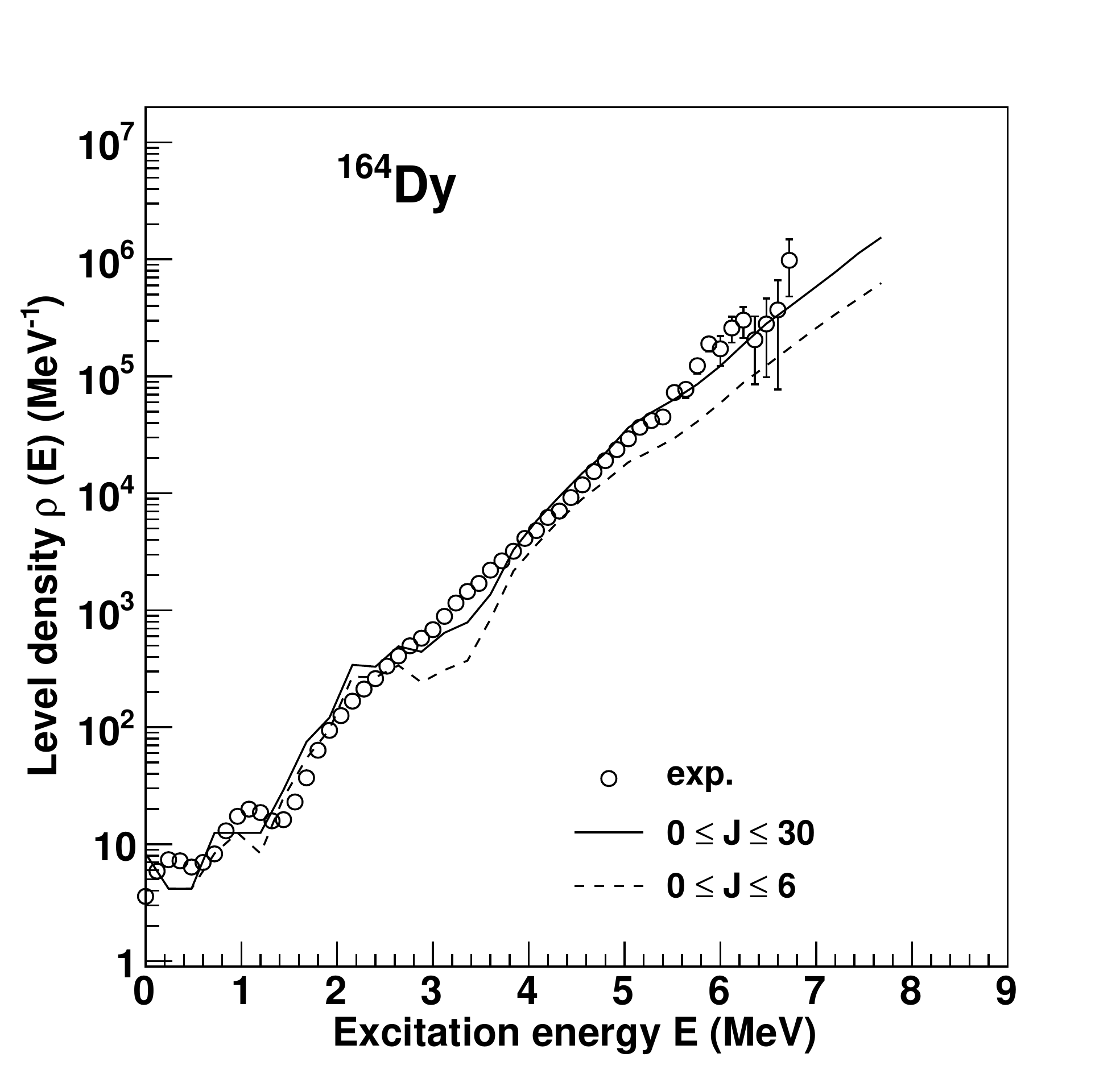}
\caption{Same as Fig.~\ref{fig:spinFe} for $^{164}$Dy, with experimental data from Ref.~\cite{hildes_DyRSF}.}
\label{fig:spinDy}
\end{figure}

The resulting calculated level densities for $^{56}$Fe and $^{164}$Dy are displayed in Figs.~\ref{fig:spinFe} and \ref{fig:spinDy},
respectively. 
%-----------------------------------------------------------------%
%-----------------------------------------------------------------%
%-----------------------------------------------------------------%
%-----------------------------------------------------------------%
For $^{56}$Fe, is it clear that most of the states are found within the spin range of $0-6\hbar$. In fact, the relative difference
between the large and the small spin window is at most $\approx 30$\%, and it is clear that there are no significant
structural differences for the two cases. One would therefore not expect any severe problems
with normalizing the level density of such light nuclei to the total level density $\rho(S_n)$. 
However, for the $^{164}$Dy case, the deviation of 
the level densities calculated within the two spin ranges can be considerable and is increasing with excitation energy, 
up to a factor of $\approx 2.5$ for $E\approx 7.5$
MeV. It is nevertheless very gratifying that, indeed, the gross structures of the level densities are very much alike, so
that extracted information from the experimental level density, such as the onset of the pair-breaking process,
is probably reliable (the larger spin window leads to a smoothing of the steps). 
However, experimental work to investigate the spin distribution further is highly desirable.

\section{Summary and conclusions}
\label{sec:sum}

In this work we have addressed uncertainties and possible systematic errors that one can encounter using the Oslo method.
The main steps of the method have been outlined, and the assumptions behind each step are investigated
in detail. Our findings indicate that although the Oslo method is in general very robust, the various assumptions
it relies on must be carefully considered when it is applied in different mass regions. The results can be summarized as follows.

\textit{The unfolding procedure.} We have investigated the effect of changing the total absorption efficiencies
of the NaI(Tl) crystals up to $\approx 20$\%. We found that the extracted level density is hardly affected at all, while the $\gamma$-ray strength function
gets a somewhat different slope. The quantitative effect on the $\gamma$-ray strength function corresponds directly to the imposed changes
in the efficiencies.  

\textit{The first-generation method.} In general, the assumption of thermalization can be 
questioned at excitation energies above $\approx 10$ MeV. Also, variation in the populated spin range as a function
of excitation energy may lead to an erroneous subtraction, especially of yrast states and other strong transitions 
at low $\gamma$-ray energies. This is particularly pronounced at low excitation and low $\gamma$-ray energies, 
where the direct reaction 
may favor or suppress the population of some levels compared to feeding from $\gamma$ decay. 
In addition, the finite, and for some energies, the mismatch of the detector resolution of the particle
telescopes and CACTUS might in principle lead to problems with determining correct weighting functions. 
However, tests on simulated spectra show that this effect is probably of minor importance. 
The two different normalization options in the method turn out to give very similar results. 

\textit{The Brink hypothesis.} Since the extraction of level density and $\gamma$-ray strength relies 
on the Brink hypothesis, one would expect problems
if there was an effective temperature dependence in the $\gamma$-ray strength function. 
From the results of simulated data, it is clear that especially the region of low $E_{\gamma}$ is affected
when the $\gamma$-ray strength function is temperature dependent, and the extracted strength is found to found to lie in between
the two temperature extremes considered. 
However, we note that there is, so far, no experimental evidence for any strong temperature
dependence of the $\gamma$-ray strength function in the excitation-energy region in question (below $\approx 10$ MeV). 

\textit{The parity distribution.} The assumption of equally many positive- and negative-parity states may 
break down, especially for light nuclei, and would then affect the normalization of both the level density and the 
$\gamma$-ray strength function. If the parity asymmetry is large, the value of the normalization point $\rho(S_n)$ might change with a factor of
$\approx 2$, which will consequently change the slope of the level density and $\gamma$-ray strength function. 
For heavier nuclei, however, the parity distribution at $S_n$ is expected to be quite even, and experimental data 
on lighter nuclei suggest that the parity asymmetry should be small also here. Therefore, we find it reasonable to believe
that the error in $\rho(S_n)$ and on the absolute normalization of the $\gamma$-ray strength function due to this assumption does not exceed $50$\%.

Another issue is the influence on the relative contribution
of $E1$ and $M1$ transitions, as a large parity asymmetry will favor $M1$ transitions. Tests with simulated data
indicate, however, that the effect is not large even with a considerable parity asymmetry.

\textit{The spin distribution.} From simulations on light nuclei with different spin ranges on the initial levels, it is seen that
the ranges including higher spins might lead to an enhanced $\gamma$-ray strength function at low $\gamma$-ray energies. This
is understood from considering the relatively low level density and the matching of spins between the
initial and final levels for dipole radiation. This could explain, at least partially, the observed
enhanced low-energy strength in experimental data for light and medium-mass nuclei. 

The spin distribution is one of the largest uncertainties in the normalization, since the determination of the 
slope of the level density and $\gamma$-ray strength function strongly depends on the relative intensities of the populated spins at $S_n$. Assuming 
a bell-shaped spin distribution with various global parameterizations of the spin cutoff parameter give up to $\approx 50$\%
change in $\rho(S_n)$; however, microscopic calculations indicate that this value can vary within a factor of 2 or more,
especially for nuclei in the rare-earth region. This will also consequently influence the $\gamma$-ray strength function. 

The experimentally reached spin 
window in OCL experiments is typically $0-8\hbar$, which means	that the extracted level-density data in fact represent this narrow spin range. This is usually
compensated for by normalizing to the total level density at $S_n$. This is not a big effect for light nuclei with relatively
few high-spin states. Even in the case of rare-earth nuclei, our calculations indicate that this scaling works well
as the main structures in the level density are indeed present in the levels of a rather narrow spin range.

\vspace{5cm}

\section{Acknowledgments}
The authors wish to thank E.~A.~Olsen and J.~Wikne for 
providing excellent experimental conditions. 
Financial support from the Research Council of Norway (NFR), project no. 180663, is 
gratefully acknowledged.

\vfill
\end{document}